\documentclass[
 aps,
prb,
  twoside,
  twocolumn,
  floatfix,
]{revtex4}
\usepackage{amsmath}
\usepackage{amssymb}
\usepackage{color}
\usepackage{graphicx}
\usepackage{tabularx}
\usepackage{dcolumn}
\usepackage{footnote}
\usepackage[normalem]{ulem}

\usepackage{xr}
\usepackage{cleveref}

\usepackage{float}
\usepackage{fontenc}

\usepackage{setspace}

\usepackage[utf8x]{inputenc}

\usepackage[toc, title, titletoc]{appendix}

\begin{document}

\newcommand{\bec}{\begin{center}}
\newcommand{\ec}{\end{center}}
\newcommand{\be}{\begin{equation}}
\newcommand{\ee}{\end{equation}}
\newcommand{\beqn}{\begin{eqnarray}}
\newcommand{\eeqn}{\end{eqnarray}}
\newcommand{\bet}{\begin{table}}
\newcommand{\ent}{\end{table}}
\newcommand{\bib}{\bibitem}

% commands for structuring
\newcommand{\sect}[1]{Sect.~\ref{#1}}
\newcommand{\fig}[1]{Fig.~\ref{#1}}
\newcommand{\Eq}[1]{Eq.~(\ref{#1})}
\newcommand{\tab}[1]{Table~\ref{#1}}

\newcommand{\angstrom}{\mbox{\normalfont\AA}}

% commands for mathmode
\renewcommand{\vec}[1]{\ensuremath\boldsymbol{#1}}
\renewcommand{\epsilon}[0]{\varepsilon}

% comments at the margin
\newcommand{\cmt}[1]{\emph{\color{red}#1}%
\marginpar{{\color{red}\bfseries $!!$}}}

%\newcommand{\beginsupplement}{%
%	\setcounter{table}{0}
%	\renewcommand{\thetable}{S\arabic{table}}%
%	\setcounter{figure}{0}
%	\renewcommand{\thefigure}{S\arabic{figure}}%
%}

% make supplementary sections in chapters be numbered with letters
\newcommand{\opensupplement}{
	\setcounter{section}{0}
	\renewcommand\thesection{\thechapter.\Alph{section}}
}
% restore the numbering system to arabic for the following chapter's sections
\newcommand{\closesupplement}{
	\renewcommand\thesection{\thechapter.\arabic{section}}
}

%\DeclareUnicodeCharacter{00A0}{ }
\nobreakspace

%\topmargin 0.5cm
%\setlength{\textwidth}{18cm} 
%\setlength{\textheight}{25.5cm} 

%\baselineskip 4mm 
%\wideabs
\title{Synergistic Self-Driven Nanoconfinement and Dynamical Heterogeneity in Deeply Supercooled "No Man's Land" Water	
}

%\LARGE{

\author{P\'eter S\"ule} 
\email{sule.peter@ek-cer.hu}

\affiliation
  {Centre for Energy Research, HUN-REN, Research Institute for Technical Physics and Material Science, Dept. of Nanostructures,\\
Konkoly Thege u. 29-33, Budapest, Hungary, sule.peter@ek-cer.hu % www.mfa.kfki.hu/$\sim$sule,\\
}

\date{\today}

%}
%\maketitle

\begin{abstract}  
The 150–235 K "no man’s land" (NML) of deeply supercooled water is a highly debated regime where rapid crystallization hinders experimental observation of the hypothesized liquid–liquid phase transition. Using TIP4P/2005 molecular dynamics simulations, we uncover a structurally and dynamically heterogeneous state at ambient pressure that exhibits novel physical features. Our results demonstrate that this low-density state ($\rho \approx 0.925$ g/cm$^3$) is a unique supercooled liquid, distinct from both glassy and spinodal states. Specifically, it emerges as a stable, structurally arrested phase driven by a self-generated "soft skeleton" of nanoclusters. This network forms an intrinsic nanoporous structure that imposes nanoconfinement on the remaining fluid, proving that NML confinement can be an entirely self-driven phenomenon. Finally, we resolve the temperature-dependent, nonlinear behavior of density and compressibility, highlighting a structurally driven anomaly that clarifies the thermodynamics of deep supercooling. Collectively, this work provides a transformative structural framework for understanding the anomalous nature of water within the NML.
%\pacs{68.35.-p, 79.20.Rf, 81.65.Cf, 61.82.Fk, 96.35.Gt}
\end{abstract}
%}

\maketitle

\section{Introduction}

 Water's anomalous properties—including its density maximum and negative thermal expansion—arise from a dynamic, two-fluid mixture of low-density and high-density liquid structures \cite{Bernal,Pauling,Debenedetti01,Nilsson04,Nilsson15,Angell08,Russo14,Clark10,Soper00,Laage15,Finney02,Mishima98,MolineroNat11,Gallo16,Palmer14,Palmer18}. The complex, partially ordered network of hydrogen-bonding nanoclusters, particularly in confined geometries, enables self-driven nanoconfinement as a strategy for developing functional materials \cite{NatMat26,Kapil22,Molinero18,Mallamace16,Kruger23,Zublin26}. Studies in supercooled "no man’s land" (150–235 K) reveal a reversible structural transition and divergence of response functions near $\sim 228$ K, reinforcing the dynamic nature of water's hydrogen-bonding network \cite{Gallo16,Kringle20,Nilsson14,Angell08,Mishima98,Kruger23,Smith99}.

%   ultraviscous
  
  Many experiments suggest that amorphous ice undergo glass-to-liquid transition at 136 K entering the {\em ultraviscous} low density liquid (LDL) regime \cite{Amann-Winkel13,Amann-Winkel16}.
  This state is characterized by extremely sluggish dynamics and high viscosity \cite{Gallo16,Amann-Winkel13,Amann-Winkel16}.
  The existence of an LDL ultraviscous state is a key aspect of the controversial liquid-liquid phase transition (LLPT) hypothesis, which posits that water can transform between a high-density liquid (HDL) and this low-density ultraviscous liquid (LDL) before vitrifying into amorphous ice \cite{Poole,Gallo16,Palmer14}.

   Simulations of liquid-liquid transition (LLT) predict its critical point to be somewhere in the middle of water's no man's land (NML), however, its very occurrence has remained speculative \cite{Nilsson14}.
   Since spontaneous freezing in NML is inevitable it is important to test various theories concerning LLT with the simulations of
   freezing.  However, the brute force (unbiased) simulation of spontaneous ice nucleation is still challenging using molecular dynamics \cite{Matsumoto,Vrbka,MolineroNat11}.
   Instead nucleation should be triggered by various specific methods to initialize freezing event within reasonable time.
   Just to mention a few, e.g. using seeds \cite{Bai,Pereyra11,Espinoza13} it is possible to follow the growth of small inserted embryos. 
   Biased but seedless approaches such as the Lattice Mold (LM) technique \cite{Mold,LM14,LMVega,LM} is a more reliable way of simulating water freezing in many respect. It is possible to induce nucleation without the loss of spontaneity as well as
   a robust speedup can be reached in the crystallization process \cite{Mold,LM14,LMVega,LM}.
   Other techniques, including  Forward Flux Sampling, Transition Path Sampling, Metadynamics and Umbrella Sampling should also be mentioned which efficiently initialize nucleation in supercooled water \cite{Parrinello08,Laio,Russo14,Parrinello19,Chen25,Pietrucci17,Pietrucci23,Nada20}.
   Recent enhanced sampling does not detect a sharp free-energy discontinuity or barrier separating LDL and HDL in the specific temperature-pressure zone of the NML \cite{Pietrucci23}.
   
   	Limmer and Chandler argue that what others misinterpret as a transition between two liquid states in the no man's land is actually a sluggish, slow-moving liquid-to-solid crystallization transition \cite{LimmerPNAS14}.
      
   The study of Offei-Danso et al. demonstrates that capturing the complex behavior of liquid water's hydrogen bond network requires a hybrid approach combining data-driven and physics-based variables. It establishes a comprehensive framework using unsupervised learning to show that water network fluctuations occur across a high-dimensional space rather than a simple pathway \cite{Hassanali22,Hassanali24}.
    
  The study of Monserrat et al. demonstrates that the local atomic environments characteristic of a wide variety of distinct ice phases are already present within liquid water \cite{Monserrat20}.

      The Lattice Mold (LM) technique is a computational approach that employs tailored potential energy wells—acting as a template of the target solid state—to induce nucleation from a metastable liquid under conditions where spontaneous nucleation is unfeasible \cite{Espinoza13}. To further enhance efficiency, we adopt a full-volume mold rather than pre-critical spherical molds \cite{LM14}, enabling the simulation of nucleation in systems exceeding $\sim 50-100k$
      water molecules within reasonable computational time.

%  dynamical heterogeneity:

  Dynamic heterogeneity (DH) represents the non-uniform spatial distribution of particle mobility, characterized by coexisting regions of fast and slow dynamics that fluctuate over time \cite{Berthier2,Royall}. It is marked by intermittent, "jump-like" motion, broad distributions of waiting times, and highly individualistic particle trajectories  \cite{Daschakraborty26}. Such systems typically exhibit non-Gaussian displacement distributions and a breakdown of the Stokes-Einstein relation \cite{Berthier2,Tanaka25}. Notably, this complex behavior, including DH, is observed in water \cite{Fitzner19,Malosso25}.
  
  In supercooled water, dynamic heterogeneity arises as molecular clusters become progressively immobilized ("stuck") at different rates, a process known as arrested dynamics \cite{Fitzner19,Malosso25,Hassanali23,Saito24}. These slow-moving regions correlate with higher angular order and act as precursors to ice nucleation \cite{Malosso25,Hassanali23,Tanaka25}. Beyond being a hallmark of the supercooled state associated with the liquid-liquid transition \cite{Gallo16,Fitzner19}, this dynamic heterogeneity also dictates the quasi-liquid state in supercooled iron \cite{psule24} .

    Deeply supercooled water (DSW) in the 150–225 K "no-man's land" exhibits an intermediate, previously unrecognized quasi-liquid state (QLS, see for brief summation in SI) arising from spontaneous nanoconfinement and structural arrest of networked, immobile water nanoclusters. This self-driven, long-lived ($\mu$s-timescales) nanoporous phase bridges the gap between liquid and glassy states by restricting particle mobility and forming a 'soft-skeleton' that persists before crystallization. 
 
 %%%%%%%%%%%%%%%%%%%%%%%%%%%%%%
 % F1
 %%%%%%%%%%%%%%%%%%%%%%%%%%%%%
 \begin{figure*}[hbtp]
 	 	
 	\includegraphics*[height=7cm,width=10cm,angle=0.]{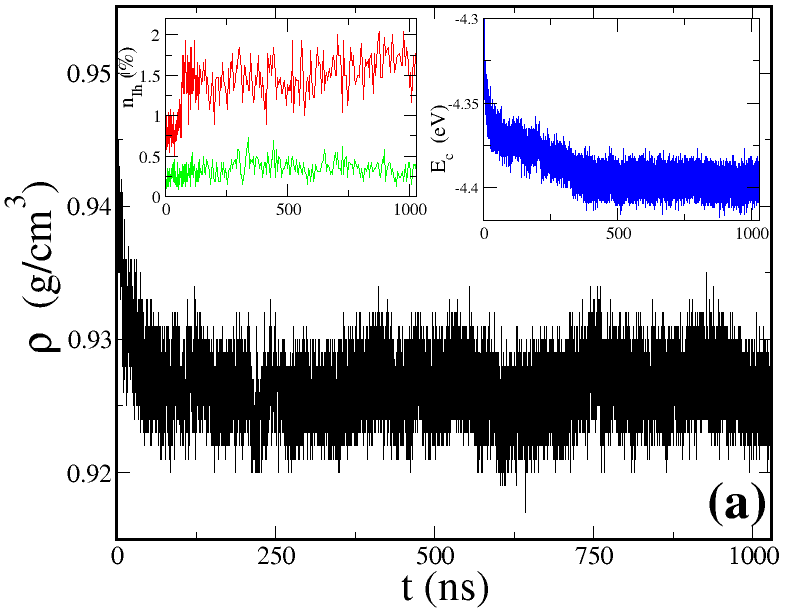}
 	
 	\includegraphics*[height=5cm,width=8.5cm,angle=0.]{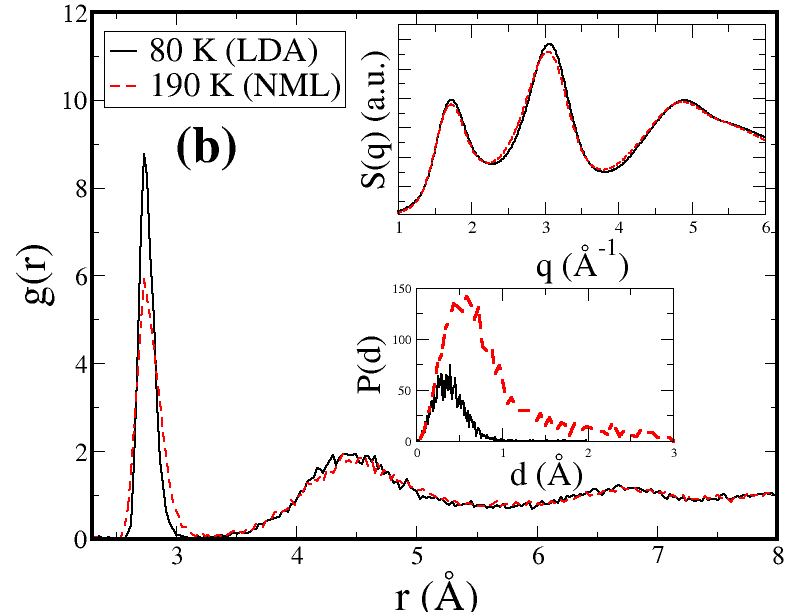}
 	\includegraphics*[height=5cm,width=8.5cm,angle=0.]{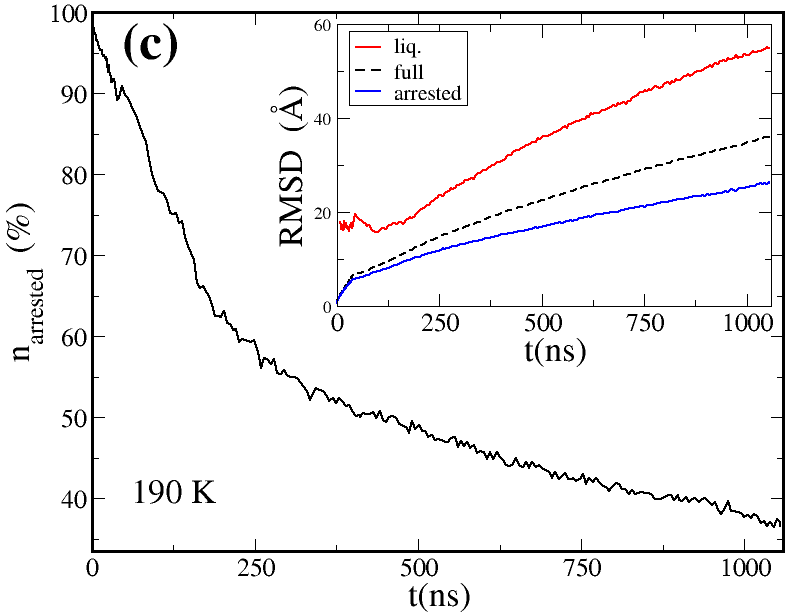}
 	
 	\caption[]{
 		Benchmark unbiased MD simulations for the low-density quasi liquid state\textbf:
 		(a) The time evolution of the density (g/cm$^3$), the proportion of Ih (ice) atoms ($n_{Ih}$, \%, red curve, Inset on the left) and potential energy ($E_c$, eV/atom, Inset on the right) at 190 K including 2592 water molecules in a periodic cubic box.
 		The proportion of cubic atoms as detected by Chill+ is also shown in Inset on the left as a green curve.
 		(b) Typical radial density function g(r), structure factor (S(q)) and the histogram of atomic displacements $d$.
 		The various curves are obtained for 80 and 190 K, respectively.
 		The structure obtained at 80 K after employing npt-MD simulation (TIP4P/2005) using slow quench rate ($\tau \approx 2 \times 10^{10}$ K/s) initialized from the carefully relaxed structure at 190 K ($\sim 500$ ns) and is considered as LDA phase ($\rho=0.942$ g/cm$^3$).
 		Simulation box includes 2592 water molecules.
 		{\em 1st Inset Fig.:} The structure factor S(q) as a function of wave number $q$.
 		{\em 2nd Inset Fig.:} The distribution of atomic displacements P(d) as a function of d(\r{A}) (self part of the van Hove correlation function).
 		(c) The fraction of the arrested particles ($n_{arrested}$(\%)) as a function of time (ns).
 		{\em Inset}: the root mean square of atomic displacements (RMSD) as a function of time $t$(ns).
 		Liq. and arrested correspond to the data collected for mobile and arrested particles using the threshold atomic displacement of $d_i=2.5$\r{A} (liq.: $d_i>2.5$ \r{A}, arrested: $d_i \le 2.5$ \r{A}). Full denotes the RMSD calculated for the full system.

 	}

 	\label{grsqrdispl80K-190K}
 \end{figure*}
 
  %%%%%%%%%%%%%%%%%%%%%%%%%%%%%%
  % F2 
  %%%%%%%%%%%%%%%%%%%%%%%%%%%%%
  \begin{figure*}[hbtp]
  	 	
  	\includegraphics*[height=3.2cm,width=4.4cm,angle=0.]{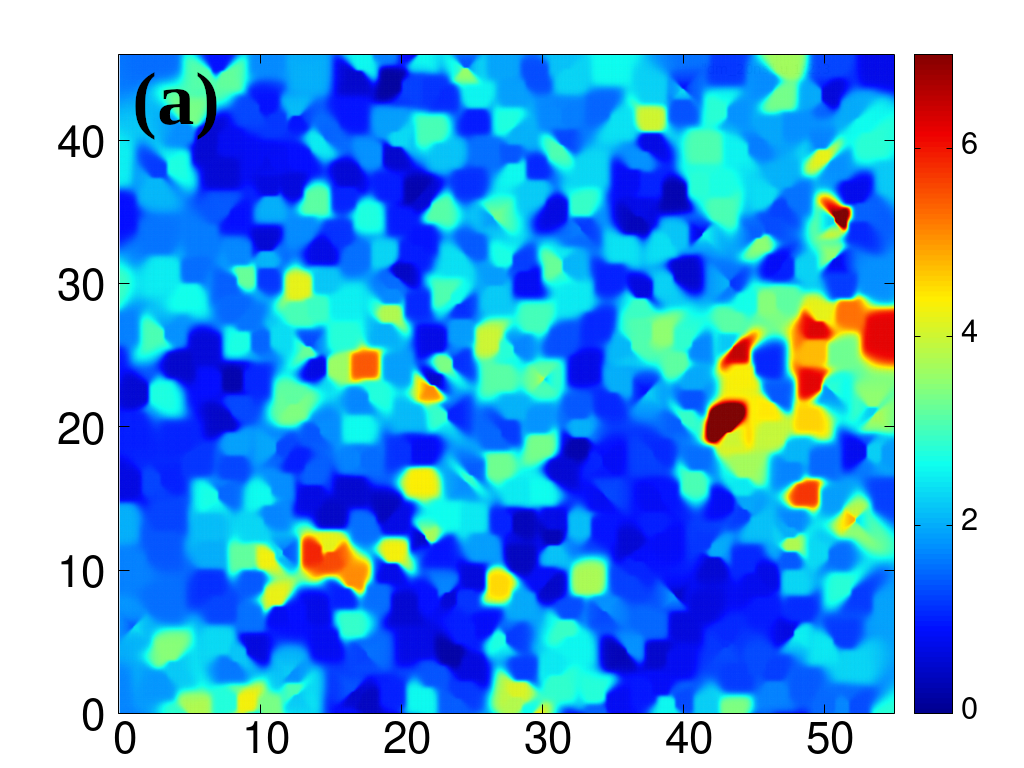}
  	\includegraphics*[height=3.2cm,width=4.4cm,angle=0.]{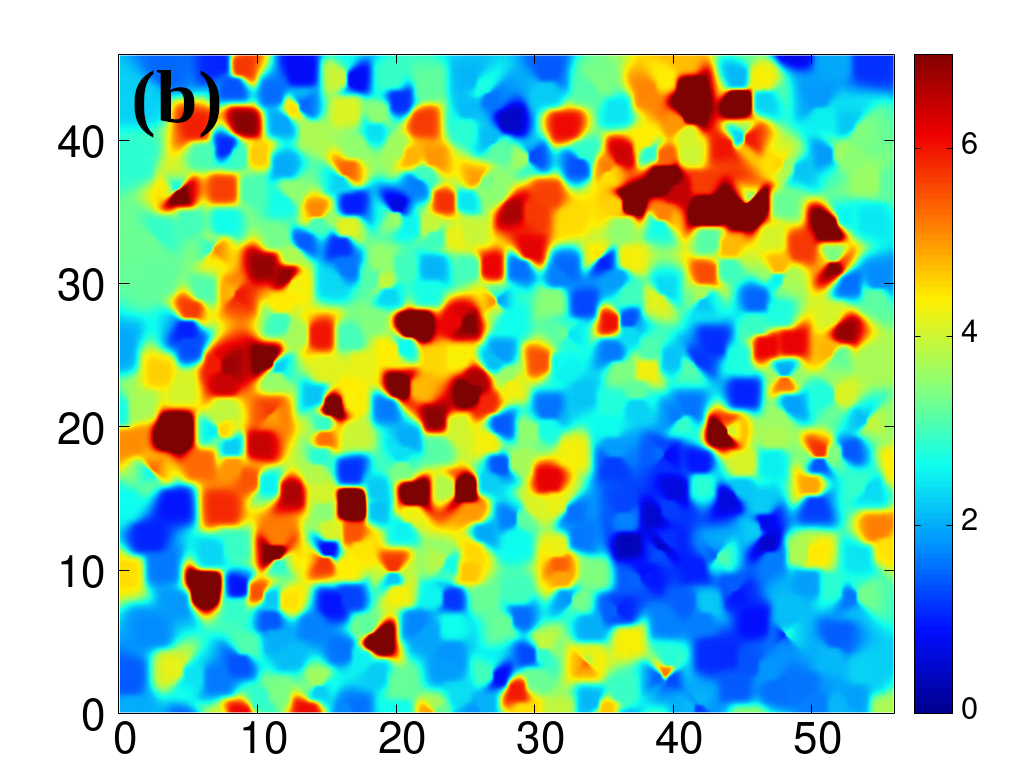}
  	\includegraphics*[height=3.2cm,width=4.4cm,angle=0.]{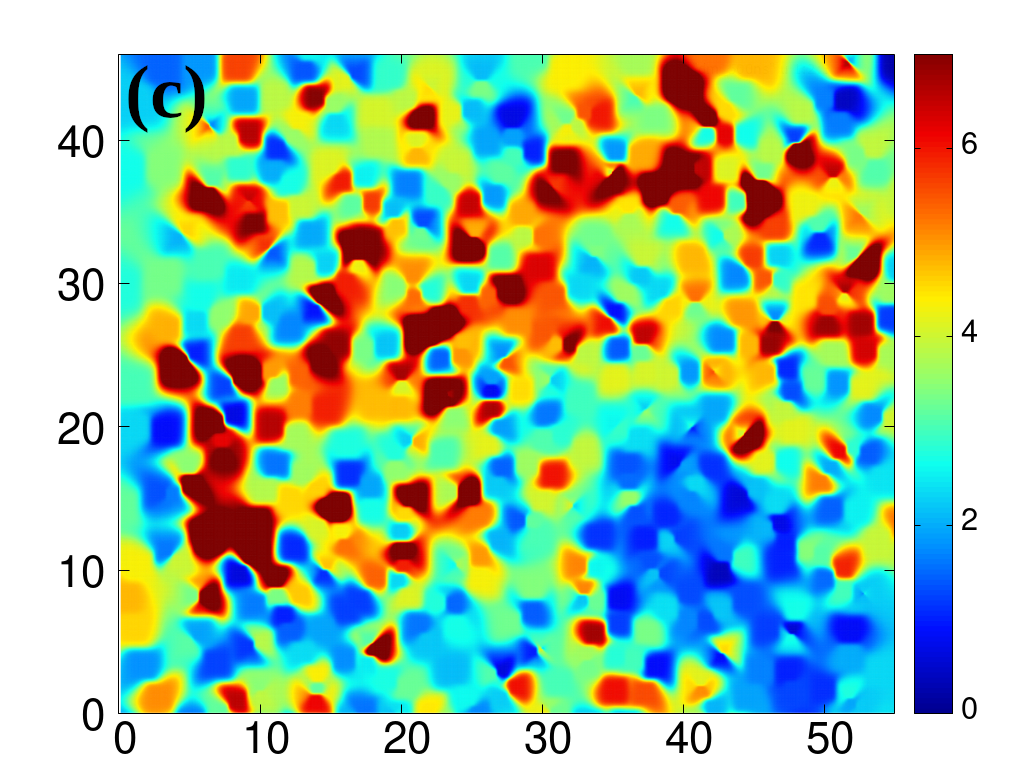}
  	\includegraphics*[height=3.2cm,width=4.4cm,angle=0.]{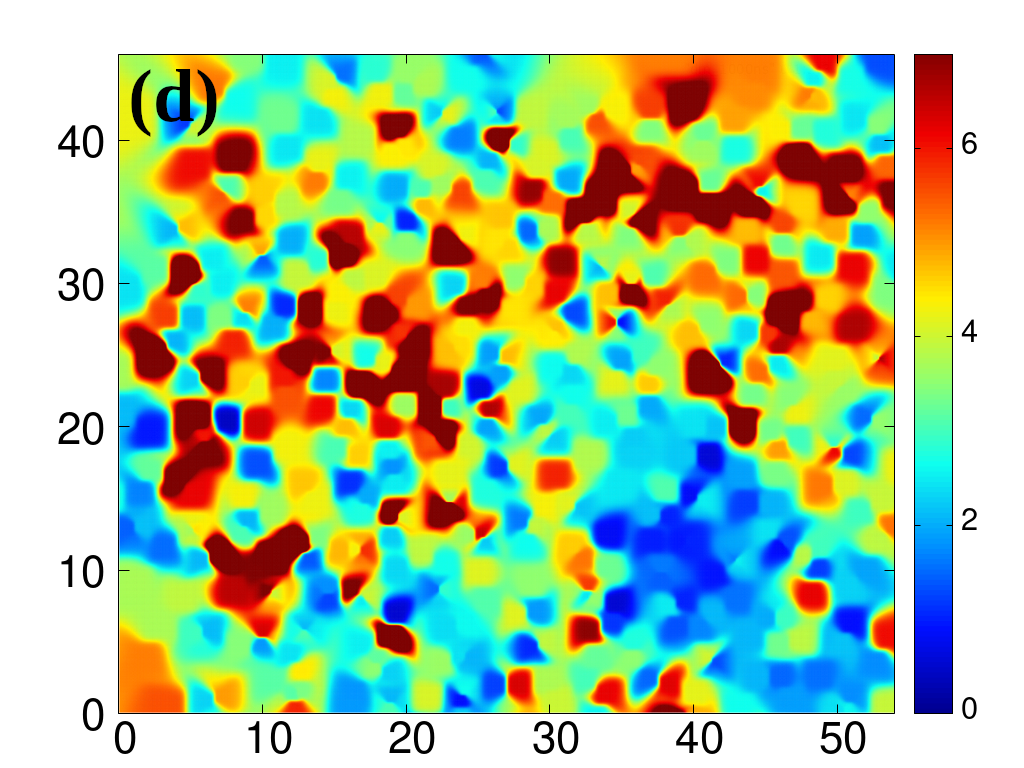}
  	
  	\includegraphics*[height=4.2cm,width=4.4cm,angle=0.]{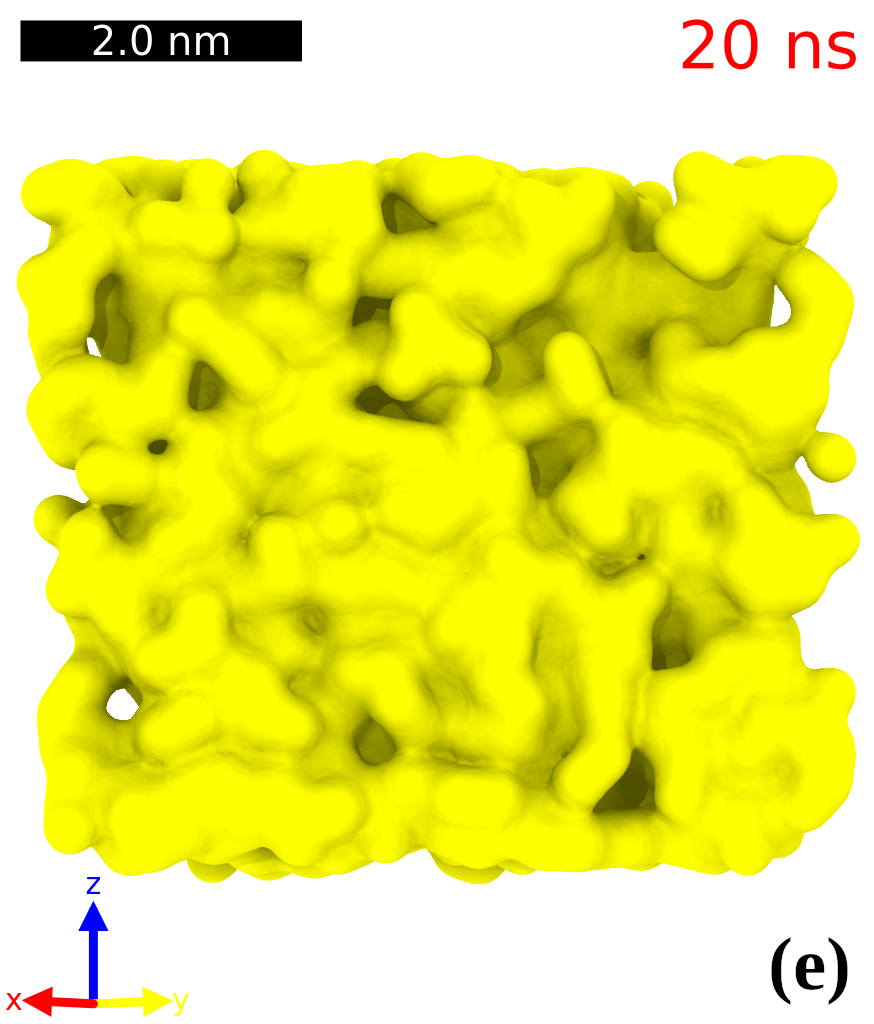}
  	\includegraphics*[height=4.2cm,width=4.4cm,angle=0.]{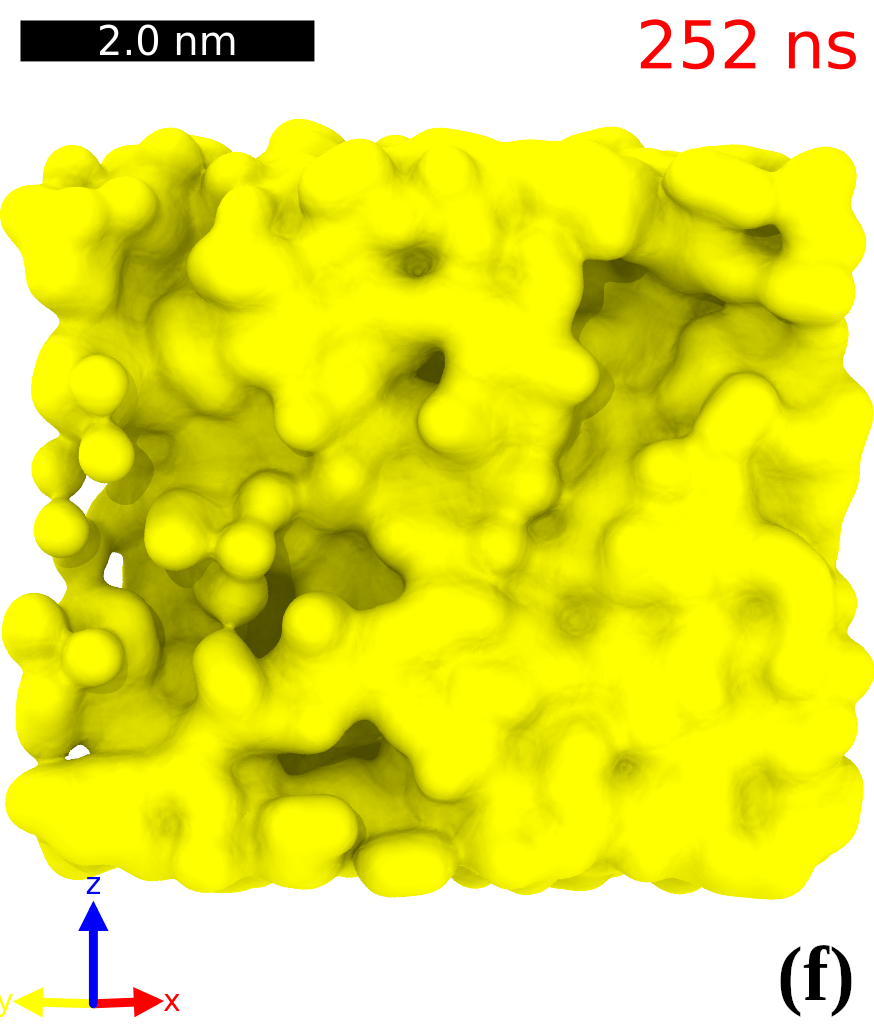}
  	\includegraphics*[height=4.2cm,width=4.4cm,angle=0.]{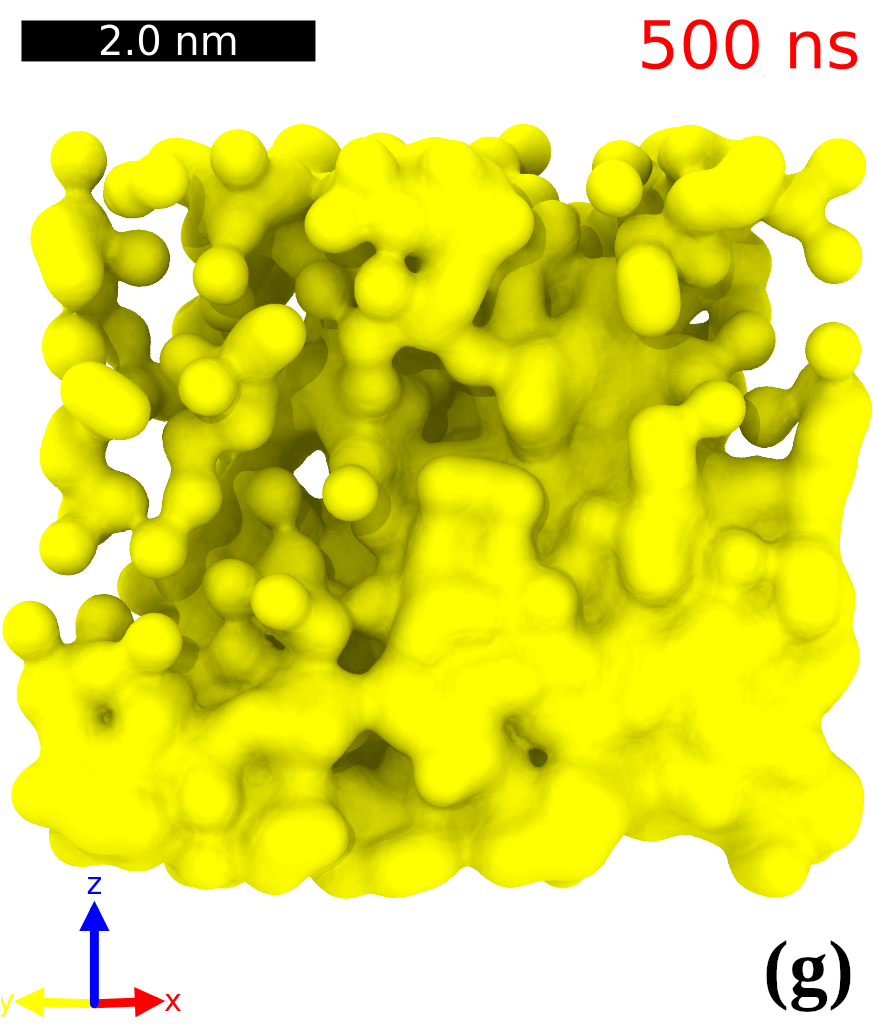}
  	\includegraphics*[height=4.2cm,width=4.4cm,angle=0.]{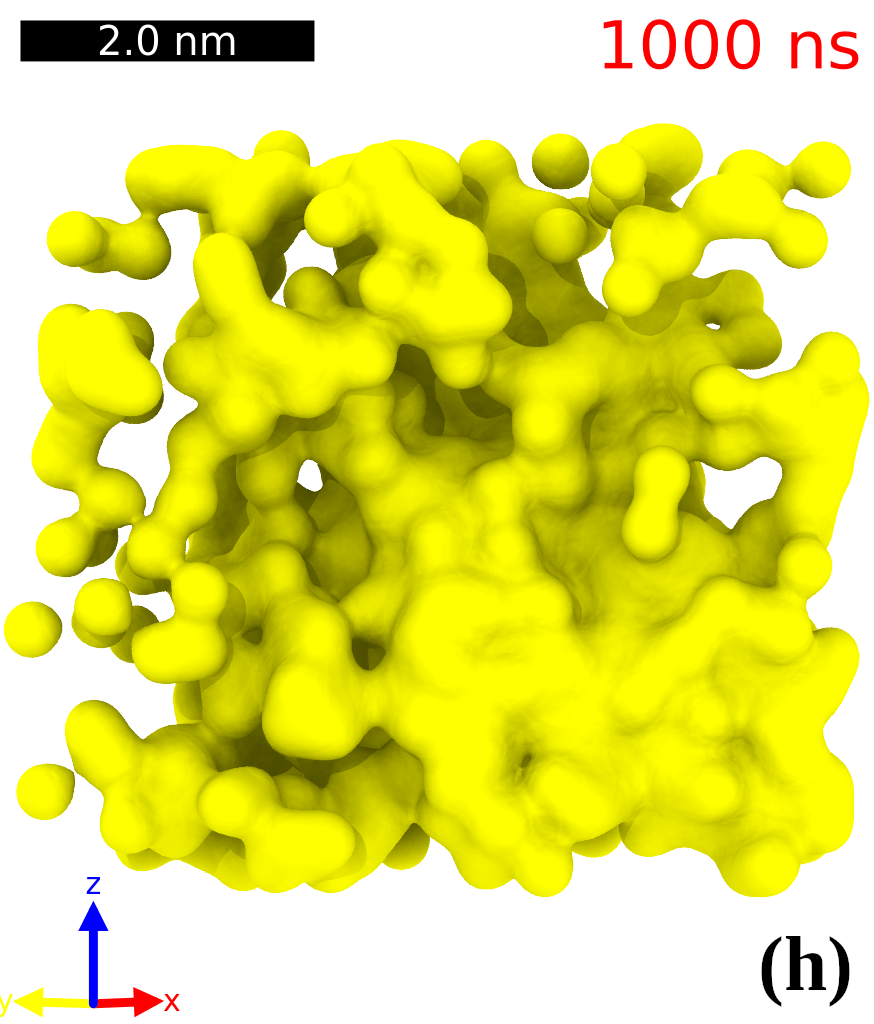}
  	
  	\includegraphics*[height=4.cm,width=5.0cm,angle=0.]{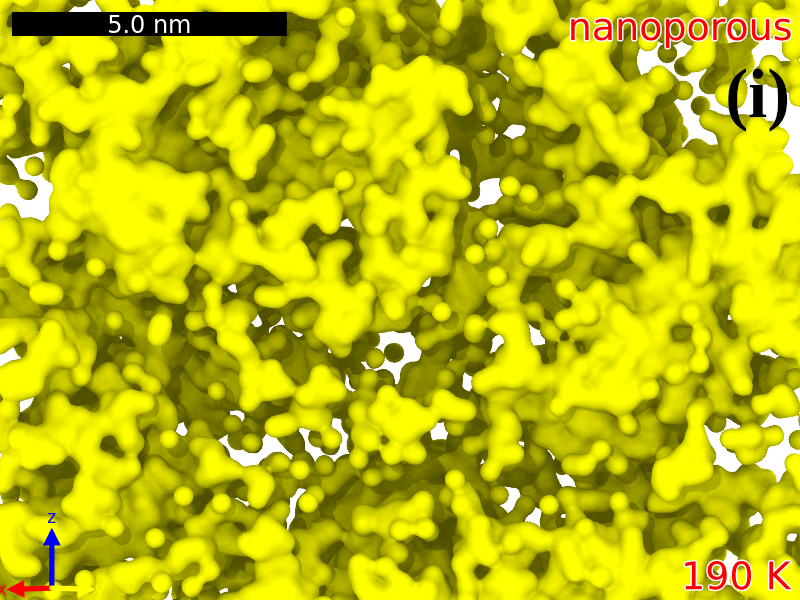}
  	\includegraphics*[height=4.cm,width=5.0cm,angle=0.]{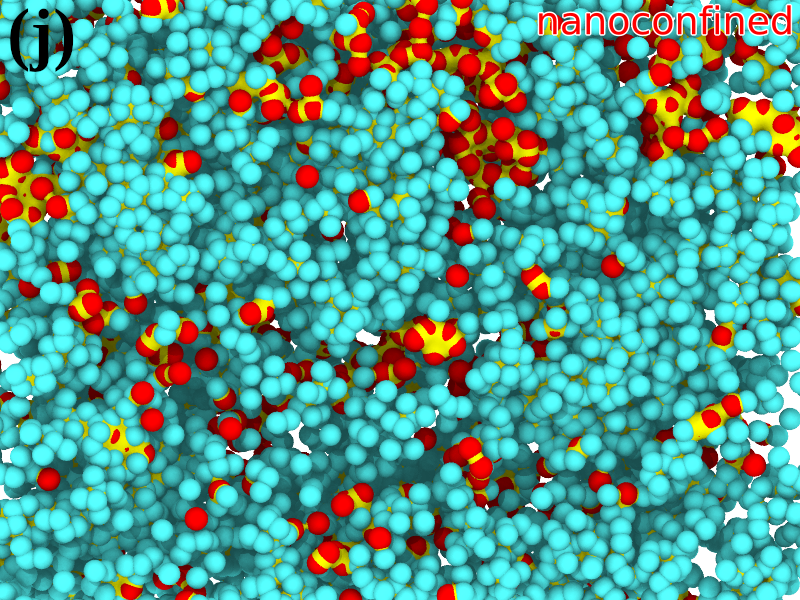}
  	\includegraphics*[height=3.5cm,width=3.5cm,angle=0.]{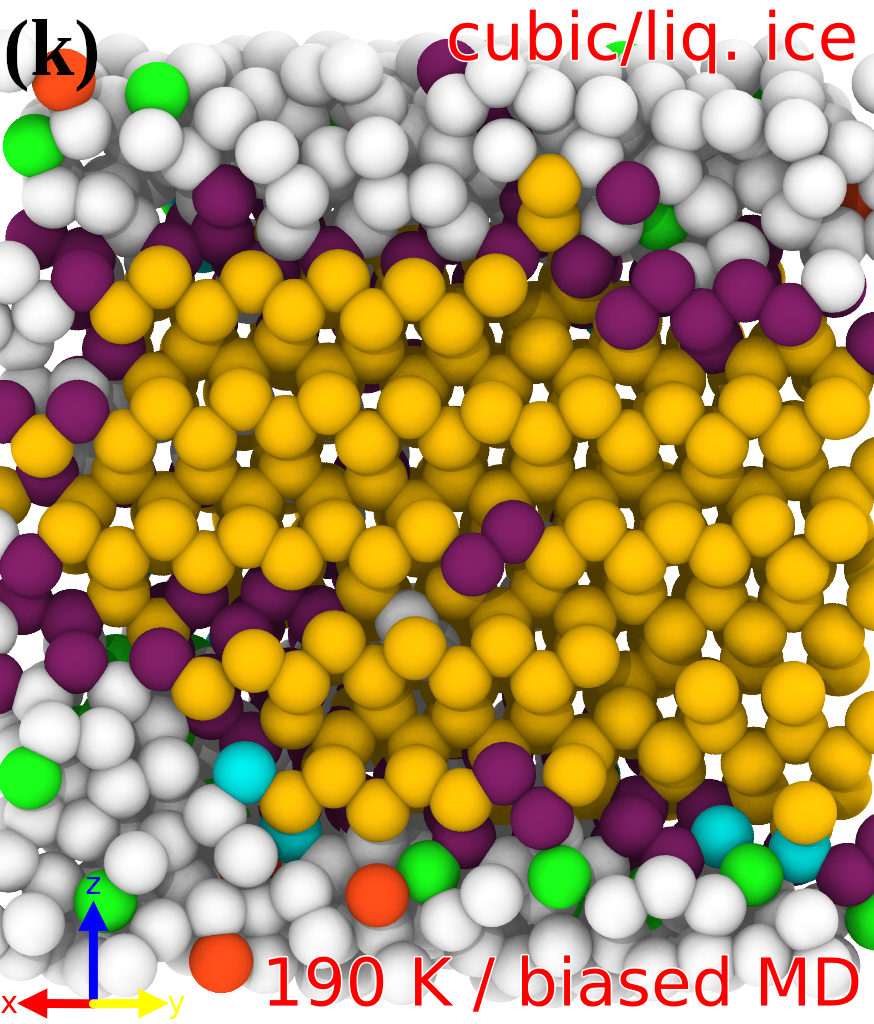}
  	\includegraphics*[height=3.5cm,width=3.5cm,angle=0.]{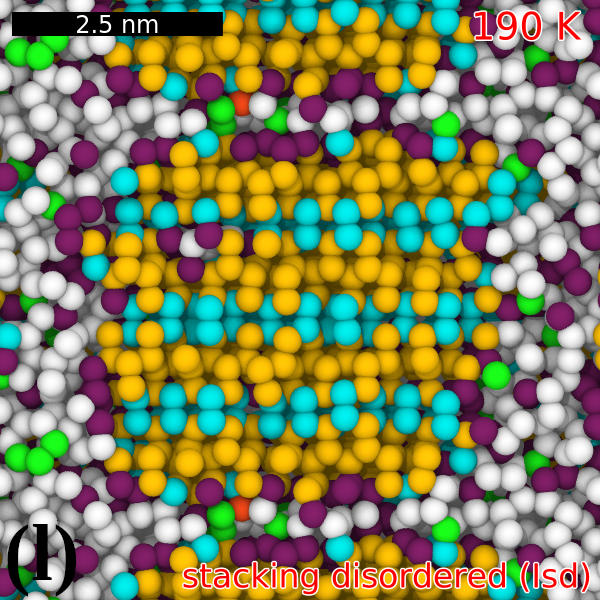}
  		
  	\caption[]{Spatially resolved atomic displacements maps and nanoporous superstructures during incubation at 190 K using unbiased MD.
  		A cubic simulation box is used including 2592 water molecules.
  		(a)-(d): D-maps recorded at 20-1000 ns and referenced to the initial state using thin slices ($\sim 7$ \r{A}) cut along the diagonal of the simulation box \textcolor{red}{(axes in \r{A})}.
  		(e)-(h): The unfolded nanoporous arrested superstructure (soft skeleton, SK) at 20, 252, 500 and 1000 ns. D-filter has been used to get rid off
  		mobile molecules (those particles are kept whose atomic displacements $d_i < 2.5$ \r{A}). Thin slices cut ($\sim 22$ \r{A}) along the diagonal of the simulation box are used.
  		The superstructures ("soft skeletons") are displayed by polyhedral surface meshes using the gaussian density method \cite{stuko}.
  		(i) The self-assembled nanoporous superstructure is shown for a larger system at 25 ns (190 K). This 'soft skeleton' comprised of arrested water molecules is plotted for atomic displacements $d_i < 1.5$ \r{A}.
  		The empty sites are filled by self-confined water molecules (not shown here).
  		(j) The nanoconfined particles are shown by red spheres together with the nanoporous skeleton (blue spheres).
  		Oxygen atoms are shown only. Thick slices ($33$ \r{A}) along the diagonal of the simulation box are plotted.
  		(k) Chill+ map for biased simulation including cubic Lattice Mold (mixed cubic/liquid final structure).
  		(l) Chill+ map for biased simulation including stacking-disordered cubic/hexagonal Ih mold.
  		The simulation leads to mixed stacking-disordered cubic/hexagonal/liquid phase.
  		Color-coding (Chill+): yellow, blue, white, purple red and green spheres are cubic, Ih hexagonal, liquid, interfacial ice, interfacial hydrate and hydrate atoms, respectively.
  		
  	}

  	\label{displ190K}
  \end{figure*}

\section{Results and Discussion}

 According to recent studies, the unbiased molecular dynamics of deeply supercooled water exhibit a profound slowdown, leading to an arrested liquid state in the NML \cite{Fitzner19,Nandi17,Donadio26}.
% In accordance with recent studies we find that the deeply supercooled water stubbornly stuck in a dynamically arrested state \cite{Fitzner19,Nandi17}.
 Despite limitations in simulating timescales well beyond the
 $\mu$s range, the deeply supercooled water system demonstrates a clear tendency to trap particle clusters within a few hundred nanoseconds.
% Although we were unable to conduct sufficiently long time simulations ($>> \mu$s or more) it is likely that % at least within few hundreds ns time scale the DSW system favors the trapping of clusters of particles.
 Recent $10$ $\mu$s lasted simulation also supports the long time stability of the arrested state of deeply supercooled water using the TIP4P/2005 model \cite{Nandi17}.
 Our benchmark simulation also confirms the lack of freezing transition up to at least $\sim 1$ $\mu$s (Fig. \ref{grsqrdispl80K-190K}(a), see also Video animations S1(a)-(b) in the Supporting Materials).
 The density, the proportion of Ih and cubic ice (left Inset Fig. \ref{grsqrdispl80K-190K}(a)) atoms and the potential energy (right Inset Fig. \ref{grsqrdispl80K-190K}(a)) show the lack of crystallization phase transition up to  $\sim 1000$ ns.
 The fundamental reason of the long time persistence of the structurally and dynamically arrested state remains, however, somewhat elusive.
 The O-O radial distribution function is shown in Fig. \ref{grsqrdispl80K-190K}(b) which displays very similar g(r) at the second peak at 80 and 190 K
 (the S(q) structure factor is also very similar at the glass forming and NML temperatures, upper Inset Fig. \ref{grsqrdispl80K-190K}(b)).
 However, in the lower inset Fig. \ref{grsqrdispl80K-190K}(b) one can see the apparent difference in the distribution of atomic displacements P(d) (which is equivalent with the self part of the van Hove correlation function, VHCF).
 This suggests that dynamical heterogeneity is robust at 190 K with a tailed P(d) and the simulation box can in principle be divided into an arrested and mobile phases.
% In the former particles slow down while in the rest of the system particles remain mobile.
 The proportion of the arrested particles ($n_{arrested}(t)$) is also shown as a function of the simulation time $t(ns)$ in Fig. \ref{grsqrdispl80K-190K}(c).
 The relatively slow decay of $n_{arrested}(t)$ is atributed to the slow, coherent motion of the superstructure formed by the arrested molecules.
 The arrested state identified in TIP4P/2005 water is validated by the more sophisticated HIPPO machine-learned potential \cite{hippo}, which similarly reveals a persistent, nearly immobile arrested phase (Supporting Fig. S1(a)-(b)).
 
 The prolonged persistence of the structural arrest can be seen in Fig. \ref{grsqrdispl80K-190K}(c).
 The persistent decay of the curve is driven by the slow transformation of the arrested structures, mediated by the limited exchange of molecules with the bulk mobile phase. 
   The root mean square of atomic displacements (RMSD) are also shown separately for mobile and the arrested particles, respectively (Inset Fig.\ref{grsqrdispl80K-190K}(c)) which supports
  that the dynamic heterogeneity is robust at 190 K. 
  
 	Based on the atomic displacement and structural analysis presented in Figs. 2(a)-(j), the nanoporous arrested region of the LDL phase in the NML forms a mechanically stable, 'frozen' skeleton over microsecond timescales. Within this network, individual molecular mobility is heavily restricted, giving rise to highly correlated, collective dynamics characteristic of a structurally arrested state.
 
The atomic displacement maps (d-map) are also shown at 190 K cut along the diagonal of the simulation box in Figs. \ref{displ190K}(a)-(d). The d-map clearly presents us the nanoscale spatial fluctuation of the dynamical heterogeneity as well as the long-term persistence of the arrested state up to 1000 ns.
These persistent blue spots highlight a residual network (soft skeleton), which we identify as innate natural nanostructures.
Application of the d-filter technique unfolds the ramified soft skeleton \cite{psule24}, exposing the previously obscured structural details (Fig. \ref{displ190K}(e)-(h)).
While the observed superstructure evolves slowly, it maintains remarkable coherence, with constituent molecules forming a supercluster that persists for at least 1
$\mu$s (see also Supplementary animations S1(a)-(b)).
Here, we denote the disordered, dynamically arrested amorphous-like state of bulk water in the no-man’s-land region as the quasi liquid state (QLS, see for short introduction in SI) \cite{psule24}, which is distinct from the low-temperature surface-melting phenomenon (quasi-liquid layer, QLL) known for ice surfaces \cite{QLL}.
 The term "quasi-liquid" is frequently used in surface physics (QLL) to describe the non-crystalline layer on the surface of crystalline ice (surface melting)
 which is, however, completely different from the term QLS used here for kinetically arrested bulk phases \cite{QLL}.
 The QLS facilitates an ultrafast, non-classical transformation of undercooled liquid iron into bulk-arrested glassy or metastable crystalline structures \cite{psule24}.

 The long-term persistence of the QLS is shown in Figs. \ref{grsqrdispl80K-190K} and \ref{displ190K} as calculated by unbiased MD. 
 The d-maps consistently display the same, largely immobile superstructural blue spot across the entire sampling sequence (Figs. 2(a)-(d)).
 %  The corresponding Chill+ maps show no characteristic features: the fraction of the Ih or cubic particles is less than 2 \% and no sign of freezing is seen.
 The proportion of the arrested particles is around $>30$ \%. Therefore we find still a large amount of liquid-like particles at around $70$ \%.
 Figs. \ref{displ190K}(e)-(h) demonstrate even more strikingly the existence of a long-lived nanoporous superstructure called soft skeleton (SK) of deeply supercooled water during the NML.
 The structural similarity between the unfolded SKs at 252, 500 and 1000 ns highlights their inherent rigidity, compactness, and robust immobility. 
 This suggests the formation of a stable, coherent, and partially ordered nanoporous superstructure stabilized by an extensive hydrogen bond network.
 Consistent atomic displacement distributions confirm that SKs correspond directly across different simulation times (Supplementary Fig. S3).
 As shown in the Inset Fig. S3, the g(r) profiles averaged over 100 ns and at t=500 ns are highly consistent, indicating that the system reaches a stable, time-invariant superstructure.
  
  The reasons for the long-term persistence of the arrested phase remain unclear; thus, elucidating the transformation mechanism from the quasi-liquid to the crystalline state is a critical objective.
 Energetic analysis indicates the arrested state lies within $0.24 \pm 0.04$
 eV/molecule of the mixed cubic/liquid (Fig. \ref{displ190K}(k)) and stacking-disordered phases (Isd, Fig. \ref{displ190K}(l) \cite{SalzmannPNAS11,Molinero11,Nandi17,Giovambattista13}) observed in biased MD at 190 K. 
 While this energy difference is significant, the arrested nanostructure is inherently a metastable state that can be stabilized by factors that do not apply to bulk crystalline materials.
 Higher in energy than the most favorable bulk phase (Isd) can exist in nature as a metastable state, especially at low temperatures like 190 K
 where thermal activation ($k_BT \approx 0.017$ eV at 190 K) is insufficient to trigger a transition to the more stable bulk phase.
 This significantly increases the kinetic stability of high-energy phases, allowing them to persist in nature indefinitely.
 At 190 K, an energy difference of few $k_BT$ is sufficient to stabilize the metastable LDL configuration, provided that a high activation barrier kinetically inhibits the transition to the nanocrystalline phase. Furthermore, comparing the hyperquenched structures via accurate energy minimization reveals a negligible energy difference (less than 0.03 eV/molecule, \textcolor{red}{}{see also Fig S9}). This indicates that the stability of the arrested system is competitive with that of the experimentally observed mixed Isd/liquid water state \cite{SalzmannPNAS11,Molinero11,Nandi17,Giovambattista13}. 
 
	Free energy calculations are essential to quantify the relative stability of the low-density liquid (LDL) phase \cite{Pietrucci23,Nada20}. Unfortunately, we were unable to capture the crystalline I$_c$ or I$_{sd}$ phases using well-tempered metadynamics simulations; instead, we observed only the LDL phase, which is characterized by a broad and deep energy minimum (see Figs. S4).
	Furthermore, the metadynamics simulations by Nada \cite{Nada20} indicate that the free energy of the LDL phase is close to those of the Isd and cubic phases.
  The arrested configuration remains stable within the NML, whereas it becomes unstable above the crossover temperature $T_s \approx 225$ K (see Video S2).
 
 This relatively high degree of stability suggests that the arrested state is competitive with the experimentally found nanocrystalline phases within the explored energy landscape \cite{SalzmannPNAS11,Molinero11,Nandi17,Giovambattista13}.
 The remarkable stability of this intermediate—rather than a mere kinetic anomaly—confirms its role as a long lived metastable structural phase.

  Visualization of the larger system at 190 K unveils that the arrested superstructure possesses an underlying nanoporous network, where the pores act as reservoirs for relatively mobile constituent particles, likely contributing to the system's stability (Fig. \ref{displ190K}(i)).
  In such an environment, water properties differ significantly from bulk water, driven by interfacial interactions and spatial constraints that rearrange its hydrogen-bond network
   \cite{ChemRev20}.
   However, the dynamically arrested state in the NML constitutes an intrinsic self-assembled nanoporous network, distinct from the extrinsic nanopores used for water confinement studies \cite{NatMat26,Kapil22,Molinero18,Mallamace16,ChemRev20} .
  Our findings indicate that the spatial restriction of mobile particles to the nanoscale pore volume is responsible for the emergence of distinct, anomalous water features in the NML (see Fig. \ref{displ190K}(j)).
  The observed retarded dynamics of the mobile particles within the pores serve as evidence of the nanoconfinement effect. 
  To demonstrate this, we compare the root mean square displacement (RMSD) of mobile particles in the nanoporous system to that in a metastable unconfined liquid at 190 K (Fig. S2a).
  Compared with the unconfined liquid, the self-confined system at 190 K shows superior structural correlation in the second coordination shell, characterized by an enhanced second peak in g(r) (Fig. S2b).  
  The unconfined system exhibits a second g(r) peak characteristic of bulk water at 300 K. (Fig. S2b).
  The  ordinary liquid at 190 K (unconfined liquid) can be simulated using the Lattice Mold technique ("pseudo biased MD" with an auxiliary template lattice) with zeroed square-well potential parameters.
  In this case the LM technique functions as a tiny perturbation which slightly disorders the system.
  A "zeroed" square-well potential in LAMMPS can still affect a simulation because, while its well depth might be set to zero, the potential still possesses a short-range repulsive core.
   Even if the attractive well depth is removed, the particles still "feel" the existence of other particles within the cutoff range through the core repulsion.
   The hard core potential ensures that fluid particles do not overlap excessively at the center of the wells, promoting the formation of a structured crystalline lattice rather than a dense, amorphous clump.
   Upon switching off the attractive part of the potential, the remaining purely repulsive interaction triggers a breakdown of the arrested hydrogen-bond network, resulting in a transition to a conventional liquid state at 190 K. 
   This mechanism highlights the structural transition from an arrested phase to a less stable, unconfined liquid state.
  However, the unconfined (ordinary liquid) state is metastable, undergoing rapid kinetic trapping into the more stable nanoporous confined-arrested state in the lack of lattice mold.
  
  Strong, self-driven interactions (such as hydrogen bonding with the pore surface) or simply space constraints induce high "crowding" of water molecules. This reduces molecular mobility (dynamical arrest), creating a "cold spot" where mobility is low but ice-like structural order is not promoted, decoupling slow dynamics from ice formation (see also video animation S3 of water self-confinement in the Supporting Material).
  Under strong confinement the entire water volume can become intrinsically slow-moving and disordered, resisting ice nucleation even at temperatures where bulk water would freeze immediately \cite{Mallamace16}.
  The combined effect of self-confinement and dynamical arrest creates a synergistically stable QLS by hindering the kinetic processes required for crystallization. 
  The high surface-to-volume ratio in nanoporous environments means surface forces dominate, inhibiting the formation of the nucleus necessary to initiate freezing, thereby enhancing the stability of the metastable supercooled liquid. 
  
  Our results demonstrate that the unique stability of the revealed quasi-liquid state originates from the synergetic interaction between self-driven nanoconfinement and emergent dynamical arrest.
  Although this intrinsic, self-driven behavior of water lacks direct experimental validation, our calculations indicate that this distinct state is robustly detectable with current or emerging experimental techniques.
 
 On the other hand it has also been suggested that ice in the NML is a mixture of stacking disordered cubic and hexagonal ice and liquid-like water as obtained by the simple mW force field model \cite{SalzmannPNAS11,Molinero11}.
 Recently it has also been suggested that LDA is not a fully amorphous phase and partially crystalline instead \cite{Davies25}. 
 Therefore, two primary scenarios are introduced to explain water's phase evolution in the NML: one, a homogeneous glassy transition, and the other, a heterogeneous system involving mixed liquid and crystalline/amorphous phases.
 The coexistence of distorted ice and liquid water, along with other heterogeneous structures, does not necessarily invalidate the dynamical arrest scenario.
 The unexpectedly stable, kinetically trapped QLS likely precedes long-term ordering or freezing that is not captured by most simulations \cite{Nandi17}.
  Simulating the transition from the arrested state (QLS) to the observed crystalline phase, however, remains challenging.
  The low-density QLS (self-confined nanoporous-arrested state) may exhibit sufficient long-term stability to allow for direct experimental characterization.

 It has already been known for a while that ice is in the stacking disordered ice I polymorph in the NML \cite{SalzmannPNAS11,Molinero11,Nandi17,Giovambattista13}.
 We also attempt to understand how the QLS transforms into crystalline ice.
 As unbiased MD simulations with the TIP4P/2005 force field cannot reach this state within a reasonable computational timeframe, our analysis focuses on the incubation period preceding crystallization.
 No freezing was observed in the smallest system ($2592$ water molecules) during the
 $\sim 1$ $\mu$s unbiased MD simulation (Fig. \ref{grsqrdispl80K-190K}(a)).
 The continuous decrease in both potential energy and density indicates a slow transformation process likely leading to a freezing transition.
  This finding is somewhat inconsistent with results from the coarse-grained mW model, which predicts a fast transition to a stacking-disordered ice phase containing a substantial amount of liquid water \cite{Molinero11}.
  {\em Alternatively, the NML supercooled system may exhibit enhanced stability, remaining kinetically arrested in a self-confined state, the specific, underlying mechanism of which warrants further investigation.}

 {\bf Freezing induced by Lattice Mold:}
 
 Furthermore, we are examining the specific, direct role of QLS in the freezing process employing biased molecular dynamics (biased-LM-MD).
 By initiating the LM-MD simulation with either Ih, cubic or stacking disordered (Isd, ice I) lattice seeds (molds), we successfully produce Ih, cubic or Isd ice (more technical details can be found in the Supplementary Material, see also the video animation S4 of the simulation for the Isd phase) \cite{Molinero11}.
% This method, when paired with a suitable mold, holds the potential to produce the experimental ice I phase.
Tracing the nucleation process remains valuable for investigating the formation of the Ih phase.
 First we show spatially resolved maps cut in the middle of the simulation cell (the diagonal of the cubic box, Figs. S5-S6)
 using LM-MD.
 Atomic displacement maps (d-maps) are shown in Figs. S5(a)-(c) at 250 K (more info on d-maps in ref. \cite{psule24}). The atomic displacements ($d_i$) are calculated with respect to the
 final crystalline state (Fig. S5(h)). 
 No obvious correspondence is seen between arrested particles (blue spots in Figs. S5(a)-(c)
 and particles detected as Ih by Chill+ (Figs. S5(e)-(h)).
 This is because exceeding $T_s \approx 225$ K promotes phase fluctuations within the prenucleation species, preventing the pinning of the pre-nucleation clusters (PNCs).
 PNCs are known to be highly dynamic in aqueous solutions at room temperature, undergoing rapid ligand exchange, mobility, and consequential aggregation \cite{Gebauer}.
 Particles are also highly mobile at 250 K in supercooled water (see the scale bar of the d-maps, atomic displacements are in the range of few tens of \r{A} (Fig. S6(a).
 At 4 ns the system becomes nearly fully crystalline (nucleation is triggered by LM) as shown by the terahedrality map (Fig. S5(d))
 and by Chill+ (Fig. S5(h)).
 
  At 150 K, however, at the lower end of NML we find a strong correlation between d-maps and Chill+ Ih spots (Figs. S7(a)-(d))). As it has been found in liquid iron \cite{psule24} the blue spots in the d-maps preindicate the
  emergence of Chill+ Ih spots. The blue spots in the d-maps (Figs. S7(a)-(d)) are the arrested regions which are absent at 250 K. No phase fluctuation of the Ih spots occurs below $T_s \approx 225$ K which is a dramatic difference with respect to 250 K (localized nucleation at 150 K, $T<T_s$). The scale bar (the magnitude of displacements) are much smaller than at 250 K
  (few \r{A}), Fig. S6(a)).

  The histogram of atomic displacements $P(d_i)$, calculated relative to the final crystalline state at the onset of prenucleation, displays a strong temperature dependence. While $P(d_i)$ remains sharp below the glass transition temperature ($T_g \approx 140$ K), it broadens and develops a significant tail in the NML region, indicating increased structural and dynamical heterogeneity (Fig. S6(a)).
  The localized fraction of particles ($n_{loc}$) and the peak position of $P(d_i)$ ($d_{peak}$) exhibit an abrupt transition within the NML (Inset Figs. S6(a)). These features signify the onset of QLS behavior inside the NML.
  
  To gain further insight into the formation of PNCs during the prenucleation stage, the temporal evolution of these clusters is displayed in Figs. S7(i)-(l). The growing PNCs are trapped from their early formation stage, consistent with observations in supercooled liquid iron \cite{psule24}. {\em Remarkably, despite being vastly different systems, deeply quenched molten iron (closely packed metallic) and water (molecular system) exhibit a similar arrested state.} This network of PNCs hinders their own motion \cite{psule24}, leading to an assembly-free growth mechanism, similar to that observed in liquid iron \cite{psule24}.

At 150 K and t=2.5 ns, the pair distribution function (Fig. S8(a)) reveals an enhanced second peak within pre-nucleation clusters compared to the liquid environment, indicating pronounced local orientational ordering. This enhanced peak suggests structural development within the clusters, contrasting with the surrounding disordered liquid phase.
Figure S8(b) illustrates the temporal evolution of the distribution at 150 and 250 K, highlighting a distinct hexagonal ice (Ih) peak emergence and profile splitting at 150 K that is absent at 250 K. These observations support the conclusion that divergent crystallization mechanisms operate on either side of the crossover temperature, $T_s \approx 225$ K.

We find a striking difference in the temperature dependence of the density $\rho$  between biased and unbiased MD simulations (Fig. S8(c)):
$\rho$ exhibits a peak in the biased case, whereas it reaches a minimum in the unbiased counterpart within the NML. This fundamental discrepancy arises from distinct nucleation mechanisms. Biased simulations facilitate the formation of networked pre-nucleation clusters (PNCs) pinned to their sites of formation within the NML \cite{psule24}. Conversely, in unbiased simulations, no such PNCs are identified; the system remains trapped in a nanoporous low-density metastable state characterized by a self-confined liquid fraction.
 
  The biased (LM) simulations proceed via a metastable high-density liquid (HDL) pathway in the NML (see Figs. S6(b),Inset S11(b)). Recent metadynamics simulations also indicated the occurrence of the HDL pathway at 200 K \cite{Nada20}, although they found no HDL minimum on the free energy surface. Our simulations using well-tempered metadynamics (WT-MetaD) with the TIP4P/2005 force field demonstrate that no distinct HDL free energy minimum exists for supercooled water at 190 K and ambient pressure, in accordance with recent findings \cite{Pietrucci23,Nada20}. Instead, the free energy landscape is dominated by a single low-density liquid (LDL) minimum that is remarkably deep and exceptionally broad along both the density and energy coordinates (see Fig. S4(a)). However, line cuts of this surface reveal that the well walls are sufficiently steep to prevent thermal fluctuations from generating higher-density phases (see SI Fig S4(c)). Consequently, the system remains deeply trapped near the expected density of the LDL phase. We also investigated how increasing pressure induces a splitting of the free energy surface into distinct LDL and HDL minima, in accordance with recent literature \cite{Pietrucci23}. We find that at around 2000 atm, the free energy profile becomes sufficiently flat to allow density fluctuations between the LDL and HDL phases, consistent with accurate umbrella sampling simulations (Figs S4(d)-(f)) \cite{Pietrucci23}.
  
{\bf The lowest density non-crystalline water: low-density QLS}  
  
  We argue that the structurally arrested QLS in the NML constitutes a distinct non-crystalline water state, characterized by the lowest density reported to date
  ($\rho \approx 0.925$ g/cm$^3$) at 190 K (see Fig. \ref{grsqrdispl80K-190K}(a)).
  Its density is lower than conventional LDA ice ($\rho \approx 0.94$ g/cm$^3$) and approaches that of ice Ih ($\rho \approx 0.91$ g/cm$^3$).
   As shown in Fig. S8(c)
   a robust density drop occurs in the middle of the NML between 150 and
   225 K (unbiased MD). 
  It is crucial, however, to distinguish between the LDA and low-density QLS phases. 
%%%%%%%%%%%%%%%%%%%%%%%%%%%%%% 
% F3
%%%%%%%%%%%%%%%%%%%%%%%%%%%%
     \begin{figure}[hbtp]
     	
     	\includegraphics*[height=7.5cm,width=9cm,angle=0.]{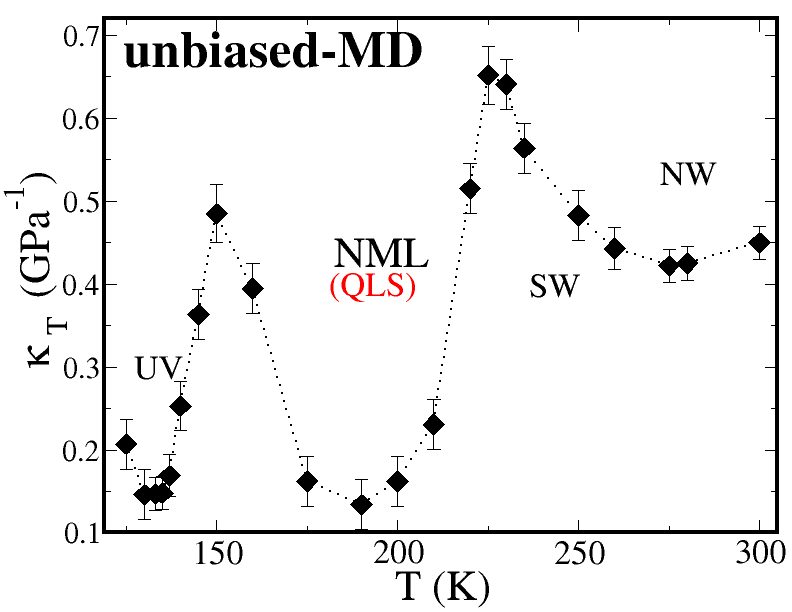}
     	
     	\caption[]{ 
     		The isothermal compressibility $\kappa$ as a function of the temperature.
     		Unbiased MD: Isothermal compressibility $\kappa_T$ (GPa$^{-1}$) as a function of the temperature (K) calculated using the TIP4P/2005 force field
     		without Lattice Mold (standard unbiased npt MD simulation). Error bars due to event sampling.
     		NW: normal water, SW: supercooled water, NML: no man's land, QLS: quasi liquid state, UV: ultraviscous state.
     		The large cubic simulation cell including 49152 water molecules has been used. 
     	}
     	
     	\label{kappaiso}
     \end{figure}
  Distinguishing between LDA and QLS is challenging due to the close structural similarities in their respective radial distribution functions,
g(r) and structure factors S(q) (see Fig. \ref{grsqrdispl80K-190K}(b)).
     These structural indicators lack the resolution to clearly distinguish between LDA and QLS.
  We therefore propose to compare atomic displacement distributions $P(d)$ (self part of the VHCF) at 80 K (LDA) and at 190 K (2nd Inset Fig. \ref{grsqrdispl80K-190K}(b)).
  Although not experimentally measurable, this quantity can be easily determined using MD simulations.
  	  A sharp, characteristic peak is observed at 80 K, signifying a structural arrest of the system in a glassy state (2nd lower Inset Fig. \ref{grsqrdispl80K-190K}(b)). 
  	However, at 190 K, becomes heavily broadened with prominent, non-Gaussian tails due to strong dynamic heterogeneity \cite{Berthier2,Royall}.
  	The clear distinction between the two data sets indicates that LDA (80 K) and QLS (190 K) constitute distinct physical phases.
  	The difference can clearly be seen and indicates that LDA (80 K) and QLS (190 K) are substantially different phases (also compare Video animations S1(a) and S5).

  	  While low-density amorphous ice (LDA) represents a largely homogeneous amorphous state \cite{Martelli18}, the QLS phase is highly heterogeneous, comprising a long-lived, preordered, and arrested nanoporous network with embedded, nanoconfined liquid-like regions.
  	The two phases (the nanoporous superstructure and the embedded liquid) at 190 K are deeply intertwinned into each other which perhaps contributes to the dynamical arrest of QLS.  
  	As no known liquid-like phase possesses similar properties, the QLS can be classified as a distinct, non-glassy novel phase.

%%%%%%%%%%%%%%%%%%%%%%%%%%%%%% 
% F4
%%%%%%%%%%%%%%%%%%%%%%%%%%%%%
\begin{figure}[hbtp]
	
	\includegraphics*[height=7cm,width=8.6cm,angle=0.]{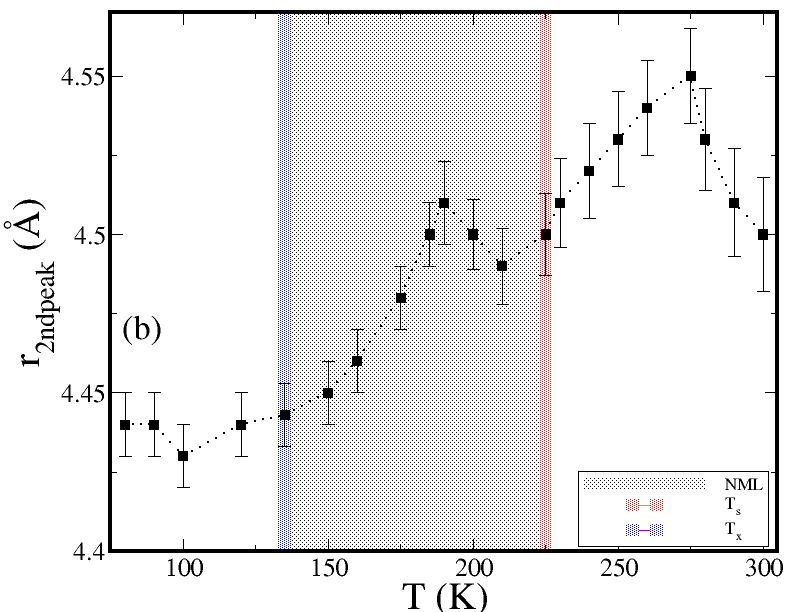}
	\caption[]{ 
		The g(r) 2nd peak position (\r{A}) as a function of temperature (K) (unbiased MD).
		The T-dependence of the 2nd peak position in the g(r) $r_{2ndpeak}$ (\r{A}).
		The calculation were carried out using the TIP4P/2005 force field and unbiased npt simulations.
		The computed values are time averaged over 5 ns simulation time.
		$T_s$ and $T_x$ are the thermodynamic divergence and glass transition temperatures.	
		NML and QLS are the no man's land and quasi liquid state.
		SW: supercooled water.
		The NML is shown in the shaded area.
		Error bar is due to the uncertainty in the determination of the peak position.
	}
	
	\label{r2ndpeak_chillp}
\end{figure}

{\bf The fingerprint of QLS in the isothermal compressibility ($\kappa_T$):}

The study of isothermal compressibility ($\kappa_T$) in water's NML at ambient pressure) is crucial because it acts as a sensitive probe for the anomalous behavior and structural changes of water, including the potential existence of a liquid-liquid critical point (LLCP). Because water's anomalies become amplified upon supercooling, measuring $\kappa_T$
helps to reveal why water behaves differently from other liquids and to map out the Widom line, which is the locus of maxima for response functions \cite{Gallo16,Kringle20,Nilsson14,Angell08,Mishima98,Kruger23,Smith99}. 
Isothermal compressibility serves as a "structural map" in the no man's land, allowing researchers to explore the mysterious, amplified properties of supercooled water and verify the existence of a second, low-temperature liquid phase.

The anomalous behavior of $\kappa_T$ as calculated by the TIP4P/2005 model is shown in Fig. \ref{kappaiso}
and demonstrates that $\kappa_T(T)$ drops whithin the NML and bounded by peaks at the lower ($\sim 150$ K) and upper ($\sim T_s$)
boundary of the NML. 
The calculated $\kappa_T(T)$ values are in line with those of reported recently \cite{Vegajcp10} and in particular
the maximum at $\sim 225$ K is consistent with the results reported experimentally \cite{Nilsson17}.
We argue that the drop of $\kappa_T$ in the NML together with the drop of the density (see Fig. S8(c)) further support the QLS model.

{\em Therefore, the transition from the liquid state to the QLS renders the water less compressible, which is attributed to the formation of a trapped, rigid superstructure within the NML}.

%%%%%%%%%%%%%%%%%%%%%%%%%%%%%% 
% F5
%%%%%%%%%%%%%%%%%%%%%%%%%%%%%
\begin{figure}[hbtp]
	
	\includegraphics*[height=7cm,width=9cm,angle=0.]{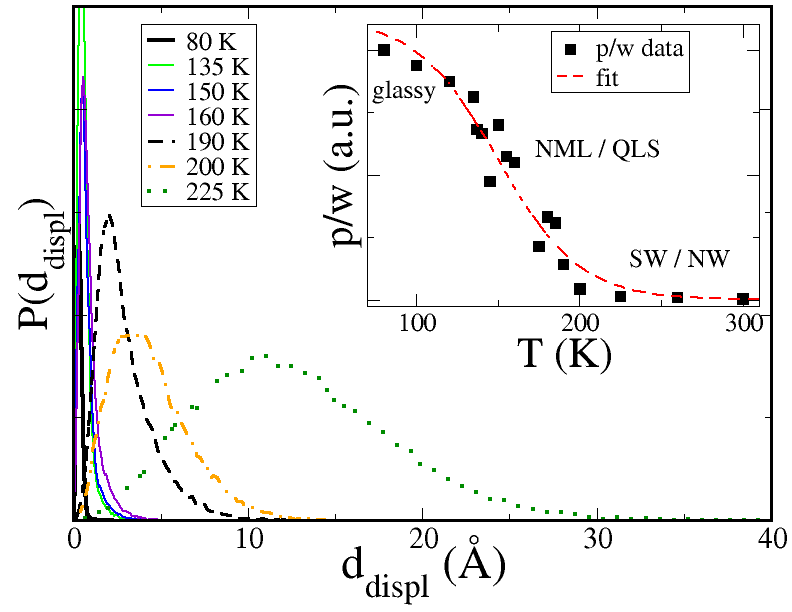}
	\caption[]{ 
		The distribution of the O-atomic displacements ($d_{displ}$, \r{A}, $P(d_{displ})$) as a function of temperature (K) (unbiased MD, 49152 water molecules in the periodic box).
		The $d_{displ}$ values have been calculated with respect to the initial stages at $\sim 10$ ns simulation time.
		{\em Inset:}
		The peak height to width ratio of the distributions $P(d_{displ})$ at various temperatures (K).
		The curves were calculated using unbiased MD simulations. 
		The width has been measured at low height (at 1/8 peak height) to account for the
		width of the tailed curves.	
		The red dashed line is a fit to the data (step function).
		NML and QLS are the no man's land and quasi liquid state.
		SW / NW: supercooled and normal water.
	}
	
	\label{ptow}
\end{figure}

A more striking feature is the peak in $\kappa_T(T)$ at approximately 150 K (see Fig. \ref{kappaiso}), which is confirmed by additional TIP4P/2005 simulations. This unexpected peak, located just above the ultraviscous regime \cite{Amann-Winkel13,Amann-Winkel16}, represents a lower bound for the NML.
However, the origin of the sharp peak at $\sim 150$  K remains somewhat unclear, though it likely signifies the emergence of a strong liquid-like state \cite{Smith99}.
It is likely that the amorphous solid undergoes a transition to a quasi-liquid state at
150 K, accompanied by a transient, not fully understood increase in the isothermal compressibility.
In phase transition studies, transient high compressibility often indicates that the substance is nearing a critical point where it will change phase, as the energy required to change its volume decreases near the critical point.
The high compressibility is a signature of the system rapidly switching between two distinct, metastable, liquid-like environments (LDL and HDL water), which is a major, yet often hidden, feature of water's anomalous behavior \cite{Gallo16}.
In Fig. S9 a robust plateau is seen around 150 K in the proportion of the LDL fraction as calculated by Voronoi's analysis.  
This is in line with the anomalous behaviour of the  $\kappa_T(T)$ curve shown in Fig. \ref{kappaiso} and supports
the emergence of a crossover point at $T \approx 150$ K which is likely related to a hidden phase transition not fully understood yet.
This satelite peak nearby the global peak of $n_{LDL}$(T) at 190 K supports the QLS scenario of the present work
as well as the complex phase development of SW in the NML not clearly revealed before.
The sharp peak of $n_{LDL}$(T) at $T \approx 190$ K with a $\sim 60$ \% LDL fraction is consistent with other Figs. such as
Fig. \ref{r2ndpeak_chillp} or with the density profile of Fig. S8(c).

{\bf The QLS in the NML is not glassy nor spinodal:}  

%%%%%%%%%%%%%%%%%%%%%%%%%%%%%% 
% F6
%%%%%%%%%%%%%%%%%%%%%%%%%%%%%
\begin{figure}[hbtp]
	\includegraphics*[height=9cm,width=9cm,angle=0.]{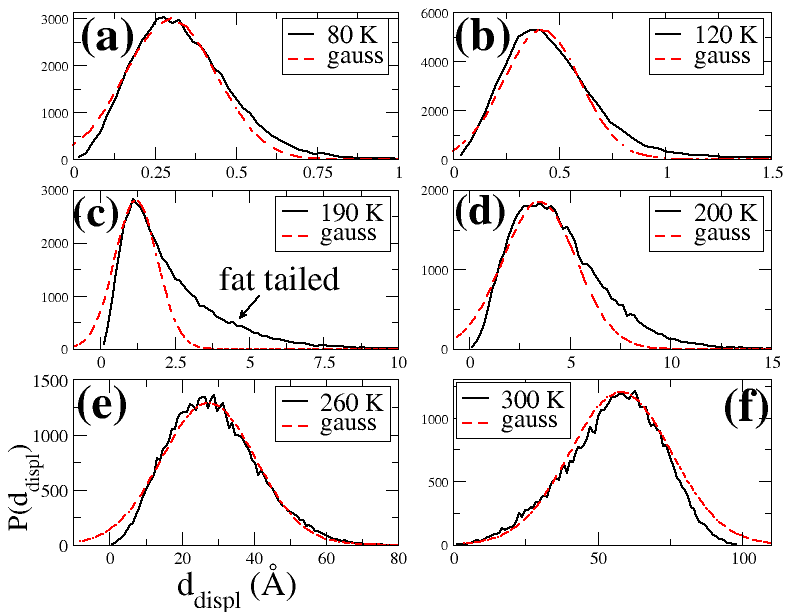}
	\caption[]{ 
		The distribution of the atomic displacements ($P(d_{displ})$) at different temperatures (K) (unbiased MD). 
		The $d_{displ}$ values have been calculated with respect to the initial stages at $\sim 10$ ns simulation time.
		Gaussian fits are also shown to the curves (dashed lines).
		The TIP4P/2005 force-field is used (brute force MD).
	}
	
	\label{PdisplT}
\end{figure}

Our findings indicate that QLS represents a specific (or unique) transformation pathway that operates outside the paradigms of standard or quasi-spinodal decomposition.
The spinodal is the limit of the liquid's stability, beyond which it must separate into two phases \cite{spinodal}. 
Spinodal decomposition (SD) is a specific type of phase separation that happens when a system is thermodynamically unstable, leading to uniform changes throughout the volume \cite{spinodal}. Dynamically heterogeneous systems are characterized by a range of local dynamics, where some regions relax faster than others, which can arise from structures like those created during spinodal decomposition, but it is not limited to it. 
In principle SD can also be the source of dynamical heterogeneity in the NML of water. 
Dynamical heterogeneity can also exist, however, in systems that are not undergoing spinodal decomposition, such as glasses \cite{Berthier2,Royall}.
This heterogeneity can be caused by structural variations, which are often formed during processes like spinodal decomposition. 
SD occurs when a system is unstable, any fluctuation will cause it to separate into two phases \cite{spinodal}.
The separation starts uniformly and simultaneously throughout the entire system, with no nucleation barrier.
Leads to the formation of intertwined structures that coarsen over time. 
In the case of QLS, however, no coarsening can be seen on the time scale reachable by our unbiased MD simulations.
Instead, the revealed hidden superstructure persists for a long time in the NML T-regime of $T_g < T < T_s$ where $T_g\approx 140$ K and shows no sign of coarsening.
Thus we do not see  
a process where smaller elements in a system merge or grow into larger ones over time, driven by a tendency to minimize energy (the lack of Oswald ripening).
As well as the influence of the Gibbs-Thomson effect is also missing in QLS: the growth of higher energy smaller atomic clusters (e.g. PNCs) in order to reduce their surface energy is also suppressed or delayed during the QLS
owing to the specific entangled and arrested state. 

Based on the structural analysis presented in Figs. 1(e)-(j), the nanoporous arrested region of the LDL phase in the NML forms a mechanically stable, 'frozen' skeleton over microsecond timescales. Within this network, individual molecular mobility is heavily restricted, giving rise to highly correlated, collective dynamics characteristic of a structurally arrested state.

Some studies observe "spinodal-like" dynamics in certain conditions (liquid–liquid spinodal points in the p-T plane under pressure), but this is not the same as a stable spinodal in the bulk liquid at ambient pressure \cite{SciortinoJCP24,Martelli19,Fonte24,Gallo07}. 
Although spinodal decomposition is mostly related to phase development under pressure \cite{SciortinoJCP24,Martelli19,Fonte24,Gallo07}
it is worth checking that the observed phase separation of trapped particles is definitely not related at least partly to spinodals at ambient pressure.
Some experiments have shown a smooth continuation of diffusion data into lower temperatures, which would rule out a spinodal and supports singularity-free model \cite{Gallo16,Kimmel16}.

{\em The signature of QLS in the O-O g(r):}
Another peculiar feature of the QLS
is the height of the 2nd peak of the g(r) ($h_{2ndpeak}$) which peaks within the NML as a function of the temperature (see Inset Fig. S11(a))
which further supports that a specific structural transition takes place within the temperature regime of the NML.   
Compared to the disordered bulk liquid, the self-confined, arrested QLS during the NML exhibits a significantly higher second peak in the pair distribution function
(Fig. S2(b)), indicating enhanced intermediate-range ordering.

It is challenging, however, to differentiate the deeply supercooled non-glassy state (above glass transition temperature $T_g$) from the true glassy state of water (below $T_g$) on the basis of the g(r) \cite{Soper06}.
Indeed, in the literature often the NML is understood as a glassy-like amorphous state which rapidly transforms into a crystalline ice \cite{Gallo16,Kringle20,Nilsson14}.
We argue that there is no clear-cut evidence that the NML is indeed glassy before freezing.
The observed tiny shifts in the g(r) (increase in 2nd peak height (Inset Fig. S11(a)), the slight increase of the peak position (Fig. S11(a)) together with other specific features (e.g. the tailed distribution of atomic displacements, Fig. \ref{PdisplT}, the fraction of the LDL phase peaks in the NML, Fig. S9) with respect to truly glassy systems
supports the expectation that a specific distinct state (QLS) appear before freezing in the NML.
The sharp change in the fraction of arrested (localized) particles as a function of temperature within the NML further supports the existence of a distinct QLS phase (Fig. S11(c)).

In particular, we find that the 2nd g(r) peak exhibits a shift towards higher peak position ($r_{2ndpeak} \approx 4.55$ \r{A}, Fig. 4) which is likely
associated with a volume expansion and/or with the decrease of the number of interstitial molecules in the 2nd coordination shell and which also explains the demonstrated density drop in the NML (see Fig. S8(c)).
This is again a specific feature of the QLS which occurs exclusively in the NML.
Below glass transition temperature of $T_g \approx 136$ K the g(r) peak position slightly shifts
in contrary direction due to volume contraction of the 2nd shell ($r_{peak} \approx 4.4$ \r{A} in the hyperquenched glassy water at 80 K) \cite{Soper06}
and/or due to the enrichment of water molecule interstitials in the 2nd shell.
This finding rules out the possibility of glass-liquid decomposition and partial immiscibility in the NML since in that case
the 2nd g(r) peak position should shift downwards in the NML.
The height variation of the 2nd g(r) peak is also sufficient to differentiate the QLS from supercooled water above $T_s$ which is an ordinary liquid (see Inset Fig. 11(a)).
The intensified peak suggests enhanced, hidden structural correlation (or orientational order) inherent to the soft skeleton.
At $T \le T_g$ the soft skeleton of the QLS undergoes a vitrification phase transition to glassy water.  

The strong temperature dependence of the 2nd g(r) peak position ($r_{2ndpeak}$)
shows a crossover point at $\sim 190$ K and further supports the QLS concept (see Fig. \ref{r2ndpeak_chillp}). 
Within the NML all of these and other properties ($h_{2ndpeak}$,$r_{2ndpeak}$, density $\rho$, isothermal compressibility $\kappa_T$) show a strong T-dependence with a crossover point at $\sim 190$ K which feature convincingly supports
the unusual structural phase transition of deeply supercooled water. 
Careful inspection of g(r) reveals it can detect subtle structural changes, providing a clear signature of a distinct liquid state in the NML that is neither purely glassy nor an ordinary supercooled liquid.

{\em A new dynamic indicator of the QLS:}
Further support for the QLS scenario emerges from the temperature-dependent evolution of the atomic displacement distribution $(P(d_{displ})$), computed from unbiased 10 ns molecular dynamics (MD) simulations of our large-scale system with 49512 water molecules.
Below $T_g$ $P(d_{displ})$ gives sharp peaks (see Figs. \ref{ptow}, \ref{PdisplT} and S12) for the glassy state.
With increasing temperature the curves become broader.
However, in the NML the broader distribution is more asymmetric and non-gaussian with a growing tail (see Fig. \ref{PdisplT}(c)-(d) and S12). 
The pronounced difference between the $(P(d_{displ})$
curves calculated for the 49152-water-molecule system at 80 K and 190 K further indicates that deeply supercooled water in the NML is distinct from typical glassy water (see Fig. S12).
At 190 K the curve is even heavy tailed (see Fig. \ref{PdisplT}(c)) indicating that
dynamic heterogeneity peaks at this temperature \cite{Berthier2,Royall}.
This is in accordance with the work of S. Daschakraborty \cite{Daschakraborty26}.
Our simulations cannot rule out translational jumps entirely. It is likely that large-amplitude molecular displacements occur within the "no man's land," which are manifested in the long-range tail of the
atomic displacement distribution (the self-part of the van Hove correlation function) shown in Fig. 6(c) at 190 K.

 In the SW/NW regime ($T>T_s$) the $P(d_{displ})$ is even more broader than in the NML with a more gaussian-like distribution \cite{Berthier2,Royall}
 (Figs. 5, 6(e)-(f)).
Plotting the peak height to width ratio (p/w) of $P(d_{displ})$ as a function of the temperature we are able to distinguish the QLS/NML regime from the glassy and SW/NW regions (see Inset Fig. \ref{ptow}) even more effectively.
The p/w ratio acts as a critical indicator, signaling the onset of a structurally distinct QLS that precedes/replaces dynamic freezing in the NML.  

 {\em The temperature dependence of the diffusion constant:} The effective diffusion coefficient was also calculated via unbiased molecular dynamics (MD) simulations and plotted as a function of temperature (Fig. S13) \cite{Skinner18}. This plot reveals a pronounced deviation in the curve at the onset of the "no man's land" (see red line in Fig. S13) Additionally, we observe a characteristic peak at $T_g \approx 136$ K, which is attributed to the onset of the ultraviscous state prior to the glass transition. Consequently, our findings indicate that within the 150–210 K temperature regime (QLS), the diffusion coefficient operates in a distinct dynamic regime compared to both the true glassy state (blue line) and Arrhenius-like supercooled water (dashed curve).

Collectively, our arguments suggest that the anomalies in deeply supercooled water within the 'no-man's land' cannot be solely described by known, metastable phases, necessitating the invocation of a distinct liquid state to account for the unique, temperature-dependent properties.
The quasi-liquid state scenario provides a robust framework consistent with the anomalous properties of water observed within the no-man's land (NML) regime.

Although we utilize only a limited number of g(r)
-based descriptors—specifically the height and shift of the second peak position—along with the distribution of atomic displacements (the self part of the van Hove correlation function), these metrics are sensitive enough to distinctly separate the glassy (below ~136 K) and NML regimes in deeply supercooled water (see Figs. 3, 4, 5, S8, S9 and S11(a) and (c)).

{\scriptsize 
 
\section{METHODS}

\subsection{Molecular dynamics}

The largest cubic simulation box consisted of 49152 water
molecules, corresponding to the approximate box side of $11 \times 12 \times 12$ nm$^3$.
The preparation of the initial Ih ice simulation cell has been done by Genice2 \cite{genice}.
Smaller cubic simulation box including 2592 water molecules has also been used for longer simulations ($< \mu$s). 

All our simulations use 3D periodic boundary conditions.
Melting a periodic cubic simulation box filled by water molecules in a few hundred thousands of steps far over the melting temperature (3000 K) has been carried out
in order to generate a completely disordered ensemble of particles.
Than first, equilibration runs were performed in the NVT ensemble for 1 ns at 500 K, at equilibrium densities determined previously for atmospheric pressure.
Than the liquid water is kept at 300 K until equilibrium is reached ($\sim 1$ ns) using the NPT ensemble at 1 bar. 
Next we supercooled the water to 250 K and kept this system at 250 K until equilibrium is reached using again NPT simulation (few ns). Than this undercooled equilibrium system at 250 K (p=1 bar) is used for further simulations as a starting system. 
 As a next step the system is cooled down (or heated up) to the target temperature using the fast cooling rate of ~10-50 K/ps (hyperquench) under NPT at ambient pressure.
 The proper choice of the coupling constant of the barostat is crucial for avoiding spurious artificial volume oscillations (reasonable volume control).
 Therefore the barostat relaxation time (damping parameter) is selected within the range of 10 to 300.0 ps depending on the size of the simulation cell
 as well as on the temperature.
Throughout the paper we employ the relatively fast cooling rate of $\tau \approx 1-10$ K/ps) in order to ensure stable isoterm conditions both for incubation and for prenucleation to avoid the onset of prenucleation during the quench process with too slowly varying temperature.
We evolve at least 3 independent $\sim 5-50$ ns long simulation events at each selected temperatures, each initialized with different sets of random velocities (with Maxwell-Boltzmann distribution) consistent with the 
selected temperature.
We collect various parameters such as the density, potential energy/atom, diffusion constant  as a lammps thermo output.
In the case of Lattice Mold simulations the mold geometry including only the Oxygen atoms (fixed at the initial minimized Ih, cubic or stacking disordered Isd ice structures generated by Genice2) is inserted into the initial data file to be read by lammps
with zero velocity.
More gentle conditions can also be applied and tested during the quench process.
The cooling process can be divided into few steps in order to reach the target temperature slowly and to avoid spurious temperature-fluctuations.
However, even using such gentle conditions, the main conclusions remain invariant.

  In this work, we employ the TIP4P/2005 water
model \cite{tip4p2005} to study homogeneous ice nucleation at normal pressure combined with the Lattice Mold technique \cite{Mold,LM14,LMVega,LM} to accelerate the simulation of nucleation (biased MD).
The TIP4P/2005 model offers an efficient platform for investigating the physics of liquid-liquid transition also close proximity to the liquid–liquid critical point, due to its useful combination of computational efficiency and accuracy relative to available experimental data \cite{Gallo16,tip4p2005_llt}.
TIP4P/2005 model in general
has shown an excellent overall performance in reproducing
basic water properties \cite{Gallo16,Vega09a}. Its ability in accounting for the more challenging “anomalous” water properties \cite{Vega09b}, prompted us to use the model to investigate the supercooled region in a more detail. 

 Simulations were performed with the Large-scale
Atomic/Molecular Massively Parallel Simulator (LAMMPS) Molecular Dynamics package \cite{Plimpton1} in the NPT ensemble (with N equal to the
number of particles, P the pressure, and T the temperature), keeping
T nearly constant with the Nosé–Hoover thermostat \cite{Nose,Hoover1} at ambient pressure. 
Using NPT ensemble we are able to relax the density of the system during the simulations fixing the pressure
(P) with the Nosé–Hoover barostat.
 We integrate the equations
of motion using the velocity-Verlet integrator. The simulation time
step of $1$ fs is chosen for the rigid TIP4P/2005
model, and the thermostat relaxation time is $100$ ps for
the TIP4P/2005 model. 
Larger time steps (2-5 fs) are working in principle below $T_s \approx 225$ K, however, our simulations became unstable after few hundreds of ns simulation time.
It could be that 2 fs time step or more violate the conservation of the energy, equipartition in the water's no man's land.
Although we carefully checked that there is no energy drift in the simulations.
The accumulated error of the velocity-verlet algorithm could also be responsible for the instability of the generated MD trajectories over long time simulations. 
Although the algorithmic error is second-order, in long simulations, floating-point roundoff errors can accumulate, especially if the time step is too large.
We conclude from this is that the lifetime of the QLS (the arrested state) is surprisingly long e.g. at 190 K and at least it is in the range of $\sim \mu$s.
The very slow decrease of the potential energy, however, indicates that the system undergoes a slugish relaxation process during the incubation period. 

The barostat relaxation time is 500 ps for TIP4P/2005
model.
In order to further speedup simulations for the relatively large system (49152 water molecules) we set the potential cutoff at $7$ $\r{A}$ for TIP4P/2005. Moreover, for the
TIP4P/2005 model, long-range Coulombic interactions are treated
with the pppm/tip4p particle–particle particle-mesh (PPPM) solver using $9$ \r{A} cutoff
in LAMMPS with a relative error in forces of $0.00001$ .
 We also check that for the TIP4P/2005 model, the neglected potential
energy contribution of employing a short cutoff ($7$ \r{A} instead of a
long cutoff such as $12$ \r{A}) is lower than $\sim 0.06$ \% when evaluating the
potential energy of a bulk liquid phase applying PPPM and long-range corrections (for further details on potential truncation effects
on ice nucleation, please see \cite{pppm}).
Precise test calculations with long-range  LJ and Coulomb cutoff ($12$ \AA) together with much slower
thermostat ($\sim 1$ ps/K) and barostat ($50000$ ps) lead to nearly the same prenucleation process and arrested state as seen with the much shorter cutoff mentioned above.
 Finally, for the TIP4P/2005
model, we keep the O–H bond length and H–O–H angle values
invariant with the Shake algorithm as implemented in LAMMPS \cite{shake}.J. Chem. Phys. 162, 244505 (2025)

{\em Smaller system size, benchmark longer runs with unbiased MD simulations:}
Using our larger system ($49152$ water molecules) we are able to reach the $\mu$s simulation time (benchmark).
In most of the event, however, few tens of ns is reached practically.
The further increase of the simulation time is challenging with the available computational resources.
According to literature data (see e.g. ref. \cite{Gallo16})
the relaxation process in supercooled water or bringing the system into equilibrium before the onset of prenucleation requires much longer MD simulations, few hundreds of ns or even more in smaller systems.
In principle increasing the system size (pre)nucleation sets in sooner \cite{Gallo16}.
Our primary purpose is, however, to study the incubation period and to point out in the NML
that the deeply supercooled water gets into a specific long lived dynamically arrested state (called QLS) even before the onset of prenucleation.
In order to further support our results obtained for the larger system
much longer MD simulations have also been conducted using smaller system size including $2592$ molecules only.
We find similar results concerning the time evolution of the fraction of arrested particles and other monitored properties (fraction of Ih particles, density etc.).
No serious difference in various properties has been found up to 1000 ns simulation time reached. 

{\em Well-tempered metadynamics:}
Enhanced sampling and free energy calculations were performed using the PLUMED v2.9.1 plugin via well-tempered metadynamics (WT-MetaD) as integrated into LAMMPS \cite{metaD1,metaD2}. An adaptive, history-dependent bias potential was constructed by depositing Gaussian kernels along selected collective variables (CVs) at regular time intervals, with the negative of the accumulated bias yielding the final free energy surface (FES).
The simulations were conducted under NPT conditions at 190 K and 1 atm, employing a 1 fs timestep and the TIP4P/2005 force field. The final configuration from a 1 $\mu$s
unbiased MD simulation served as the initial structure for the WT-MetaD runs including 2592 water molecules. The coupling constants for the thermostat and barostat were set to 1 ps and 10 ps, respectively.
The metadynamics simulations utilized the following parameters:
System volume (density) and total potential energy were selected to monitor and drive structural transitions as collective variables (CVs).
Gaussian Depositions: An initial Gaussian height of
$\omega=1.1$ kJ/mol, Gaussian widths of $\sigma_{den}=0.5$ and $\sigma_{ene}=50.0$ and a deposition stride of 0.5 ps.
Well-Tempered Scaling: A bias factor of $\gamma=50$
was chosen to scale down the Gaussian heights over time, ensuring smooth convergence of the FES.
This relatively large bias factor is necessary to efficiently fill the deep potential energy well of the low-density liquid (LDL) phase. Care was taken to avoid overbiasing; test simulations with a higher bias factor ($\gamma=100$) resulted in an artificial, poorly converged "puddled" potential energy surface. To guarantee proper free energy convergence, the time evolution of the FES was rigorously monitored to ensure that the added energy hills did not prematurely "surf" a rough, non-converged landscape characteristic of glassy dynamics.

{\em Structural Analysis:}
The atomic configurations obtained from simulations were analyzed by the OVITO code \cite{stuko}. Chill+ \cite{chillp}  
has been used to identify various phases (Ih, cubic ice and others). 
The displacement of each atoms ($d_i(t)$ where $i$ is the atomic index, $t$ is the instantaneous time) with respect to the corresponding reference configurations (the final Ih ice crystal) were calculated within the OVITO code. 
In many cases we also calculated displacement maps with respect to the initial state (unbiased MD). . 
As far as solidification is concerned $d_i(t)$ is the distance what atom $i$ yet to be displaced to take its final position in the Ih ice crystal at $t$ instantaneous time. The proper indexing (sorting) of atoms during MD simulation were ensured by the specific "dump\_modify" command in LAMMPS.

}

\normalsize

\section{Conclusions}

While low-density liquid-like water is well-characterized in its glassy state (low-density amorphous ice, LDA), the behavior of water within the deeply supercooled 'no man’s land' (NML) remains a subject of intense debate.
Here, we report novel structural and dynamical features of the low-density liquid (LDL) phase identified within this region.
% Here, we report the identification of novel features of the low-density liquid (LDL) phase within this region, exhibiting unique structural and dynamical properties. Our findings bridge the gap between amorphous solid water and the liquid-liquid transition hypothesis, offering new insights into the anomalous properties of water at the nanoscale.
Our results demonstrate that the complex heterogeneity of this low-density quasi-liquid state (QLS) arises from the nanoconfinement of mobile particles within a spontaneously formed, soft, self-assembled nanoporous skeleton. Furthermore, we establish a robust synergy between the self-confinement of the NML water fraction and the resulting dynamical and structural arrest. The QLS exhibits hallmarks of a nanoconfined fluid within this self-assembled superstructure, maintaining a distinct confined-state signature.

Outside the NML, dynamic heterogeneity and confinement vanish: below $\sim 140$ K the systems become glassy, while above the crossover temperature $T_s \approx 225$ K, the undercooled systems behave like ordinary liquids. However, within the NML, the supercooled water is both dynamically and structurally heterogeneous, with trapped particles forming networked clusters. We argue that within the NML, the supercooled liquid is neither a frozen glass nor a spinodal (unstable) liquid, but a structurally distinct, metastable liquid phase that exists prior to crystallization.
Our results indicate that the NML supercooled liquid possesses significant non-crystalline local order, distinguishing it from both a simple liquid and a fully arrested glassy state. 

This remarkably stable quasi-liquid state can be sustained under tailored conditions, providing an ideal, long-term opportunity for direct in situ experimental characterization. 
Furthermore, the striking features of self-confined, arrested LDL water provide a valuable model for studying other aqueous solutions and hydrated biomolecules under supercooling conditions, such as in cryopreservation.

%\section{Visual Abstract}

\begin{figure*}[hbtp]
	\includegraphics*[height=8cm,width=14cm,angle=0.]{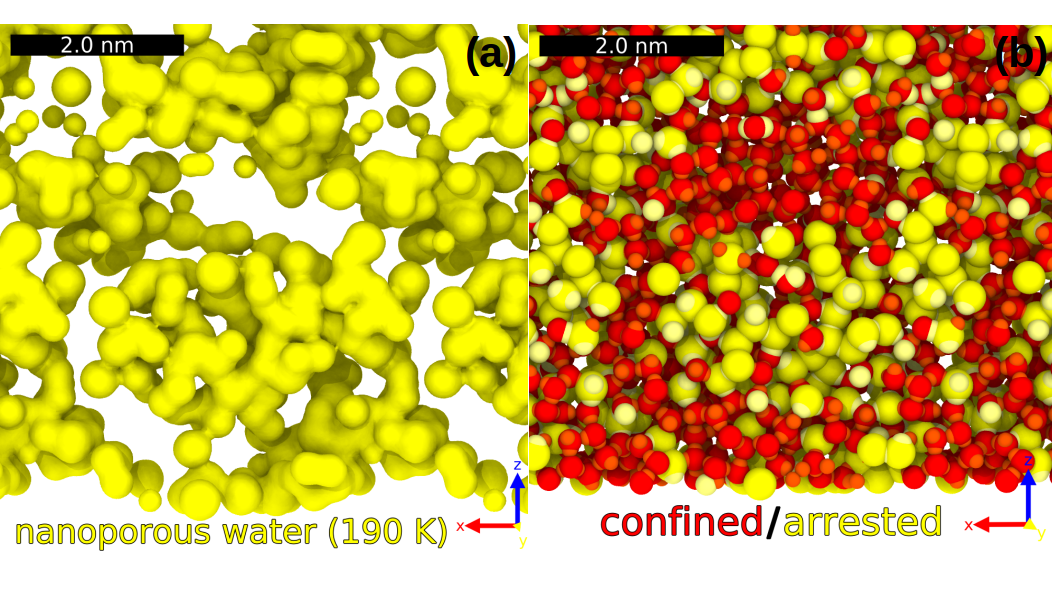}
	\caption[]{
		Visual Abstract:
		Structural representation of the unfolded nanoporous water network embedded within supercooled low-density liquid (LDL) water at 190 K (on the left)
	  and the same superstructure accommodating self-confined, liquid-like water molecules (on the right).
%		A molecule is classified as arrested if its average atomic displacement was less than 2.5 \r{A} over the final 100 ns of the 1 $\mu$s simulation.	
%		Confined molecules are those which displaced more than 2.5 \r{A}.
%		Thin slices of the simulation box along the x-axis are shown for the periodic system.
	}
	\label{TOC}
\end{figure*}

%\vspace{2cm}

%\section{Supplementary Material}

\section{Supporting Information Available}

Further Figures. Brief overview of the quasi liquid state. Further details on Methods (Inherent structures, Hippo force field, Lattice Mold technique, isothermal compressibility). 
Video animations are shown for the time evolution of various systems at 80, 150, 190 and 225 K using unbiased MD.
The nucleation of the stacking-disordered Isd phase has also been shown using biased (LM) MD (S3). 
The sluggish glassy dynamics experienced at 80 K is also animated (S4).
The quasi liquid state is animated at 150 K (S5).
The more liquid-like dynamics at 225 K can also be seen (S6).
In the animations the Gaussian density method has been used to construct the surface mesh which fits the best the most likely molecular shape of the water molecules \cite{stuko}.

Video S1(a): Animation of the nanoporous arrested system ("soft skeleton") at 190 K ($1 \mu$s simulation time) shown in Fig. \ref{displ190K}. Nanoconfined liquid atoms are not shown. The thin slice cut in the diagonal of the simulation cell is shown (unbiased). 

Video S1(b): The same as S1(a) with full volume (unbiased).

Video S2: Animation of the dynamics at 225 K (liquid, unbiased)

Video S3: The animation of the self-confined water (red oxygen atoms) within the nanoporous "soft-skeleton" (yellow oxygen atoms).

Video S4: Animation of the nucleation of the mixed stacking-disordered (Isd) phase (mixed cubic/hexagonal/liquid). 
Thin slice of the simulation cell is shown along the diagonal (biased MD).

Video S5: Animation of the dynamics at 80 K (glassy system, unbiased).

Video S6: Animation of the dynamics at 150 K (QLS, unbiased).

%{\bf Author Information}\\

%{\bf Authors}
%Péter S\"ule, Dept. of Nanostructures, Centre for Energy Research, HUN-REN, Budapest, Hungary,
%\email{sule.peter@ek-cer.hu}

\section{Data availability}
All data supporting the findings of this study are available within the paper and its Supplementary Information.
The datasets generated and/or analyzed during the current study are not publicly available due to the large size of the raw simulation trajectories and intermediate files and the associated storage/transfer limitations but are available from the corresponding author on reasonable request.
%The data that support the findings of this study are available from the
%corresponding authors upon request.

\section{author contributions}
P. S. designed the theoretical model and performed the
simulations. P. S. wrote the manuscript.

\section{Competing interests}
The author declares no competing interests.

\section{additional information}
%Supplementary information is available for this paper at https://doi.org/10.1038/
%s41563-019-0514-1.
Correspondence and requests for materials should be addressed to P. S.

\section{Code availability}
Custom gnuplot codes for image processing of d-maps as well as the
scripts for LAMMPS MD simulations or OVITO settings are available from
the corresponding authors upon request.

\section*{Acknowledgements}
This work is supported by National Science Foundation of Hungary grant no. 
KKP-138144, K-134258, FK-142985.
The simulations were performed on the national HPC resources of KIFÜ (Cray-EX: Komondor).

\appendix
\clearpage
%\onecolumn

\section{Supplementary Material}
% Reset equation/figure counters with 'S' prefix if desired
\setcounter{equation}{0}
\setcounter{figure}{0}
\renewcommand{\theequation}{S\arabic{equation}}
\renewcommand{\thefigure}{S\arabic{figure}}

% Input your supplementary text or write it here
%\input{supplement_content}

%\appendix
%\renewcommand\thefigure{\thesection.\arabic{figure}}    
%\setcounter{figure}{0}   
%\renewcommand{\appendixname}{S}}

%\begin{appendices}
%	\addtocontents{toc}{\protect\setcounter{tocdepth}{1}}
%	\setcounter{chapter}{1}
%	\chapter*{Appendices}
%\setcounter{figure}{0}
%\renewcommand{\appendixname}{S}

%\appendix
%\renewcommand{\appendixname}{S}
%\renewcommand{\thesection}{S\arabic{section}}
%\setcounter{figure}{0}
%\renewcommand{\figurename}{FIG. S}

%%%%%%%%%%%%%%%%%%%%%%%%%%%%%% 
% S1
%%%%%%%%%%%%%%%%%%%%%%%%%%%%%
\begin{figure*}[pt]
	
	\includegraphics*[height=4cm,width=3.5cm,angle=0.]{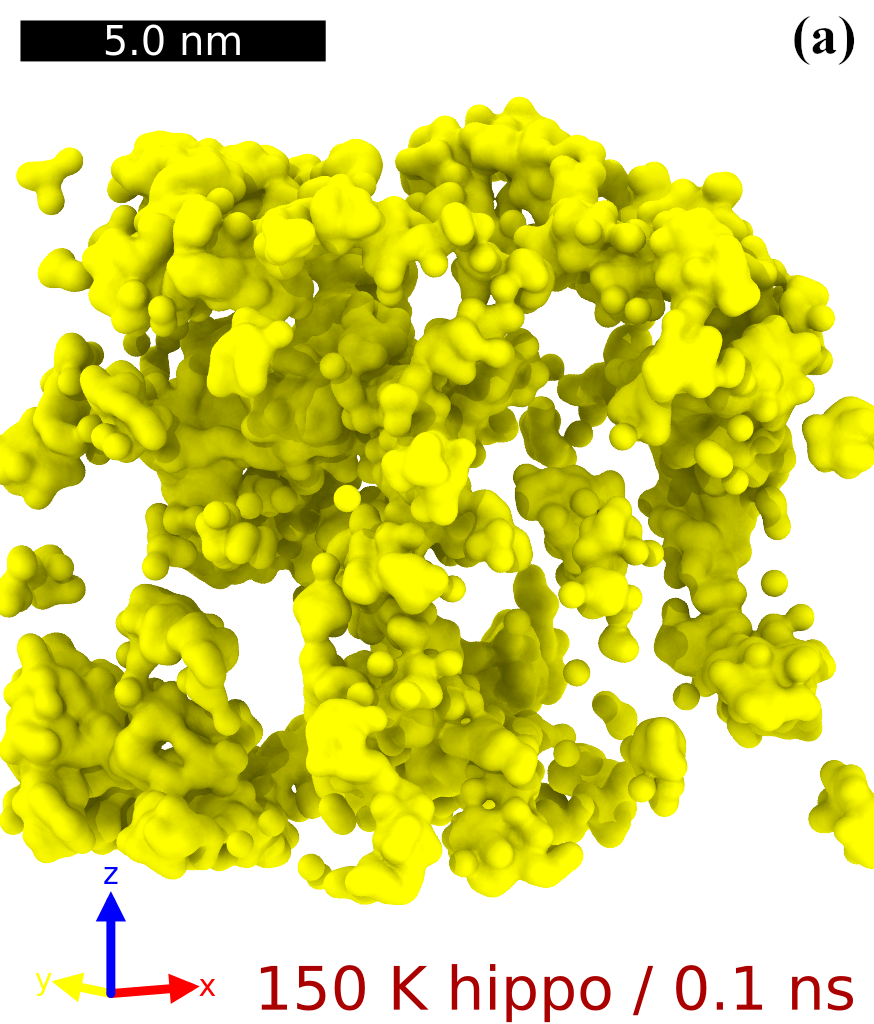}
	\includegraphics*[height=4cm,width=3.5cm,angle=0.]{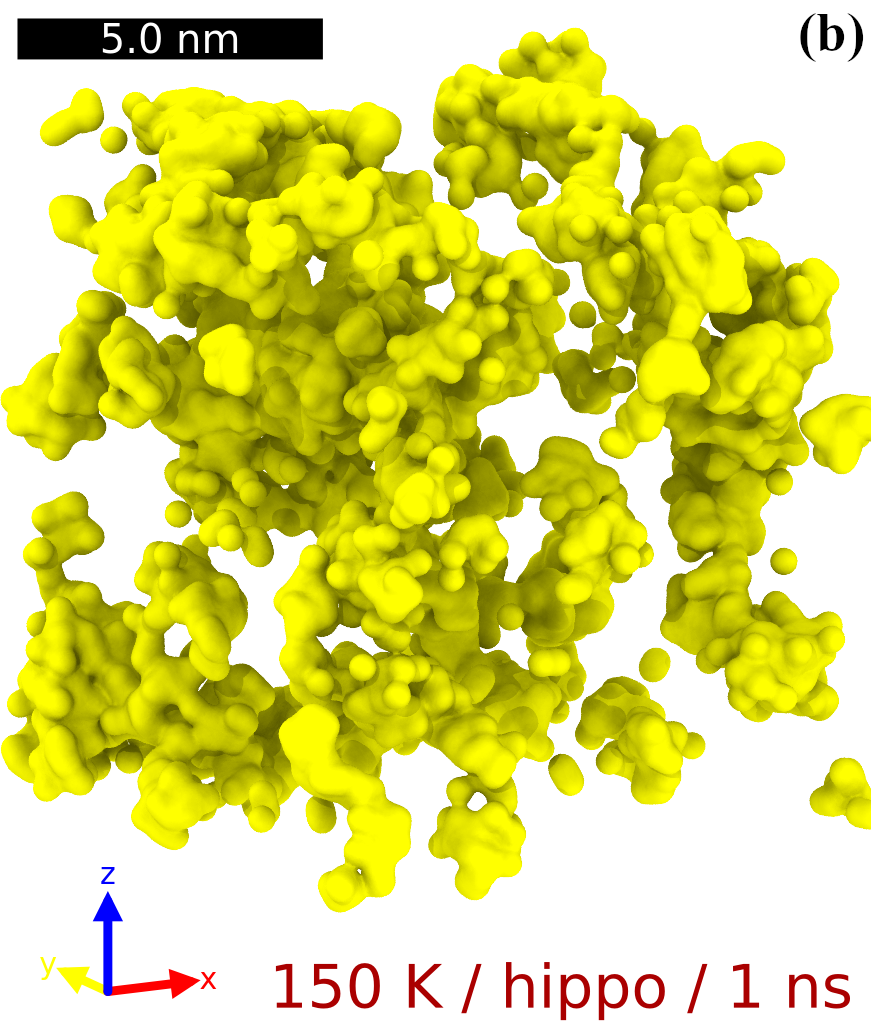}
	\includegraphics*[height=4cm,width=4.8cm,angle=0.]{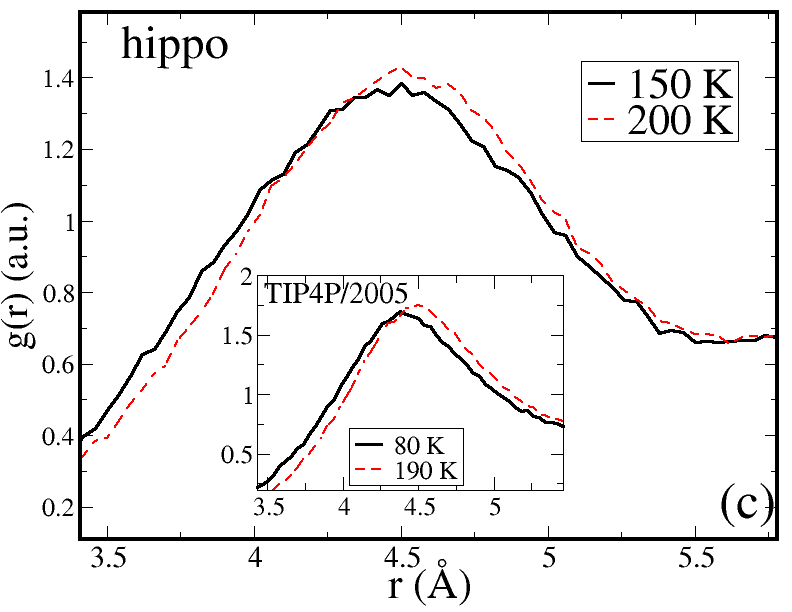}
	
	\caption[]{ 
		{\bf The unfolded trapped superstructure calculated by hippo and the comparison of the g(r)
			at 0.1 ns (a) and at 1 ns (b).}
		Unbiased MD simulations using the hippo polarizable force field \cite{hippo}.
		The particles which form the unfolded soft skeletons
		are selected using the d-filter, $d_i \le 2 \r{A}$ referenced to the nearly initial state at $0.1$ ns (b)
		and to the final state at 1 ns (a).
		Note the similarity between the instantaneous structures at 0.1 and 1 ns which suggests that the hidden internal superstructure persists at least for a ns.
		The structures were unfolded using the dc-filter. The details of the filtering process is given
		in the caption of Supplementary Fig S9.
		(c) The 2nd peak of the OO g(r) calculated by unbiased MD using hippo and TIP5P/2005
		at various temperatures, respectively. Time averaged structures ($t_{ave}=5$ ns) used for the g(r).
		Note the tiny shift of the peak position with respect to the glassy state at 80 K.
	}
	%\end{center}
	%	\footnotemark{
	%	}
	\label{hipposkelet_gr}
\end{figure*}

%%%%%%%%%%%%%%%%%%%%%%%%%%%%%% 
% S2
%%%%%%%%%%%%%%%%%%%%%%%%%%%%%
\begin{figure*}[pt]
	
	\includegraphics*[height=6cm,width=8cm,angle=0.]{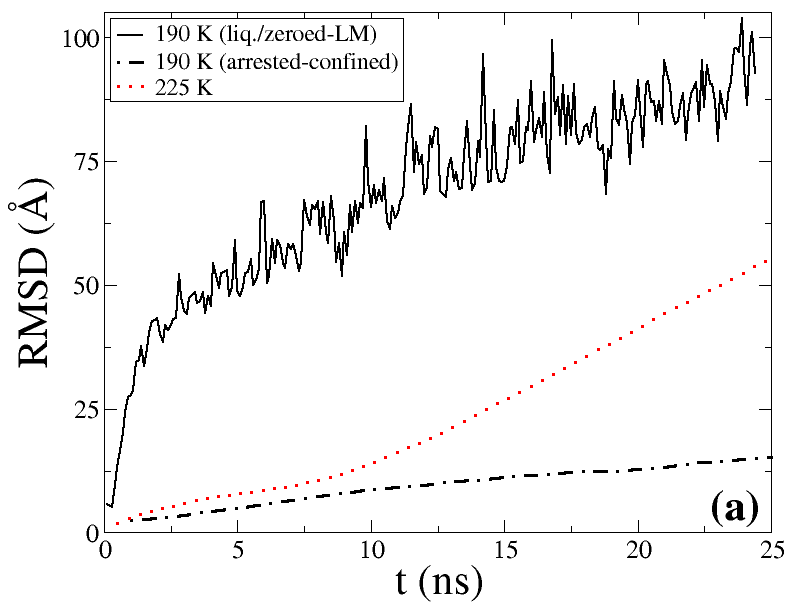}
	\includegraphics*[height=6cm,width=8.5cm,angle=0.]{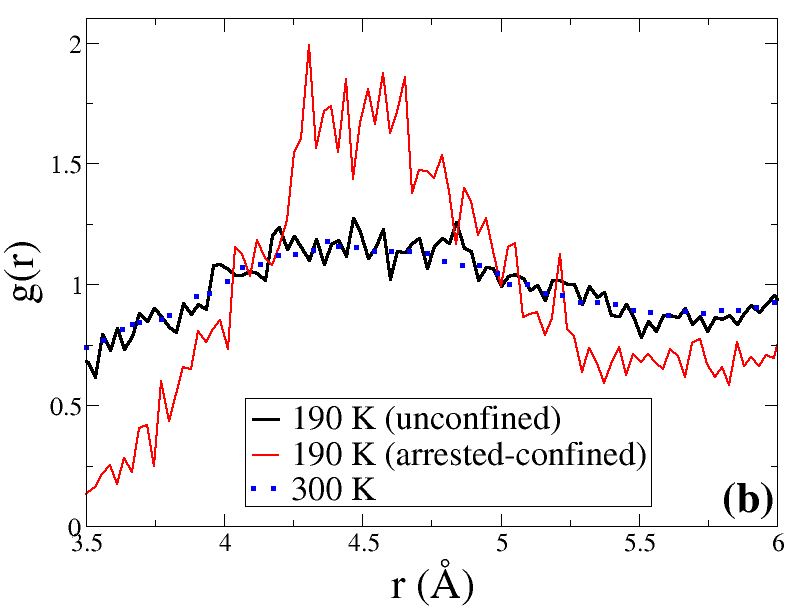}
	
	\caption[]{
		{\bf Root mean square of atomic displacements (RMSD) of the liquid phase.}
		(a) The comparison of the RMSD of the self-confined arrested system (the RMSD of the confined particles, the contribution of the arrested particles is excluded) with the unconfined (ordinary liquid, liq./zeroed-LM) system at 190 K.     
		In the unconfined liquid system particle arrest is destroyed by the core repulsion of the zeroed-square well potential leading to a
		conventional liquid state. 
		The RMSD at 225 K is also shown: particle arrest is also missing (or reduced significantly) at this crossover temperature.
		(b) The 2nd peak of the g(r) of the self-confined and unconfined systems at 190 K. The peak is also shown for 300 K. 
	}
	
	\label{figS1}
\end{figure*}

%%%%%%%%%%%%%%%%%%%%%%%%%%%%%% 
% FS3
%%%%%%%%%%%%%%%%%%%%%%%%%%%%%
\begin{figure}[pt]
	
	\begin{center}
		
		\includegraphics*[height=6.5cm,width=8cm,angle=0.]{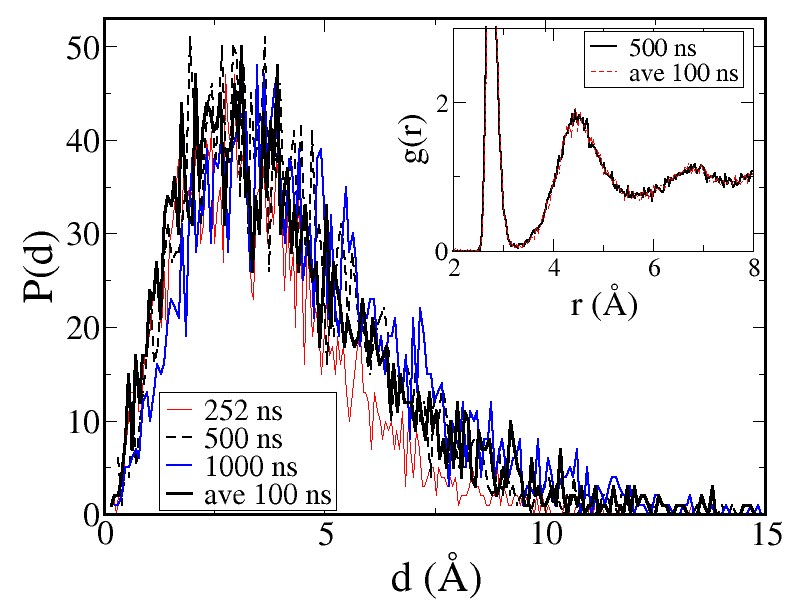}
		
		\caption[]{ 
			{\bf The distribution of the atomic displacements ($P(d_{displ})$) at 190 K at different elapsed times.}
			The $d_{displ}$ values have been calculated with respect to the initial stages at $\sim 10$ ns simulation time.
			The TIP4P/2005 force-field is used (unbiased MD).
			{\em Inset:} The radial distribution function g(r) at 190 K at 500 ns and for the time-averaged system over 100 ns (900-1000 ns).
		}
		
	\end{center}
	
	\label{figS3}
\end{figure}

%%%%%%%%%%%%%%%%%%%%%%%%%%%%%%
% S4
%%%%%%%%%%%%%%%%%%%%%%%%%%%%%
\begin{figure*}[pt]
	
	\includegraphics*[height=4cm,width=5.5cm,angle=0.]{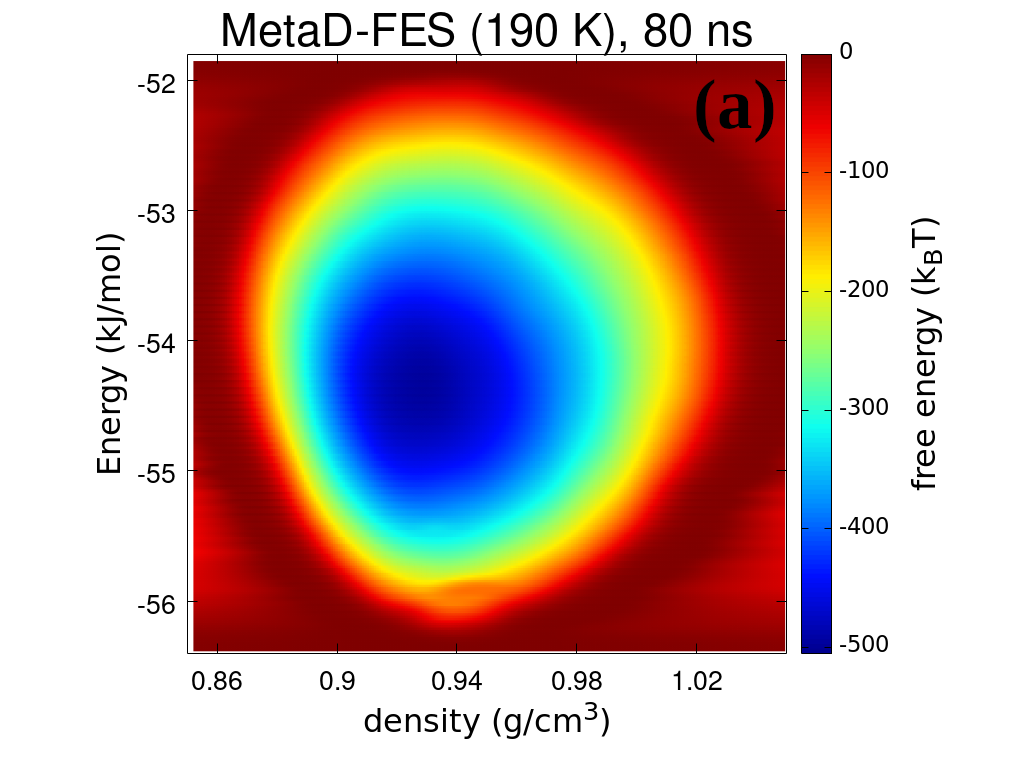}
	\includegraphics*[height=4cm,width=5cm,angle=0.]{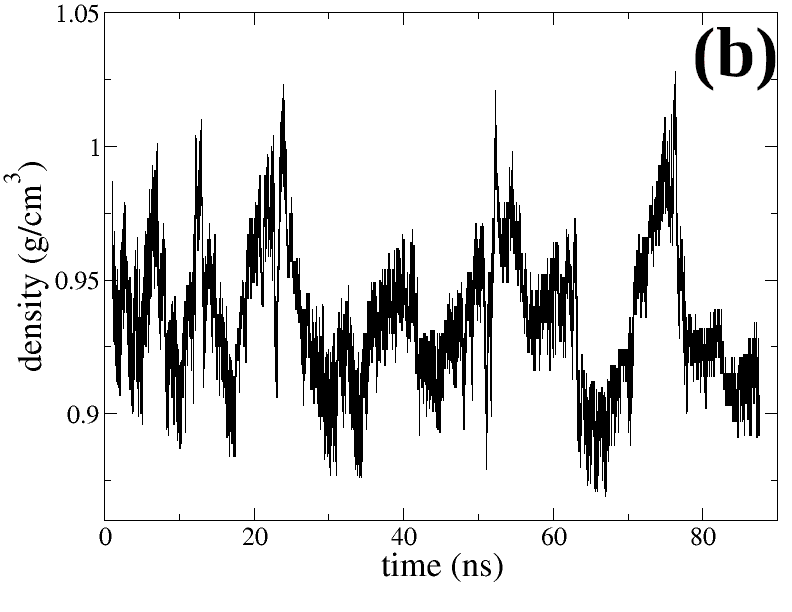}
	\includegraphics*[height=4cm,width=5cm,angle=0.]{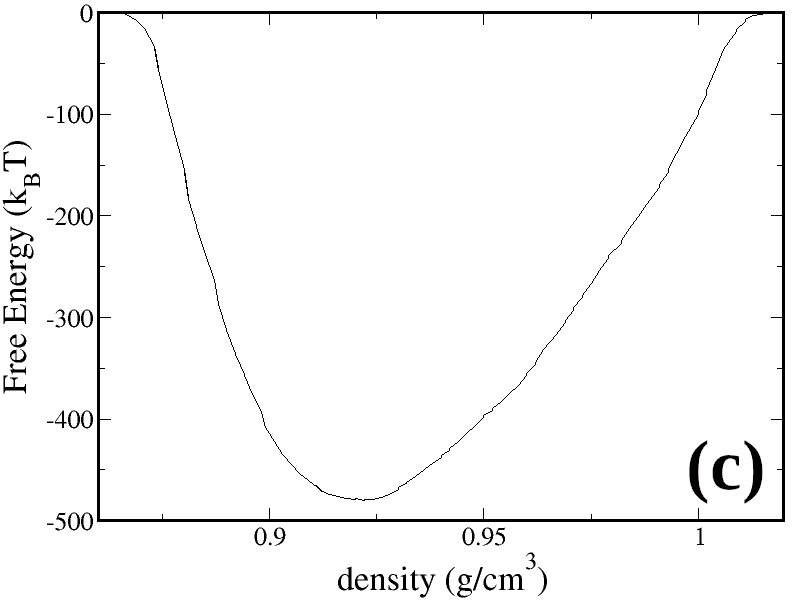}
	
	\includegraphics*[height=4cm,width=5.5cm,angle=0.]{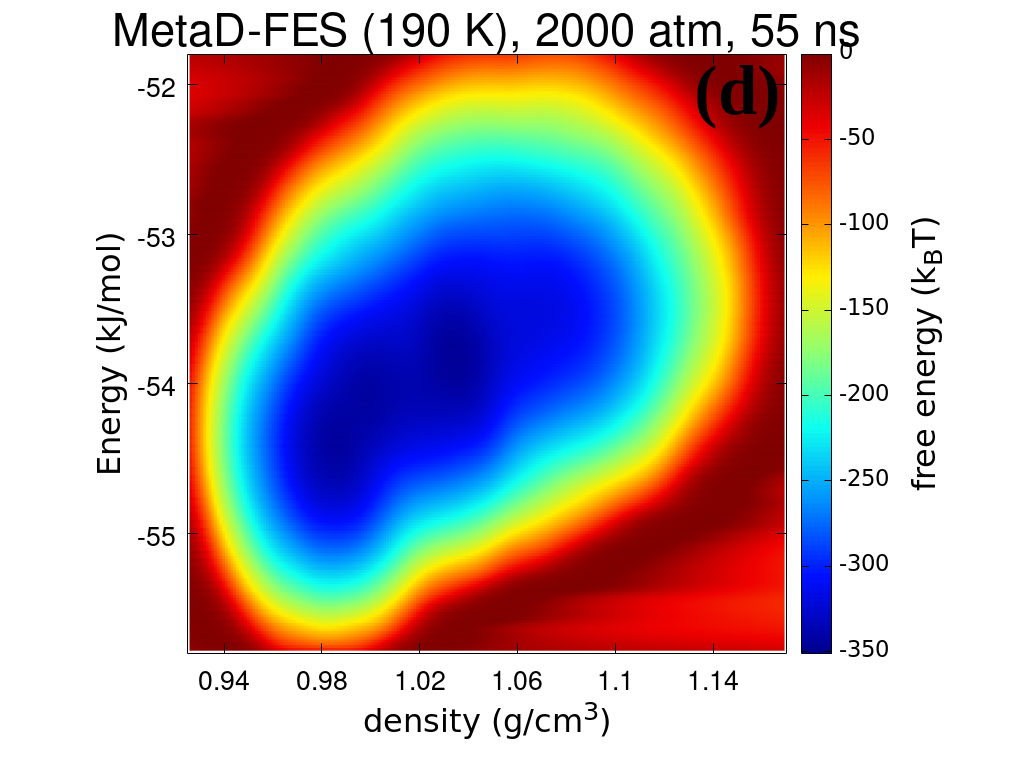}
	\includegraphics*[height=4cm,width=5cm,angle=0.]{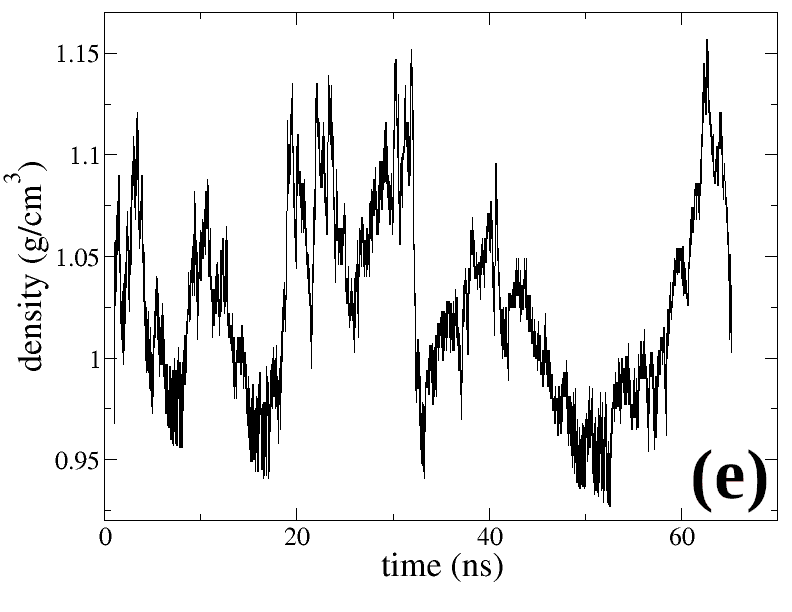}
	\includegraphics*[height=4cm,width=5cm,angle=0.]{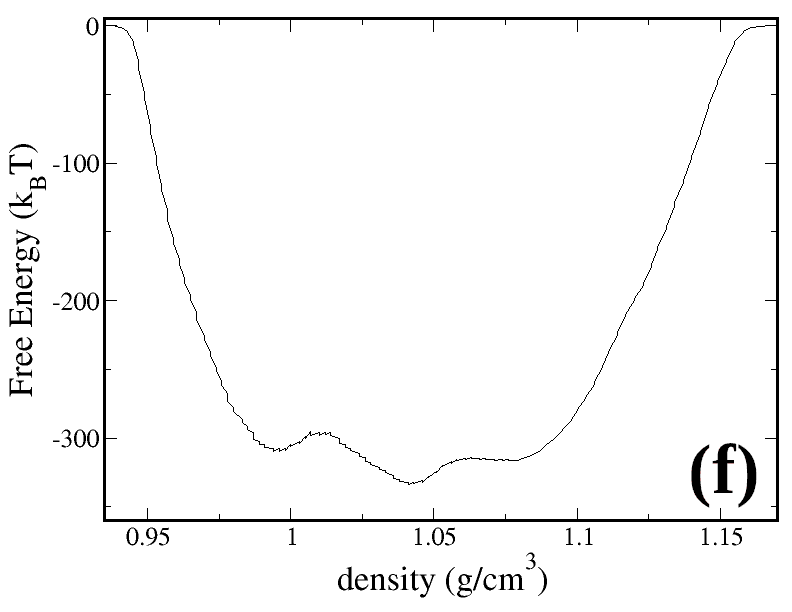}
	
	\caption[]{ 
		{\bf WT-MetaD simulation results at 190 K}
		(a)-(c) At ambient pressure, ((d)-(f) at pressure p=2000 atm (2.026 kbar).
		(a) The Free energy surface (FES) at 80 ns in thermal energy ($k_BT$).
		(b) The density ($\rho$) profile as a function of the simulation time $t$ (ns).
		(c) Line cut along the diagonal of the FES. 
		(d) FES at 55 ns (p=2k atm, $k_BT$).
		(e) $\rho(t)$ at 2k atm pressure.
		(f) Line cut at 2k atm pressure.
	}
	\label{figS4}
\end{figure*}

%%%%%%%%%%%%%%%%%%%%%%%%%%%%%%
% S5 
%%%%%%%%%%%%%%%%%%%%%%%%%%%%%
\begin{figure*}[hbtp]
	%	\begin{center}
	
	\includegraphics*[height=3.2cm,width=4.3cm,angle=0.]{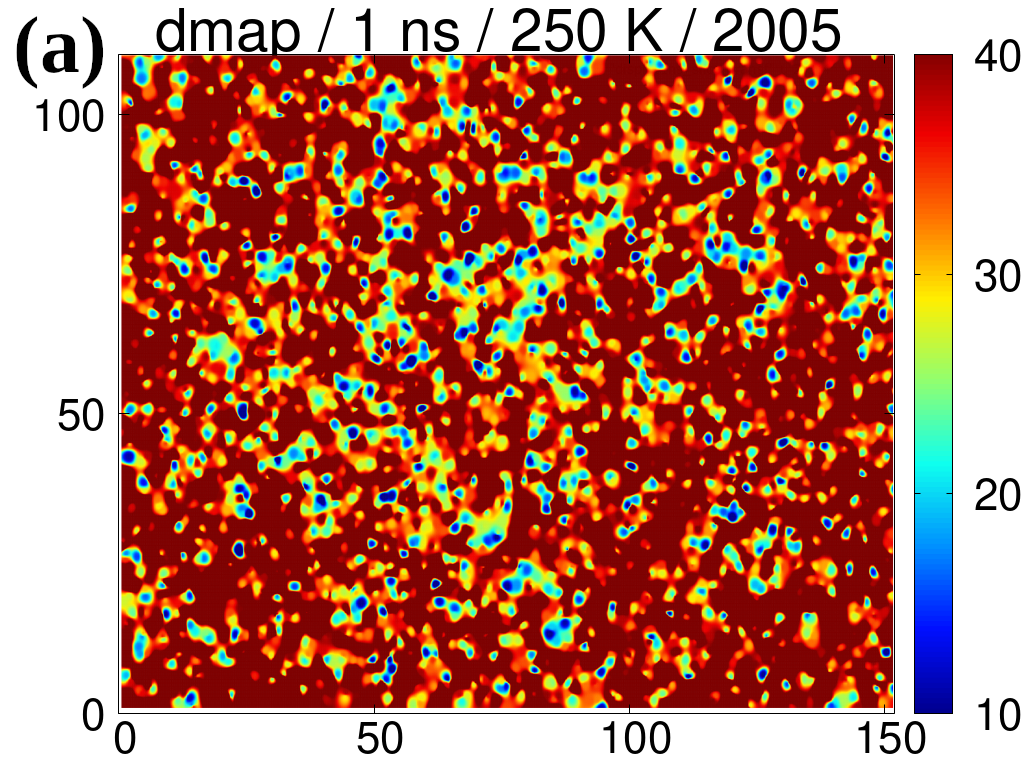}
	\includegraphics*[height=3.2cm,width=4.3cm,angle=0.]{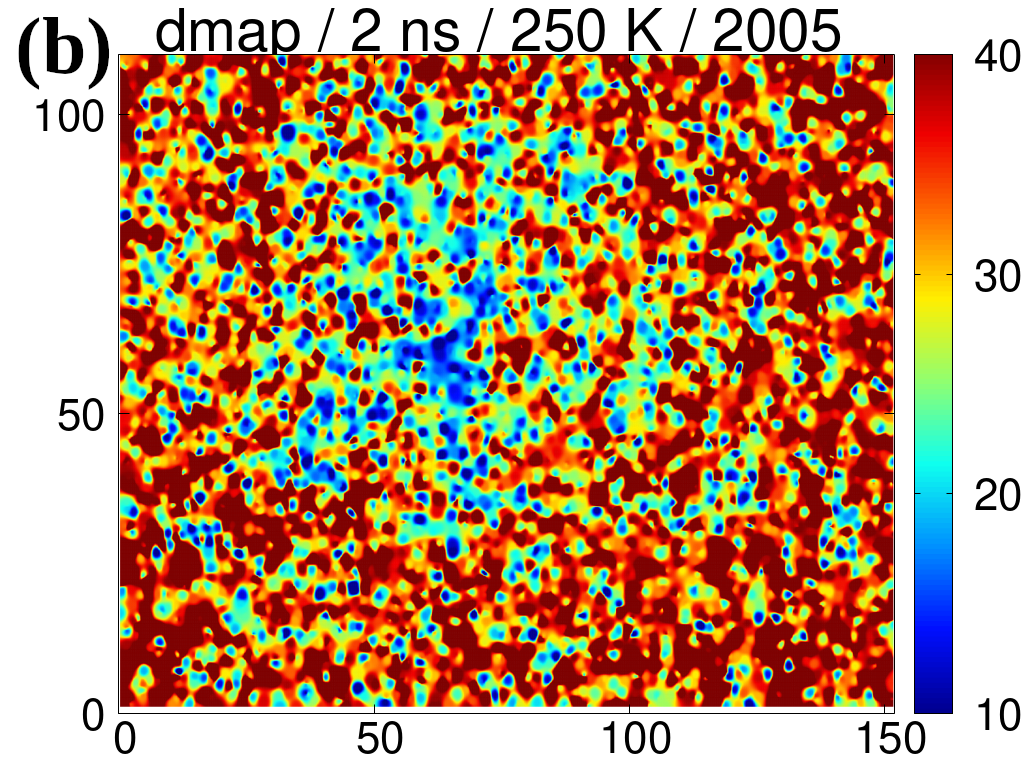}
	\includegraphics*[height=3.2cm,width=4.3cm,angle=0.]{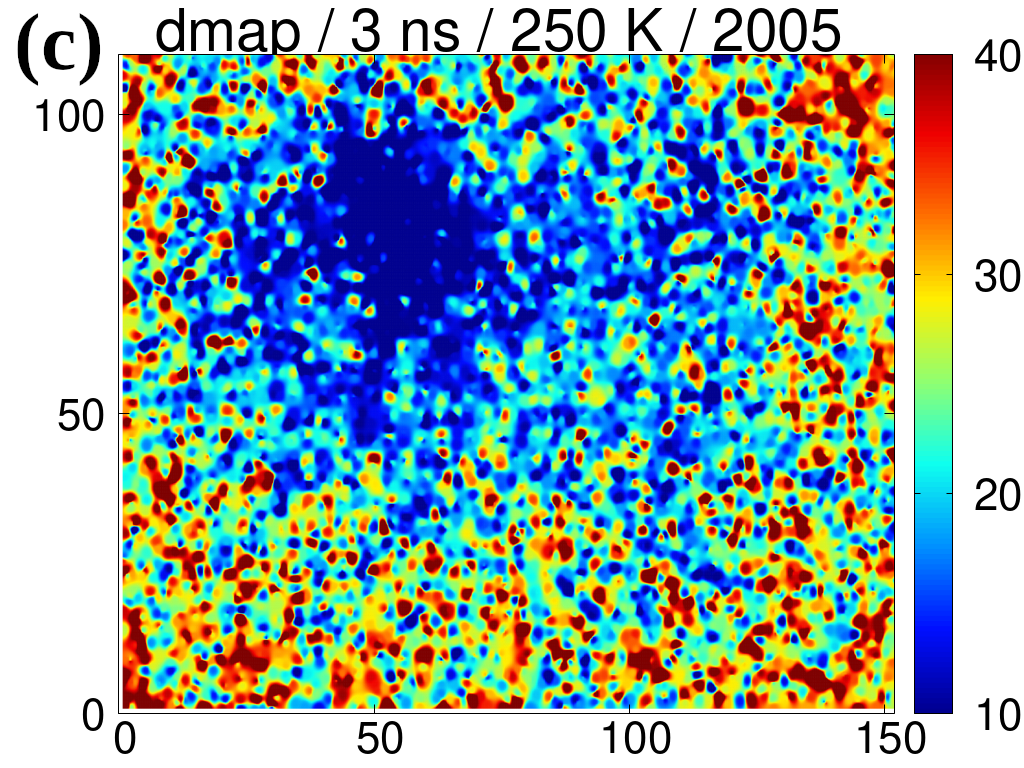}
	\includegraphics*[height=3.2cm,width=4.3cm,angle=0.]{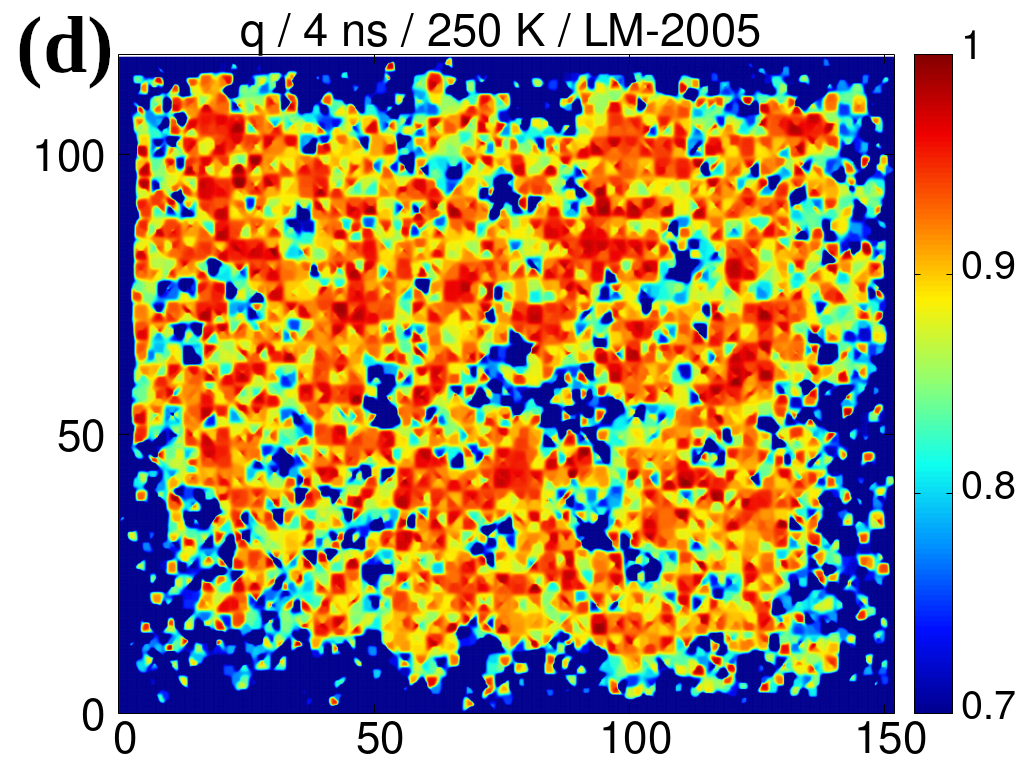}
	
	\includegraphics*[height=3.2cm,width=4.3cm,angle=0.]{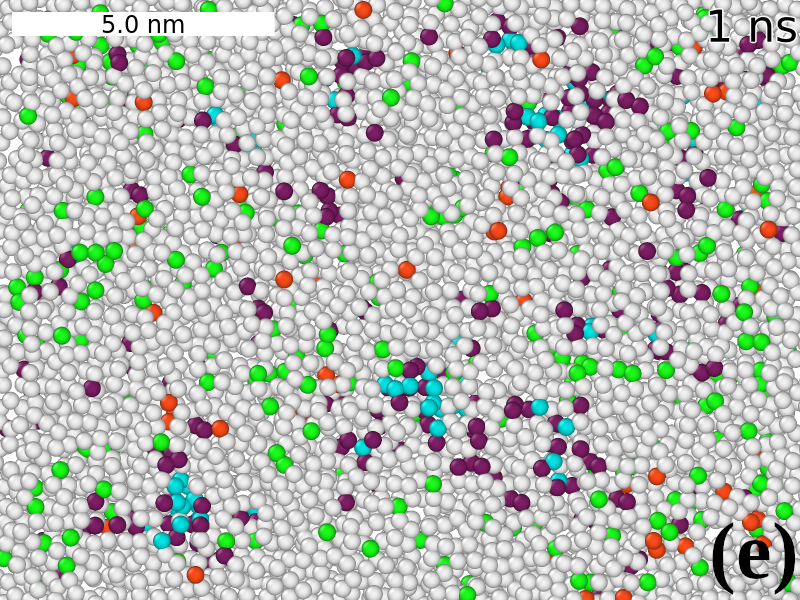}
	\includegraphics*[height=3.2cm,width=4.3cm,angle=0.]{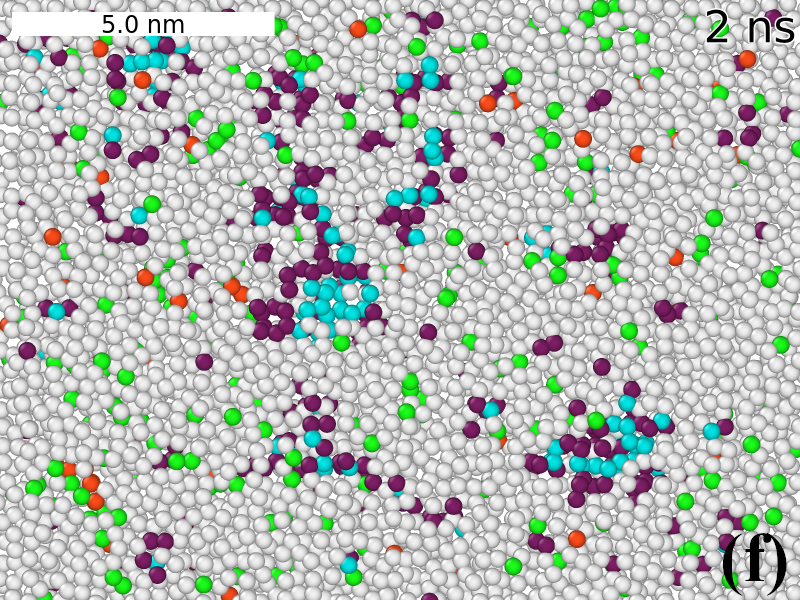}
	\includegraphics*[height=3.2cm,width=4.3cm,angle=0.]{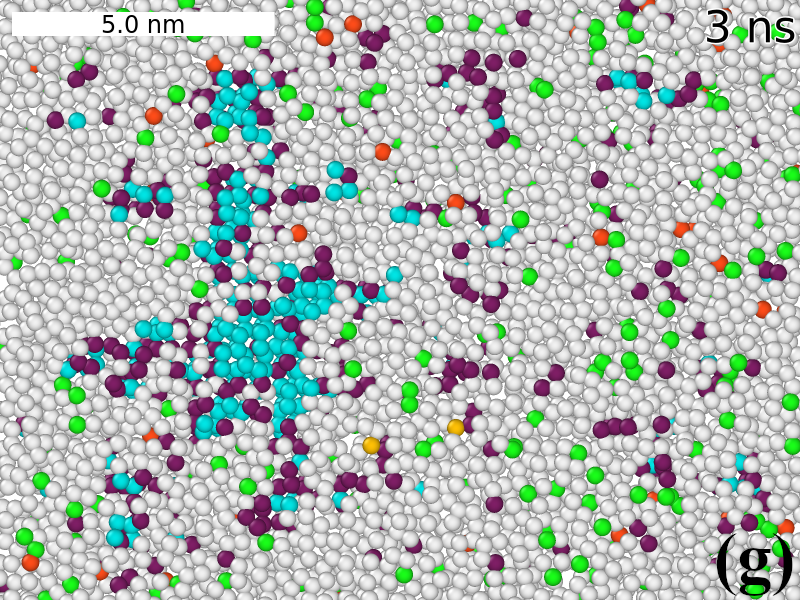}
	\includegraphics*[height=3.2cm,width=4.3cm,angle=0.]{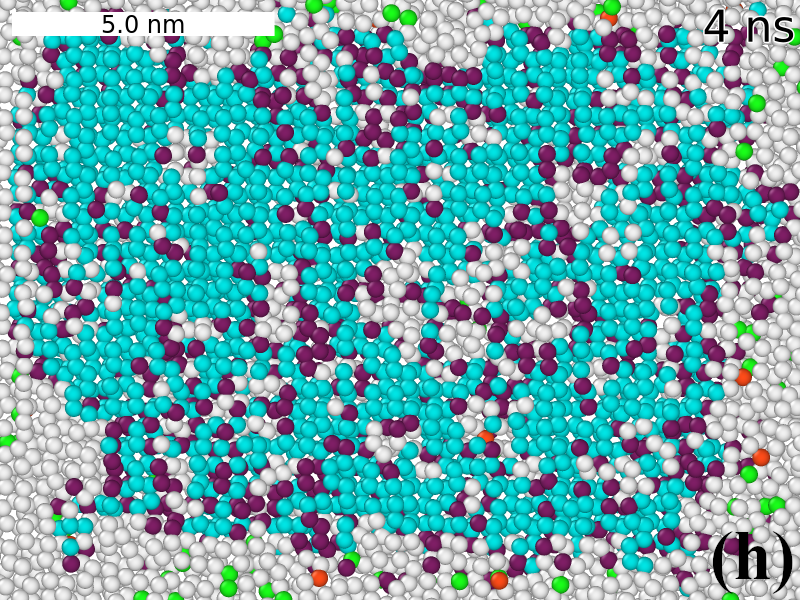}

	\caption[]{
		{\bf Spatially resolved maps for (pre)nucleation at 250 K using biased MD simulations.}
		(a)-(d) Atomic displacement (with respect to the final state shown in (h)) maps at 1, 2, 3 ns simulation time.
		At 4 ns the tetrahedrality (q) map is shown. 
		The phase development of freezing has been followed at 250 K using the Lattice Mold technique (LM-TIP4P/2005) by MD.
		(e)-(h) Corresponding panels with Chill+ plots as obtained using OVITO. Structure types: blue = Ih, red = hydrate, green = interfacial hydrate,  purple = interfacial ice.
		Thin ($\sim 1$ nm) slices cut along the diagonal of the simulation cell were used to display the various maps by gnuplot.
	}
	
	\label{maps250K}
\end{figure*}

%%%%%%%%%%%%%%%%%%%%%%%%%%%%%%
% S6 (anno F5) 
%%%%%%%%%%%%%%%%%%%%%%%%%%%%%
\begin{figure*}[hbtp]
	
	\includegraphics*[height=5.5cm,width=7cm,angle=0.]{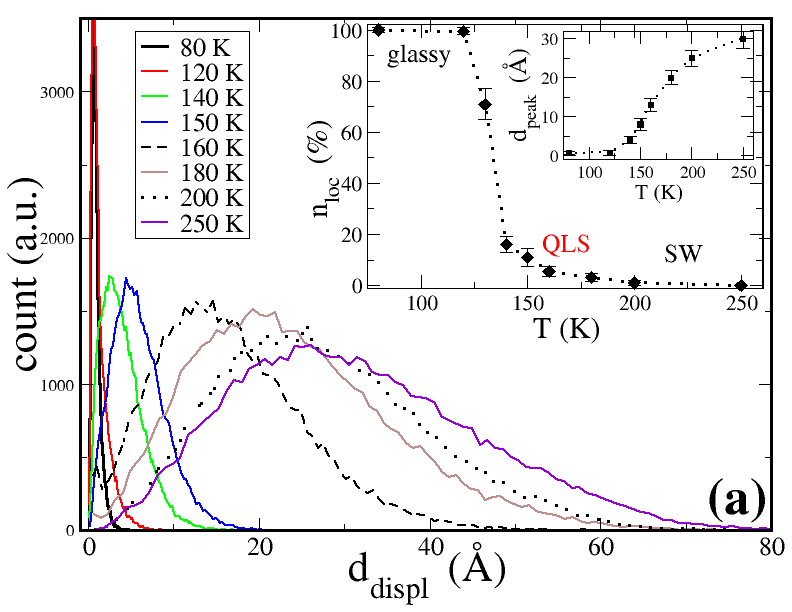}
	\includegraphics*[height=5.5cm,width=7cm,angle=0.]{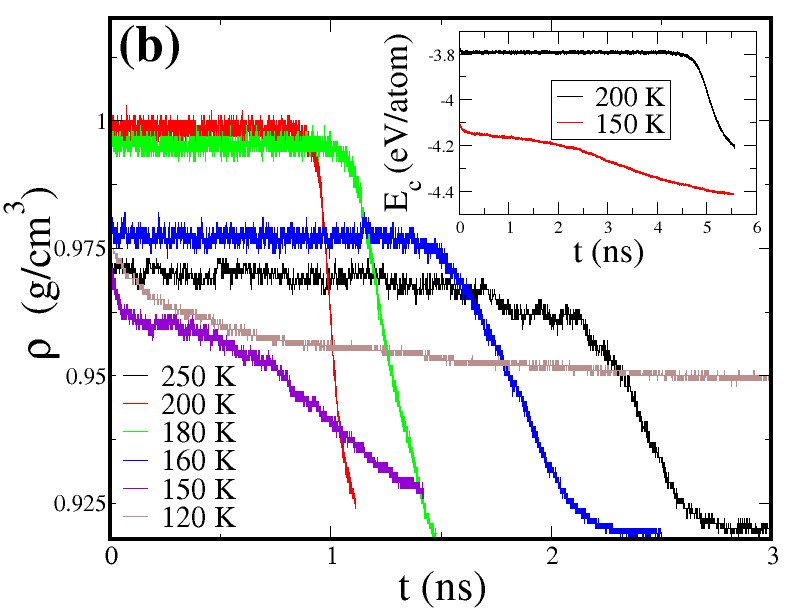}
	\includegraphics[height=5.5cm,width=7cm]{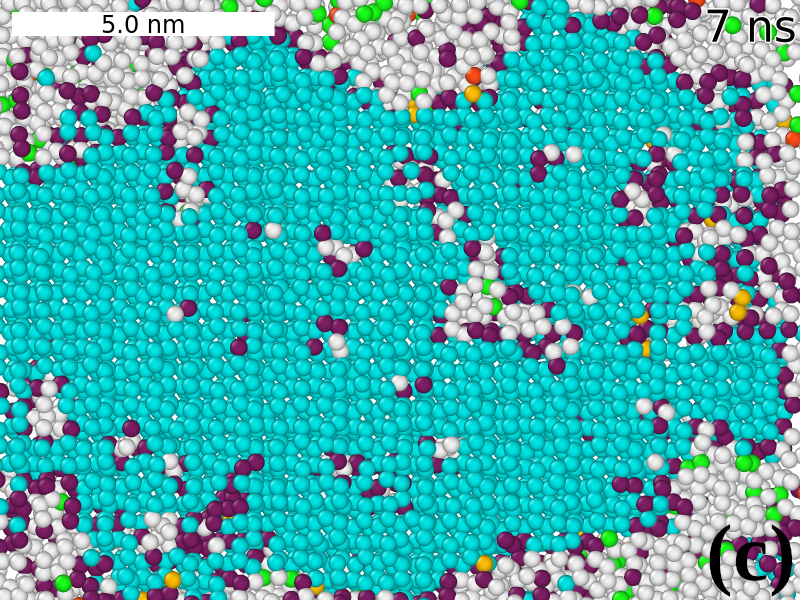}
	
	\caption[]{
		{\bf P(d), the fraction of immobile particles (Inset (a)), the time evolution of the density (b) (biased MD).}
		(a) The histogram of atomic displacement distribution P(d) with respect to the final crystalline state.
		At 80 and 120 K no Ih ice phase is reached, therefore the final glassy state at around 50 ns is used as a reference state.
		The curves were sampled close to the end of the prenucleation phase where the proportion of the Ih ice phase is around $0.5$ \%
		(as detected by Chill+).  
		{\em Inset:} The fraction of localized particles ($n_{loc}$, \%) as a function of the temperature (K) before the onset of nucleation.
		Those particles are considered nearly immobile which displaced less than $d=4$ \r{A} to the final crystalline position
		from the end of the prenucleation period. 
		{\em Inset in the Inset:} The peak positions of the histogram of atomic displacements ($d_{peak}$, $\r{A}$) at the end of prenucleation.
		Error bars are due to the uncertainty in sampling as well as to the 3 events considered at each temperature.
		(b) Time-evolution of the density (g/cm$^3$) at different temperatures (K).
		(c) Chill+ map at 7 ns (150 K). The final state of prenucleation/nucleation shown in Figs. S6.
		
	}
	
	\label{PdLM}
\end{figure*}

%%%%%%%%%%%%%%%%%%%%%%%%%%%%%%
% S7
%%%%%%%%%%%%%%%%%%%%%%%%%%%%%
\begin{figure*}[hbtp]
	
	\includegraphics*[height=3.2cm,width=4.4cm,angle=0.]{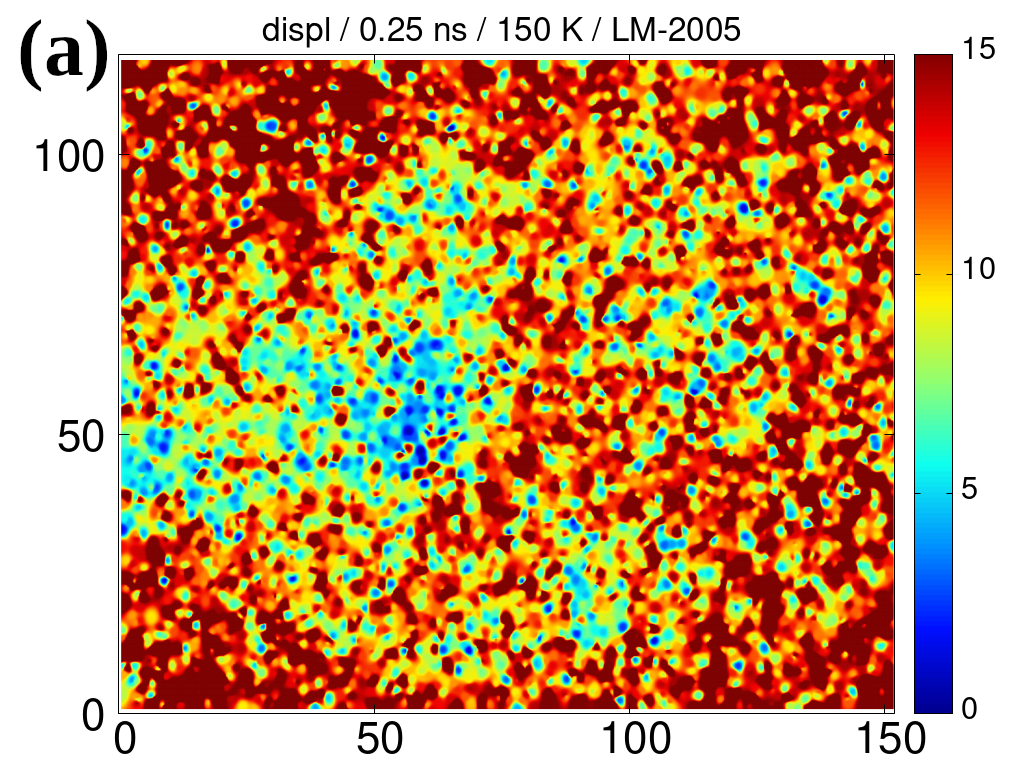}
	\includegraphics*[height=3.2cm,width=4.4cm,angle=0.]{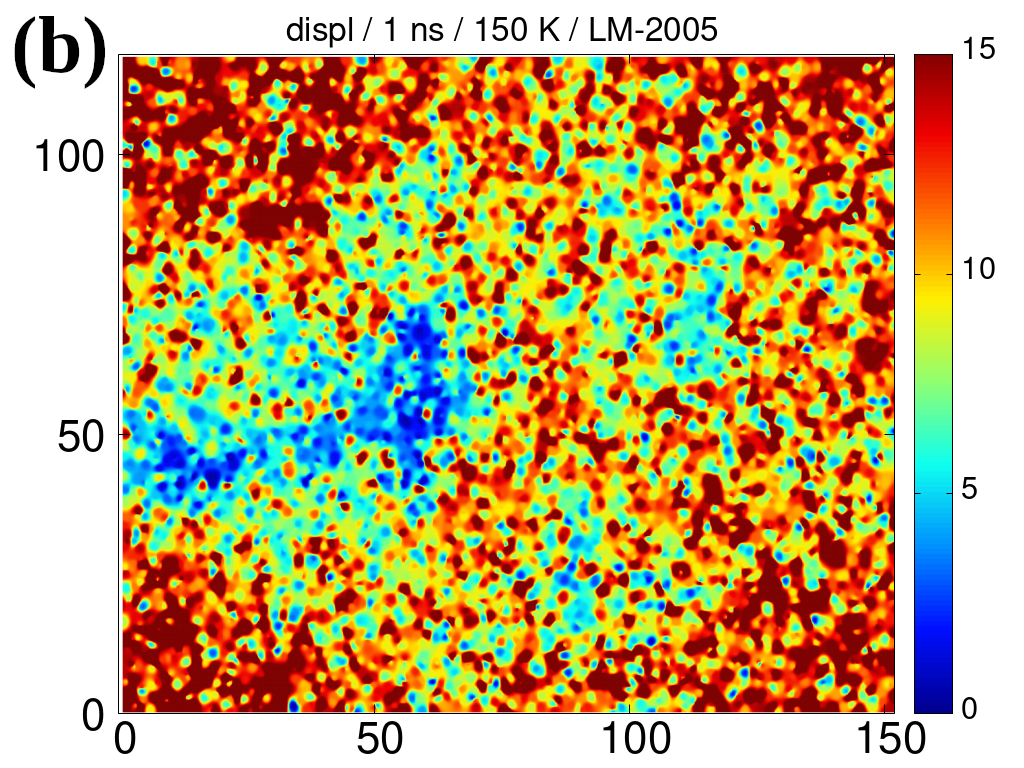}
	\includegraphics*[height=3.2cm,width=4.4cm,angle=0.]{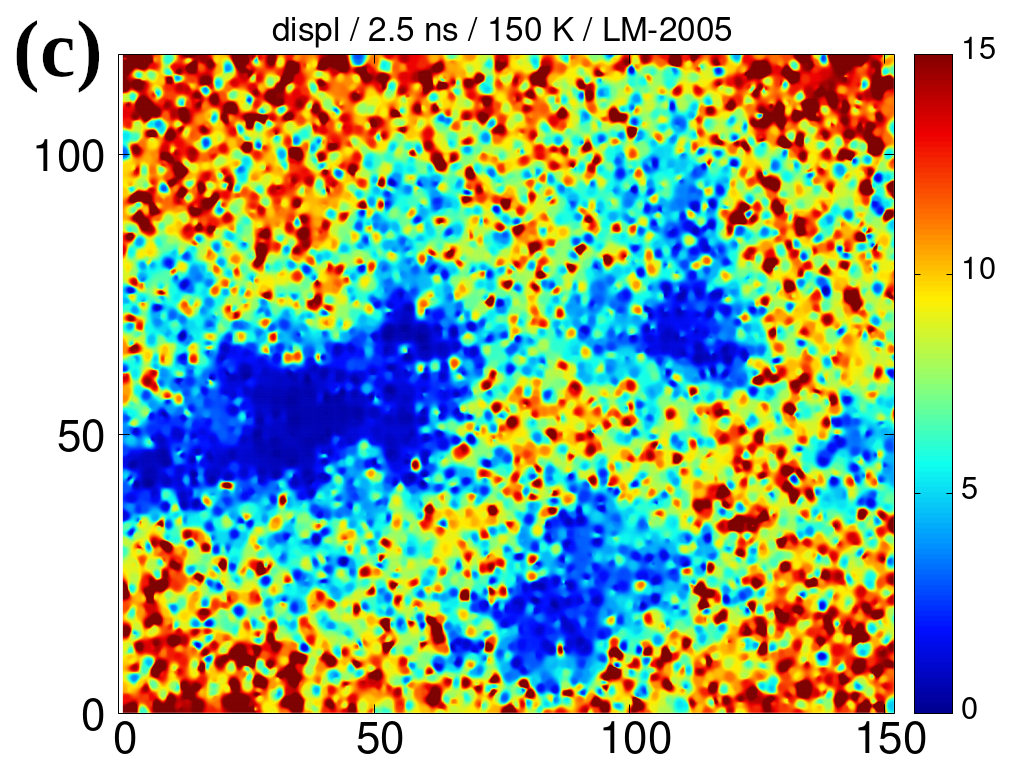}
	\includegraphics*[height=3.2cm,width=4.4cm,angle=0.]{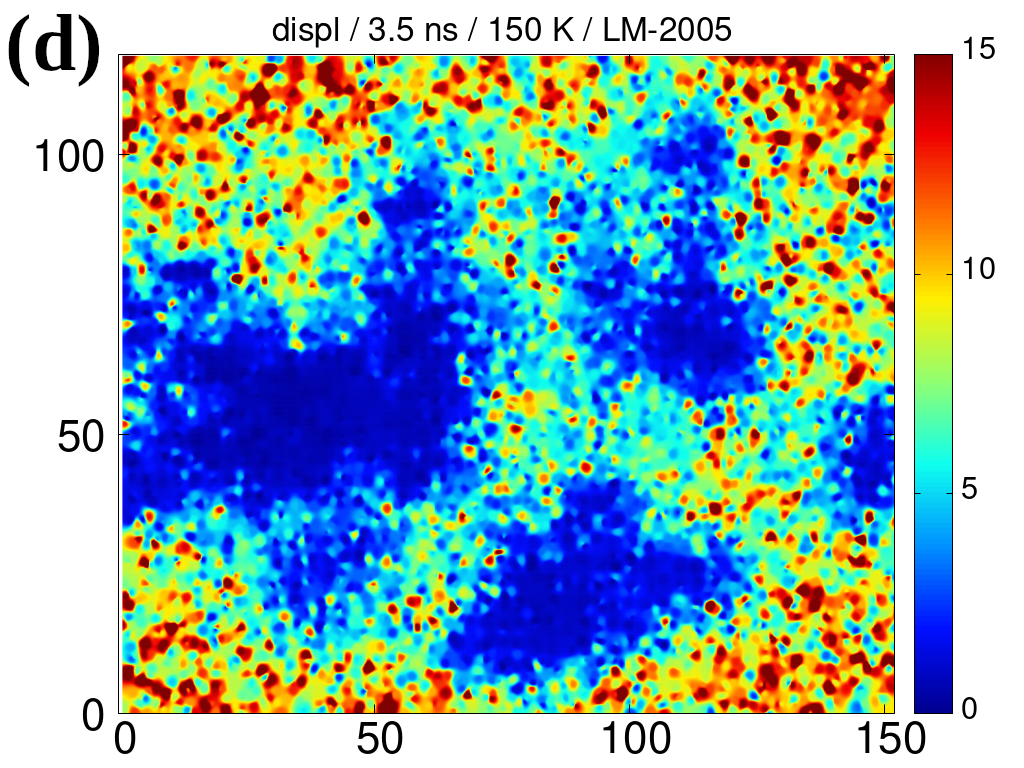}	
	
	\includegraphics*[height=3.2cm,width=4.4cm,angle=0.]{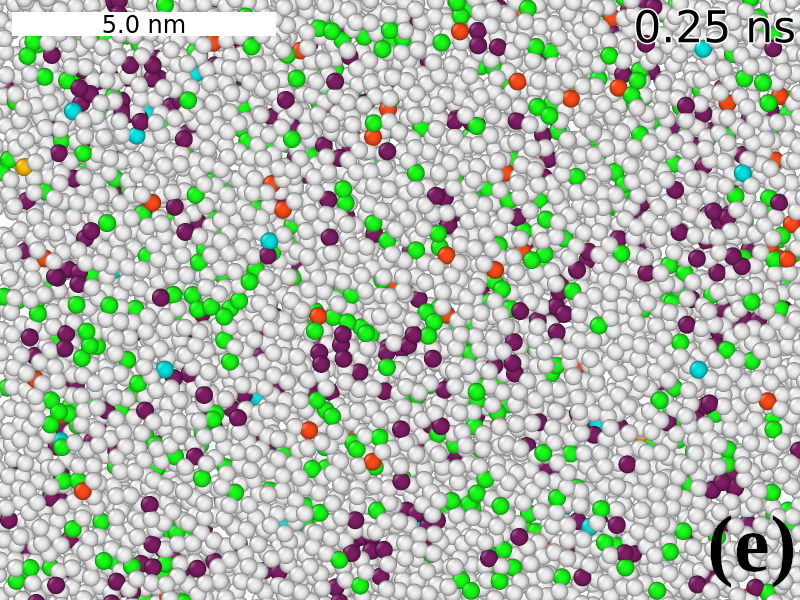}
	\includegraphics*[height=3.2cm,width=4.4cm,angle=0.]{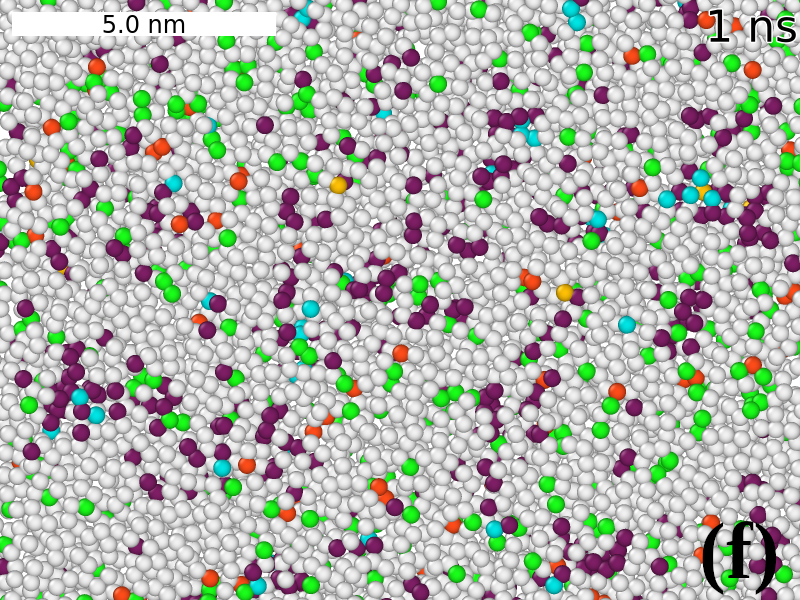}
	\includegraphics*[height=3.2cm,width=4.4cm,angle=0.]{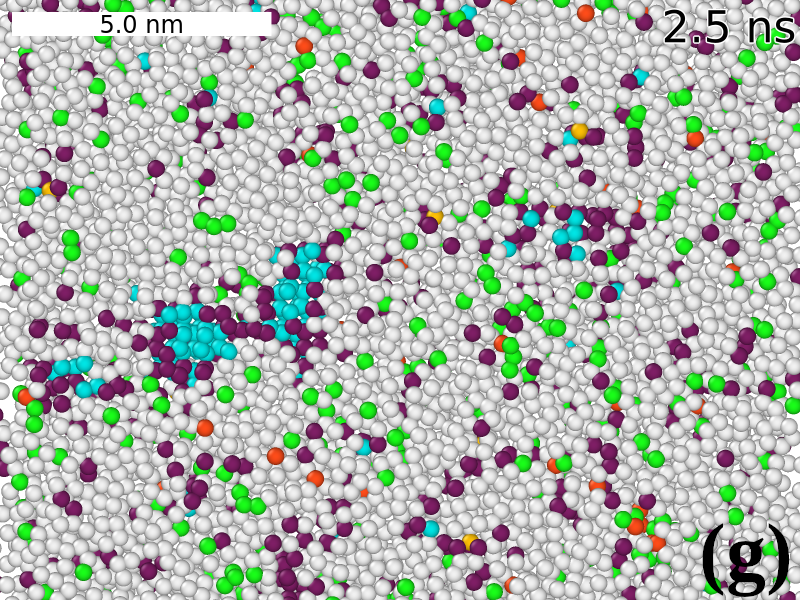}
	\includegraphics*[height=3.2cm,width=4.4cm,angle=0.]{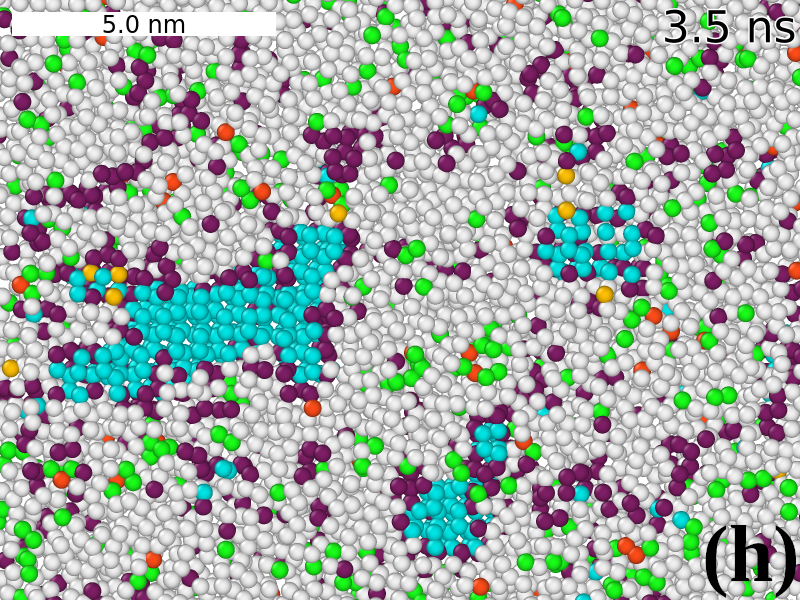}
	
	\includegraphics*[height=4.2cm,width=4.4cm,angle=0.]{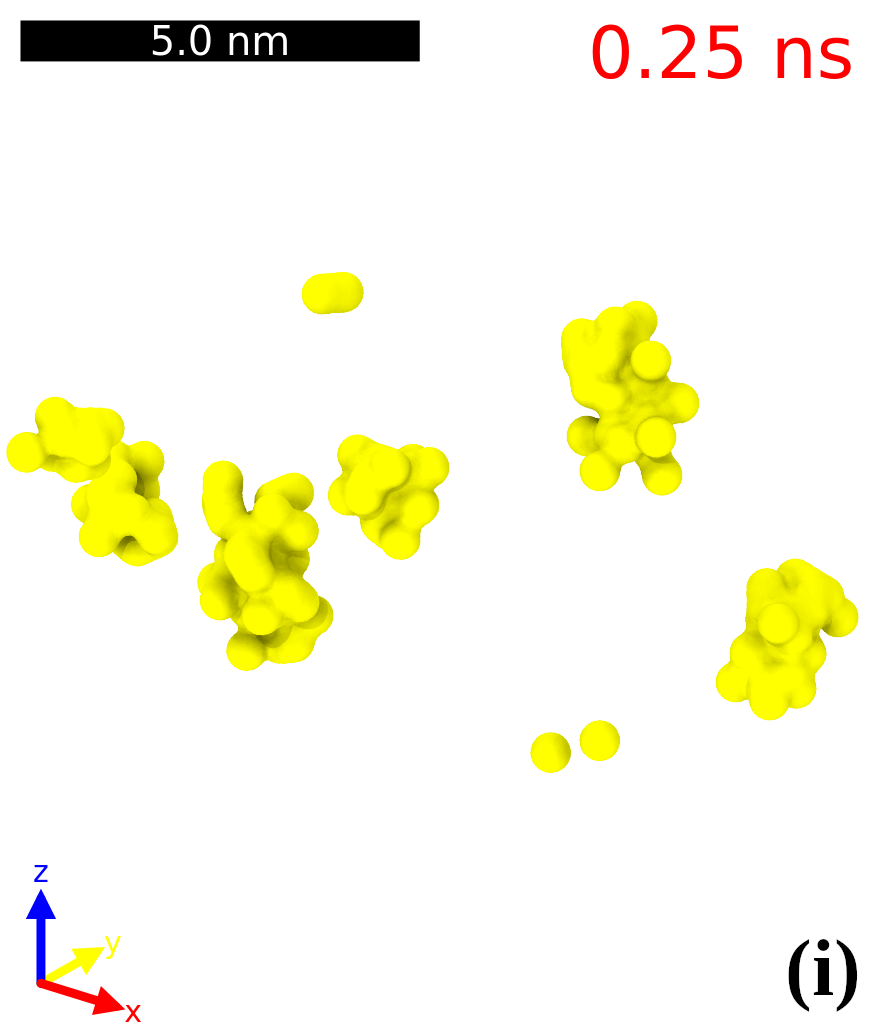}
	\includegraphics*[height=4.2cm,width=4.4cm,angle=0.]{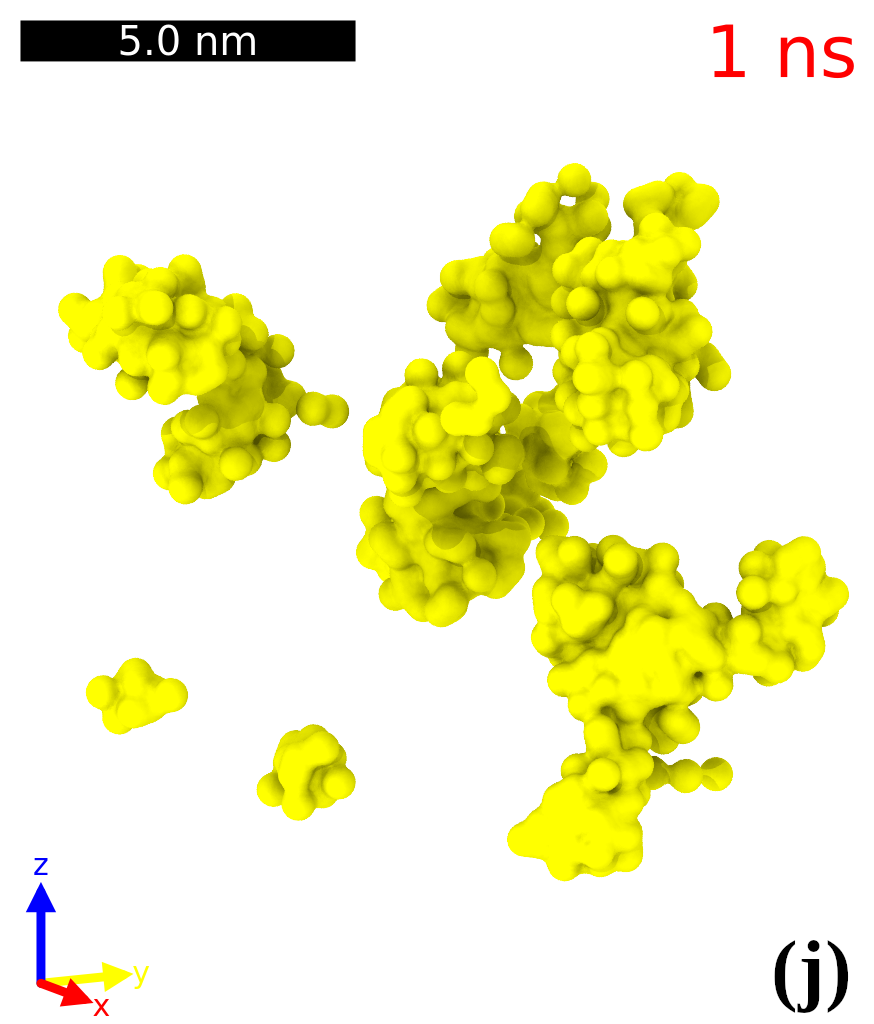}
	\includegraphics*[height=4.2cm,width=4.4cm,angle=0.]{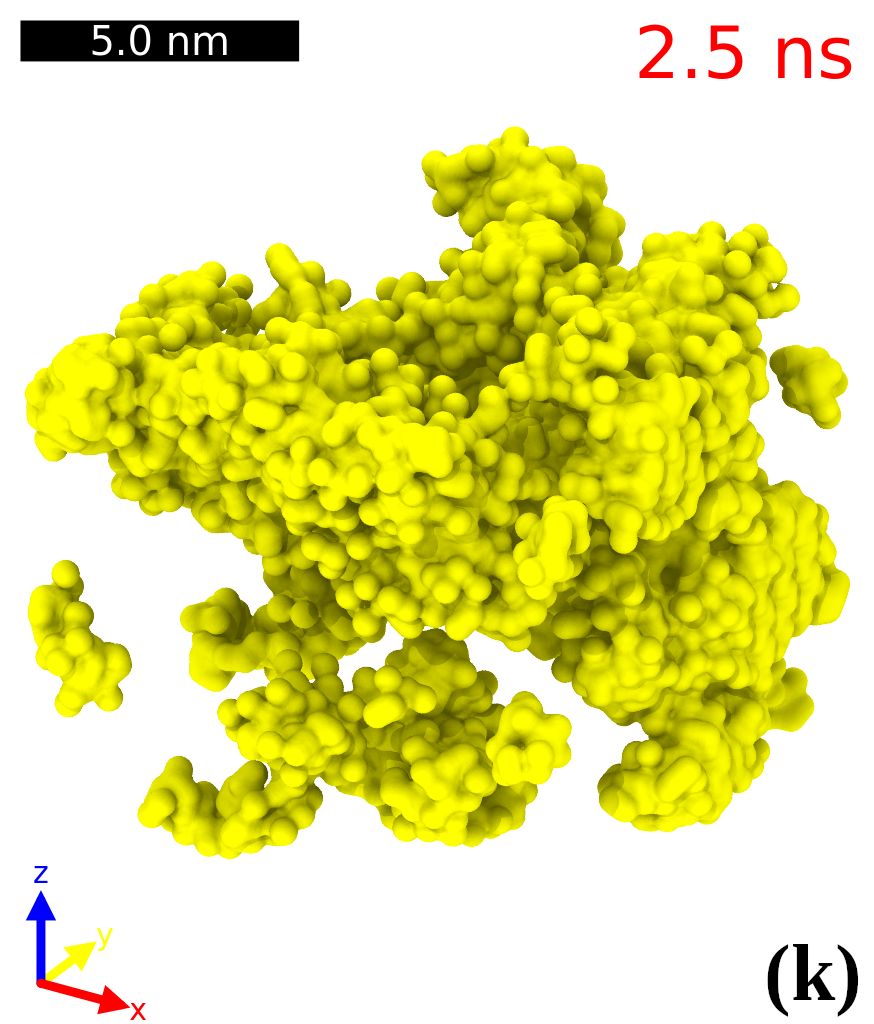}
	\includegraphics*[height=4.2cm,width=4.4cm,angle=0.]{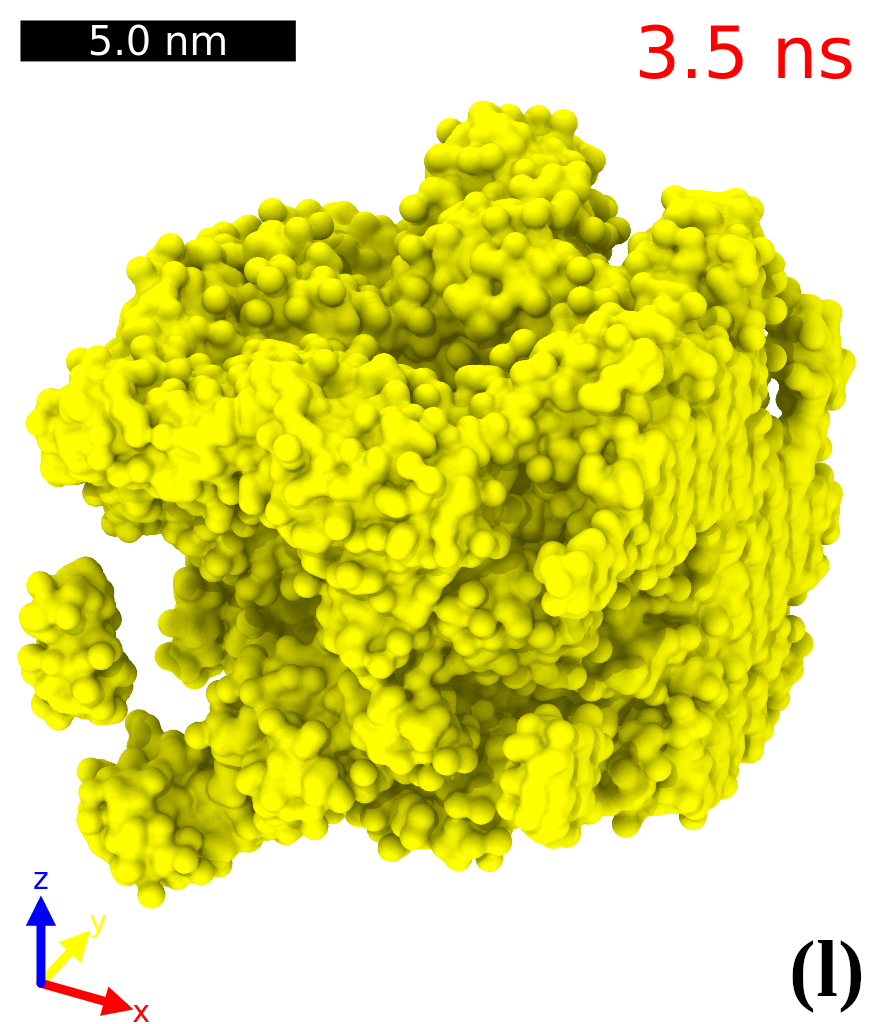}
	
	\caption[]{
		{\bf Spatially resolved displacement and chill+ maps during (pre)nucleation at 150 K (biased MD).}		
		The phase development of freezing has been followed at 150 K using the Lattice Mold technique (LM-TIP4P/2005) by MD.
		(a)-(d) Atomic displacements (\AA) with respect to the final nearly complete Ih phase (see at Fig. S11(c), reached at $\sim 7$ ns).
		(e)-(h) Corresponding panels with Chill+ plots as obtained by OVITO. Structure types: blue = Ih, red = hydrate, green = interfacial hydrate,  purple = interfacial ice.
		Thin ($\sim 1$ nm) slices cut along the diagonal of the simulation cell were used to display the various maps by gnuplot.
		(i)-(l) 3d view of the pre-nucleation cluesters (PNCs) at 0.25, 2, 2.5 and 3.5 ns.
		Clusters of trapped particles are shown only (atomic displacements to the final state at 7 ns is $d \le 2.5$ $\r{A}$, see at Fig. S11(c)).
		Polyhedral surface meshes (gaussian density method) were used to display the PNCs \cite{ovito,stuko}.
	}
	
	\label{dmap150K}
\end{figure*}

%%%%%%%%%%%%%%%%%%%%%%%%%%%%%%
% S7 (anno F6)
%%%%%%%%%%%%%%%%%%%%%%%%%%%%%
\begin{figure*}[hbtp]
	
	\includegraphics*[height=5.5cm,width=7cm,angle=0.]{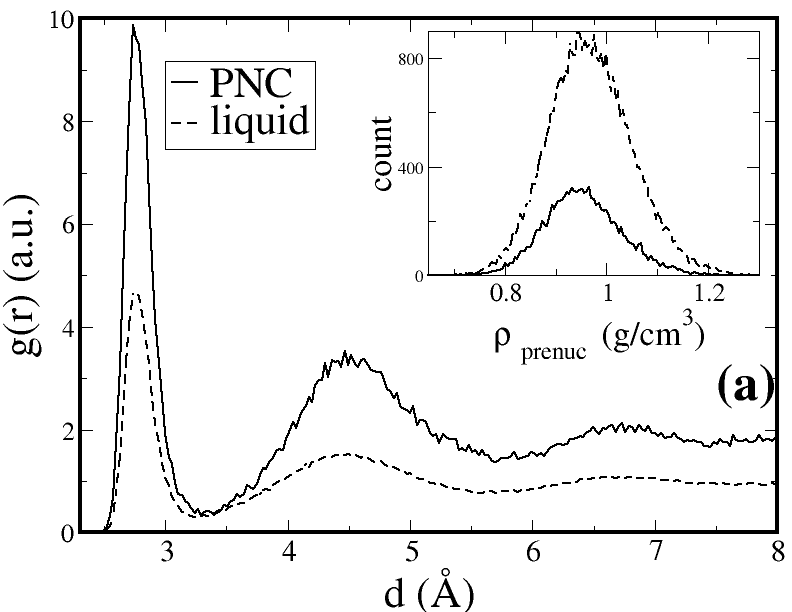}
	\includegraphics*[height=5.5cm,width=7cm,angle=0.]{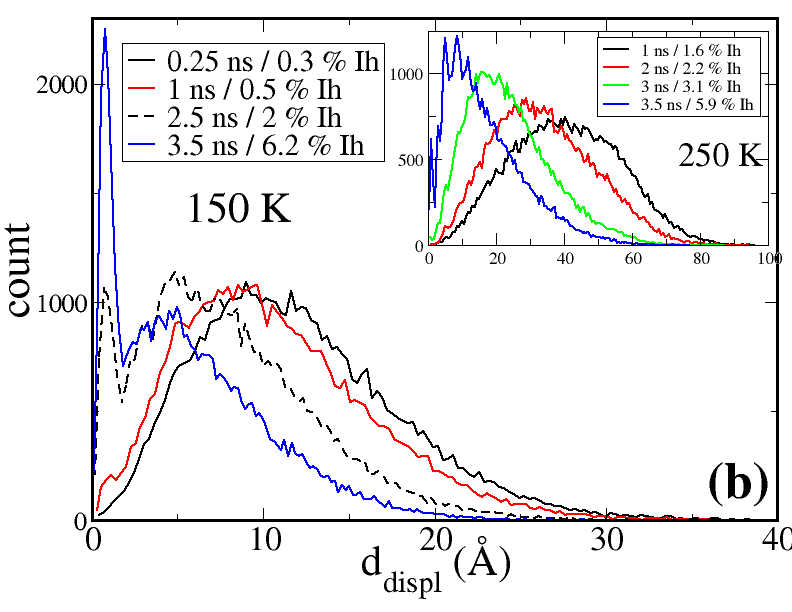}
	\includegraphics*[height=5.5cm,width=7cm,angle=0.]{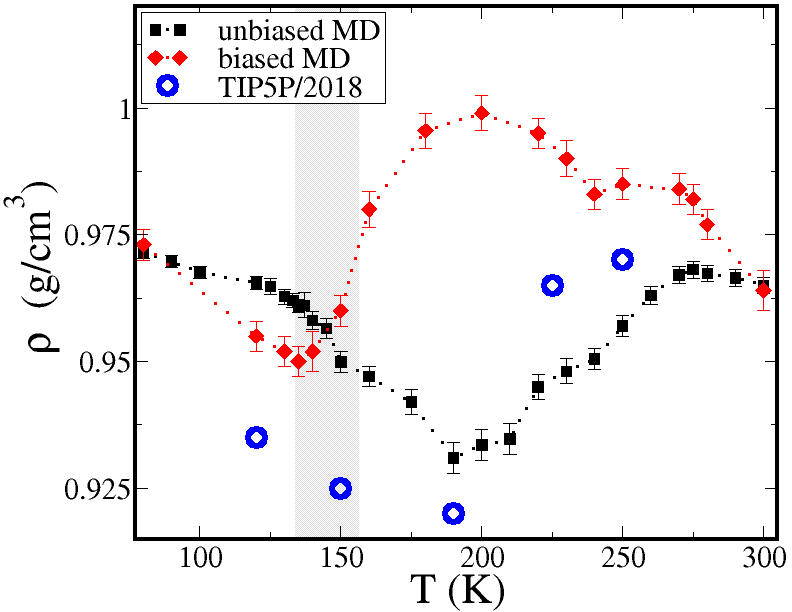}
	
	\caption[]{
		{\bf The radial density function g(r) of prenucleation cluster (PNC) at 150 K (a), the histogram of atomic displacements (b),
			The density ($\rho$) as calculated by MD simulations as a function of T (K) (c)	}
		(a) Biased MD: The OO g(r) for the PNC and its liquid environment. The PNC phase is obtained at 150 K by
		using filtering with respect to the atomic displacements (d-filter, $d_i \le 4$ $\r{A}$) referenced to the final Ih state at 2.5 ns.
		%		The PNC phase is shown in Fig. \ref{PNC}(f).
		The rest of the system is denoted as liquid which includes mostly mobile particles.
		Biased (LM) npt-MD is used TIP4P/2005.
		(a) {\em Inset}: 
		The histogram of the atomic density values as calculated using Voronoi analysis (g/cm$^3$) for the
		PNC (nearly immobile particles) and liquid phases, respectively. 
		(b) Biased MD: Atomic displacements ($d_i$) with respect to the final crystalline (Ih) state at 150 and 250 K (Inset).
		(c) The density is calculated by biased (red) and unbiased MD (black).
		biased MD (LM): the data points were sampled close to the end of the prenucleation phase where the proportion of the Ih ice phase is around $0.5$ \%
		(as detected by Chill+).
		Note the crossing point of the two curves at $\sim 145$ K is denoted by
		the shaded area. Few density values are also shown using the unbiased TIP5P/2018 \cite{tip5p2018} MD simulations (blue circles). 
		
	}
	
	\label{grhistodkappa150K}
\end{figure*}

%%%%%%%%%%%%%%%%%%%%%%%%%%%%%% 
% S8
%%%%%%%%%%%%%%%%%%%%%%%%%%%%%
\begin{figure}[hbtp]
	
	\includegraphics*[height=6.5cm,width=8cm,angle=0.]{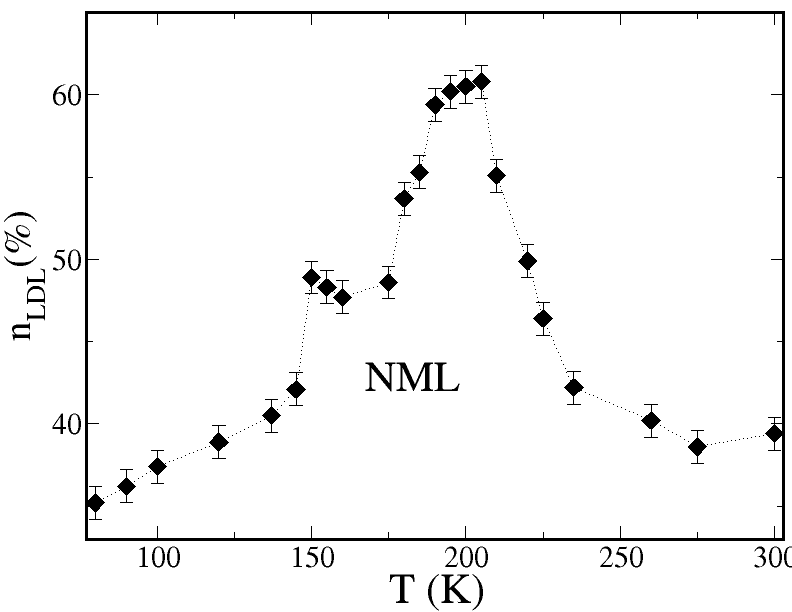}
	
	\caption[]{ 
		{\bf The proportion of the low density liquid (LDL) fraction ($n_{LDL}$, \%) as a function of the temperature (T). (unbiased MD)}
		The LDL fraction has been averaged for few ns npt runs and the Voronoi's atomic density criterium of	
		$0.95$ g/cm$^3$ is applied to separate the LDL and HDL fractions.
		The error bar is due to the fluctuation of $n_{LDL}$(t) as a function of the simulation time $t$.
	}
	
	\label{figS3}
\end{figure}

%%%%%%%%%%%%%%%%%%%%%%%%%%%%%%
% S9
%%%%%%%%%%%%%%%%%%%%%%%%%%%%%
\begin{figure*}[hbtp]
	
	\includegraphics*[height=5.5cm,width=6.8cm,angle=0.]{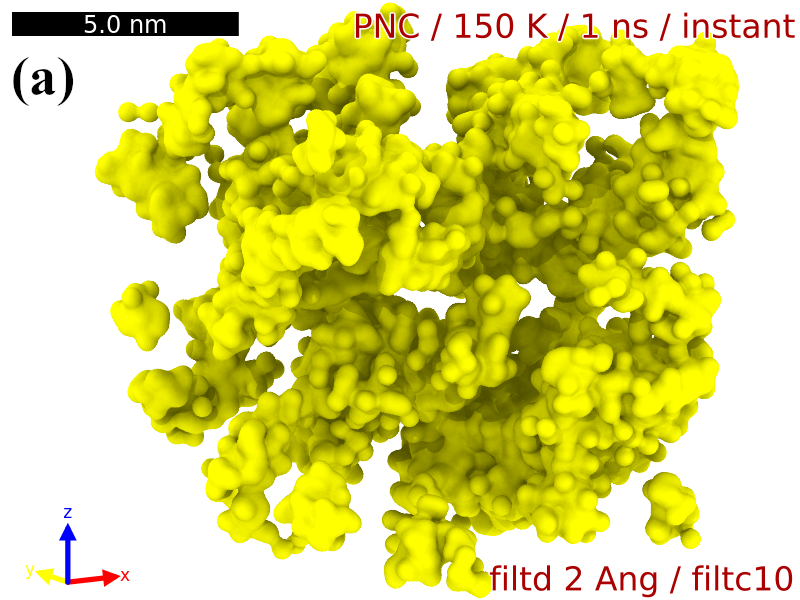}
	\includegraphics*[height=5.5cm,width=6.8cm,angle=0.]{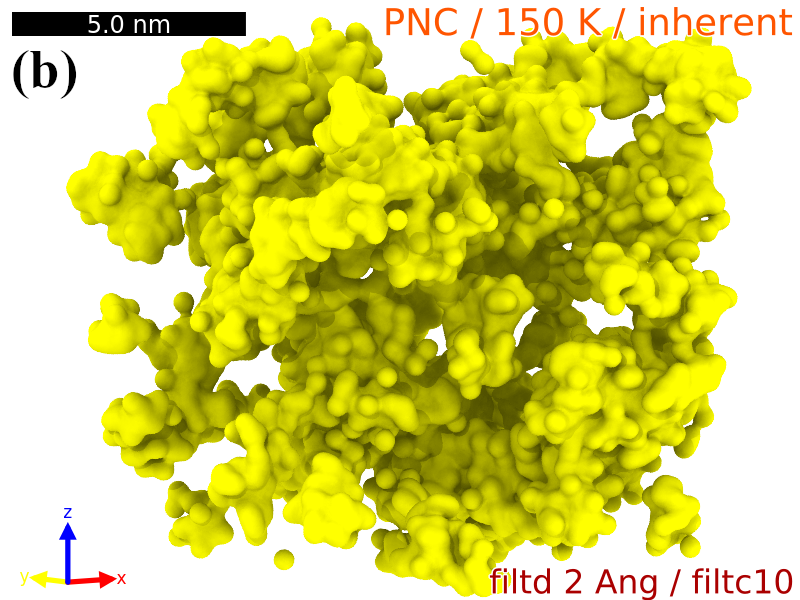}
	
	\caption[]{ 
		{\bf Comparison of network-like PNCs (arrested soft-skeleton).} PNCs as extracted from an instantaneous structure of Lattice Mold biased MD simulation using the TIP4P/2005 force field ((a), 150 K, at 1 ns)
		compared with its inherent counterpart (b).
		The inherent structure is obtained by hyperquenching the instantaneous structure to nearly 0 K using npt simulation
		in order to filter the thermal noise.
		The PNC phases in both cases are obtained 
		using double filtering (dc-filter) with respect to the atomic displacements (1st filter: d-filter, $d_i \le 2$ \r{A}) referenced to the final Ih state
		and a subsequent c-filter (with respect to the self-coordination of the d-filtered phase) is also used to get rid off dangling and undercoordinated atoms.
		Oxygen atoms are shown only. The structures are visually very similar indicating that the inherent structure barely perceptibly alters
		from the instantaneous structure.
		Structures were extracted from biased LM-MD-TIP4P/2005 simulation (a) and subsequent hyperquenching (b). 
	}
	
	\label{compareskeleton}
\end{figure*}

%%%%%%%%%%%%%%%%%%%%%%%%%%%%%%
% S11
%%%%%%%%%%%%%%%%%%%%%%%%%%%%%
\begin{figure*}[hbpt]
	
	\includegraphics*[height=7cm,width=9.5cm,angle=0.]{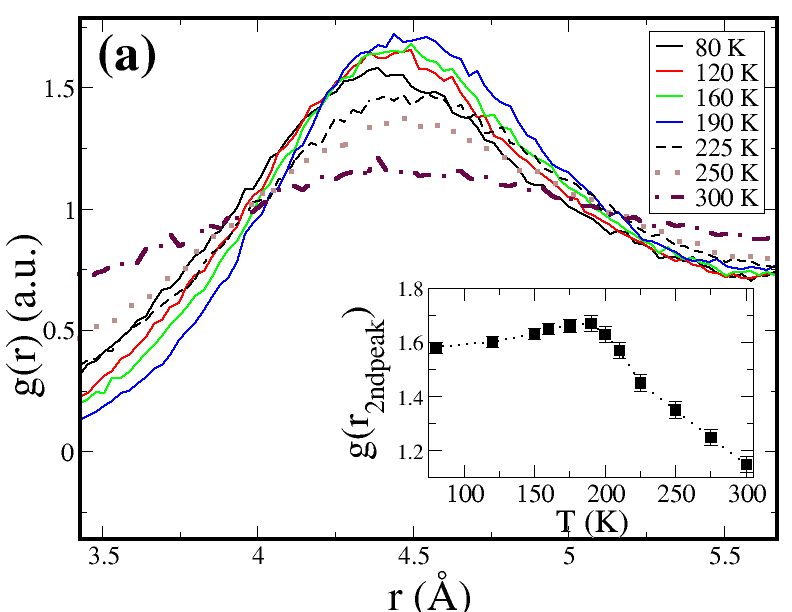}
	
	\includegraphics*[height=5cm,width=7.cm,angle=0.]{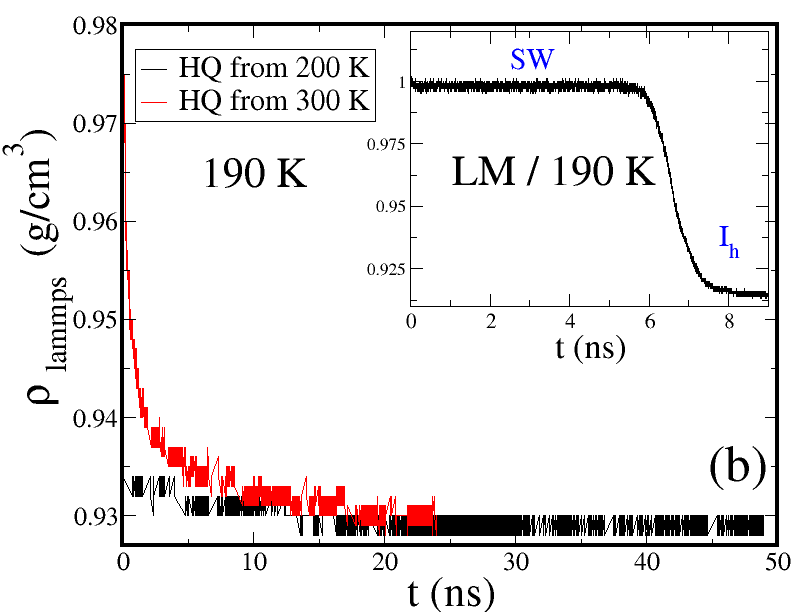}
	\includegraphics*[height=5cm,width=7.cm,angle=0.]{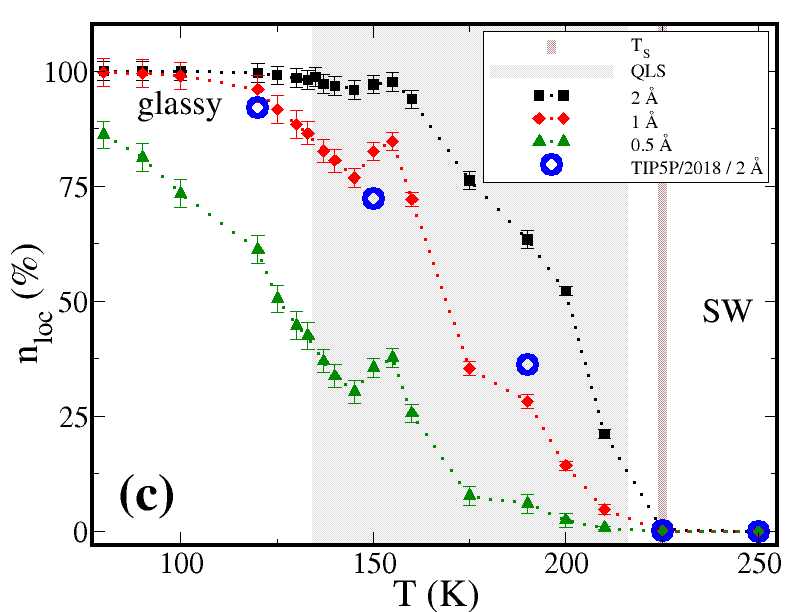}
	
	\caption[]{ 
		{\bf The broad 2nd peak of the O-O radial density function g(r) as a function of the distance r ($\r{A}$) (a).} 
		Brute force MD is used (unbiased MD). The g(r) has been timelapsed over 5 ns.
		{\em Inset:} The demonstration of the crossover point at $T \approx 190$ K. The height of the 2nd g(r) peak as a function of temperature T(K).
		Error bar is due to sampling events.
		(b) Unbiased MD: The simulated density $\rho_{lammps}$ as a function of the time $t$ (ns) using brute force MD and with Lattice Mold (LM, Inset) at 190 K.
		Using different initial states (relaxed liquids at 200 and 300 K) we demonstrate that the simulations lead to the same
		low density amorphous (glassy) state (LDA) (hyperquenched (HQ) from 200 and 300 K, respectively).  
		{\em Inset:} The $\rho_{lammps}$ as a function of t (ns) using LM technique. We initialize freezing at 190 K in this way.
		Note that the initial SW is a relaxed high density phase. 
		The cubic large scale systems were used with $49152$ water molecules. 
		(c) Unbiased MD: The fraction of immobile particles ($n_{loc}$, \%) as a function of temperature T(K)) using unbiased MD.
		$n_{loc}(T)$ has been calculated at different atomic displacement filtering criteria $d_{loc}$ (0.5, 1.0 and 2.0 $\r{A}$).
		Those atomic displacements ($d_i$) with respect to the final state are kept for which $d_i \le d_{loc}$. 
		Blue circles denote $n_{loc}(T)$ results coming from unbiased TIP5P/2018 \cite{tip5p2018} simulations $d_i \le 2.0$ $\r{A}$).
		The wide shaded area is the expected regime of QLS.
		Outside this regime $n_{loc}(T)$ decays to zero or reaches nearly 100 \%.
		The narrow vertical shaded area denotes the temperature regime ($T_s \approx 225$ K)
		where thermodynamic response functions diverge.
		Dotted line are for guiding eyes only.
	}
	\label{gr_grpeak}
\end{figure*}

%\begin{center}

%%%%%%%%%%%%%%%%%%%%%%%%%%%%%%
% S12 
%%%%%%%%%%%%%%%%%%%%%%%%%%%%%
\begin{figure*}[hbpt]
	
	\includegraphics*[height=7.5cm,width=9cm,angle=0.]{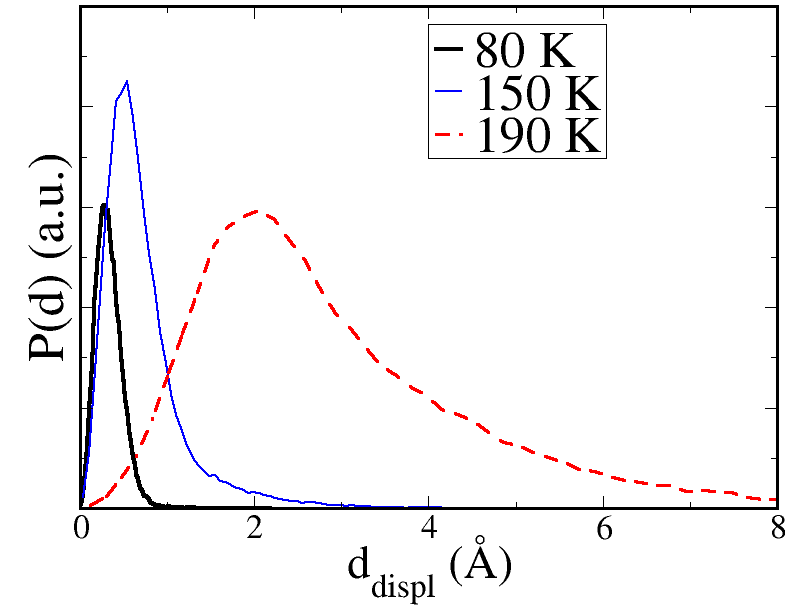}
	
	\caption[]{
		{\bf The distribution of atomic displacements (P(d)) at various temperatures.} 
		The histogram of atomic displacement distribution P(d) is calculated with respect to the initial state
		(equivalent with the self-part of the van Hove correlation function).
	}
	
	\label{S12}
\end{figure*}

%%%%%%%%%%%%%%%%%%%%%%%%%%%%%%
% S13
%%%%%%%%%%%%%%%%%%%%%%%%%%%%%
\begin{figure*}[hbpt]
	
	\includegraphics*[height=7.5cm,width=10cm,angle=0.]{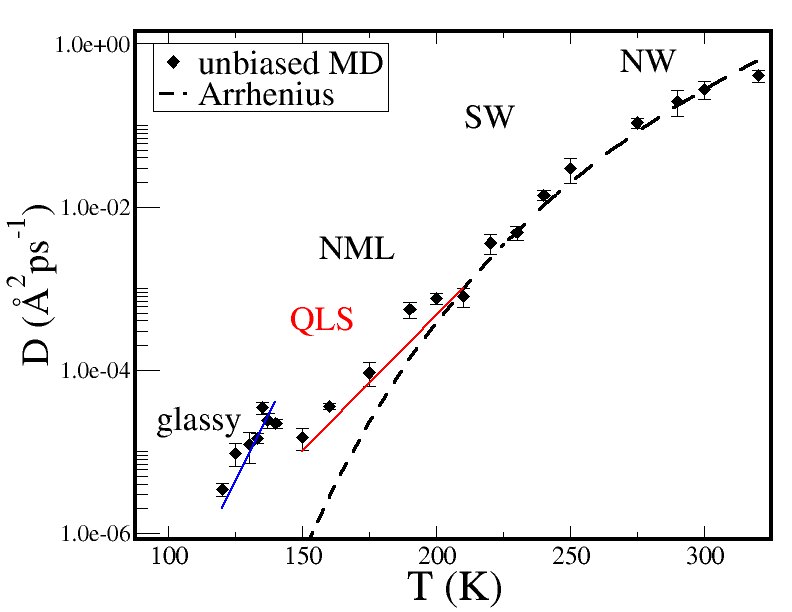}
	
	\caption[]{
		{\bf Calculated diffusion constant $D (\r{A}^2/ps)$.} Vertical axis in log scale as a function of the temperature (K). MD-LM: with Lattice Mold, unbiased MD: without LM (brute force MD). The various $D$ values are time averaged over a 1 ns period during the unbiased MD run.The fitslope approach is used which estimates the slope of the mean square of displacements from its two end-points, and uses it to compute the diffusion coefficient. The Arrhenius plot is also given (dashed line): $D=D_0exp(-E_a/RT)$, where $D_0=0.00143$ and $R=8.314$ J/(mol.K) is the gas constant. The activation energy is $E_a=32.8$ KJ/mol. 
		Lines denote the QLS (red) and glassy regimes (blue). NW: normal water, SW: supercooled water, NML: no mans's land, QLS: quasi liquid state, UV: ultra viscous state.
		
	}
	
	\label{D}
\end{figure*}

\subsection{Quasi liquid state}

The quasi-liquid state (QLS) represents a more general liquid phase that may exist across a wide range of supercooled liquids \cite{psule24}. But what exactly is the QLS ? Characterized as a distinct liquid state, it is neither glassy, spinodal, nor an ordinary liquid. Within the QLS, long-range correlations and collective behaviors are significantly amplified, while dynamical heterogeneity reaches its peak. Furthermore, the QLS exhibits a unique networked system where particle nano-clusters form a nanoporous superstructure. Perhaps its most intriguing feature is the occurrence of, synergistic, self-driven nanoconfinement, which distinguishes it from other liquid-like states. This mechanism culminates in an arrested state that persists for at least several microseconds in NML water. In contrast, other systems—such as deeply supercooled molten iron (QLI)— exhibit a much shorter QLS lifetime of approximately 150 picoseconds \cite{psule24}. Ultimately, further verification is required to determine whether the QLS is a widespread phenomenon in nature.

Distinguishing between a truly glassy state (e.g., below $T_g \approx 136$ K) and the glass-like state of no-man's land (NML) water is non-trivial due to the progressive transition that occurs with decreasing temperature. However, as demonstrated here, selected structural descriptors suggest that NML water is distinct from "ordinary" glassy water in several respects. Beyond differences in indicators derived from the radial distribution function (g(r) 2nd peak peaks in the NML, tiny shift in the 2nd peak position), several other factors support the anomalous nature of NML water (the nonlinear behavior of the isothermal compressibility, the self part of the van Hove correlation function (VHCF) is fat tailed (Fig. 6(c)), the sudden increase of the peak to height ratio of the VHCF (Inset Fig. 5). Beyond these signatures, the most intriguing is the spontaneous self-nanoconfinement of the liquid water within its own nanoporous, arrested superstructure—a feature entirely absent below the glass transition temperature or above $T_s \approx 225$ K. Within the glassy regime (below $T_g$ at ambient pressure), the system approaches a nearly homogeneous amorphous state. This is accompanied by a severe reduction in dynamical heterogeneity and a diminished population of liquid-like particles, which ultimately suppresses self-confinement. Thus, evidence increasingly supports the distinction of the LDL water in the NML between truly glassy behavior below $T_g$ and liquid-like supercooled water above $T_s$. To fully capture the distinction between the glassy and non-glassy behavior of NML water, however, we must look toward more advanced structural descriptors yet.

\subsection{Further details on methods}

{\em TIP4P/2005f:}
The flexible version of TIP4P/2005 (TIP4P/2005f \cite{tip4p2005f}) has also been employed for checking the importance of internal degrees of freedom of water molecules
in the development of various properties of supercooled water in the NML.
However, no significant deviation has been found with respect to the results obtained by the rigid TIP4P/2005 potential.
{\em TIP5P/2018:}
For comparative purpose we also employ the reoptimized version of the in principle more sophisticated five-point rigid potential TIP5P \cite{tip5p2018}.
This force-field provides results very similar to that of TIP4P/2005 for the densities in the NML.
By taking the density and diffusion dependence on temperature as target properties, significant improvements to the behavior of isothermal compressibility were achieved along with improvements to other thermodynamic and rotational properties \cite{tip5p2018}.
We argue that the 4 and 5 point models exhibit very similar features for the density and diffusion constants as well as for dynamic heterogeneity which
further support the observed anomalous dynamical behavior of deeply supercooled water. 

\subsection{Inherent structures}

Inherent structures (IS), obtained by quenching to 0 K, remove thermal fluctuations to reveal the underlying potential energy surface (PES). However, when studying temperature-dependent properties of a liquid characterized by rapidly fluctuating, nearly degenerate phases (such as LDL/HDL in water), IS can be misleading. Because the system exists mostly between minima at finite temperatures ($\le kT$), the physical picture is better represented by time-averaged instantaneous configurations rather than hyperquenched structures. Furthermore, energy minimization often artificially alters the structural composition (e.g., increasing Ice
fraction). Thus, we avoid IS for T-dependent studies, using them only for structural comparison.
The effect of hyperquenching from different temperatures on the time evolution of the density (unbiased MD) is shown in Fig. S11(b):
initializing supercooling leads basically to nearly the same quenched state.  
Furthermore, we find that the instantaneous and inherent structures are remarkably similar. (see Fig. S10(a) and (b)).

\subsection{Hippo potential}
We briefly summarize the most important aspects of the HIPPO potential which is in principle a more reliable
force field than the rigid water potential of TIP4P/2005.
Unfortunately at present yet the Lattice Mold approach does not work with HIPPO as implemented in LAMMPS.
Therefore we use it for comparison with unbiased TIP4P/2005 for SW without the initialization of nucleation.
In particular, we compare the calculated densities as a function of temperature to demonstrate that 
a simple density shift can effectively account for the difference between the models \cite{Espinoza23}.
Although TIP4P/2005 tends to underestimate the density, however, it is still an effective model to simulate
systems with a more realistic size. 

The HIPPO (Hydrogen-like Intermolecular Polarizable POtential) water model, which is a polarizable force field for water molecules derived from a simple, model electron density. This model aims to accurately represent the many-body effects and electronic polarizability in water, leading to a balanced description of gas-phase clusters, liquid water, and ice properties across various temperatures and pressures \cite{hippo}. 
HIPPO incorporates several key energy terms: Charge Penetration (Damped permanent electrostatics), 
Polarization (Damped polarization effects), 
Charge Transfer, Pauli Repulsion and Dispersion (Accounts for van der Waals interactions). 
Parameterization:
Initial parameters were fitted to Symmetry Adapted Perturbation Theory (SAPT) energy components, then refined for bulk phases of water. 
Accuracy:
The model achieves comparable or superior accuracy to other polarizable force fields for properties of water in different phases. 
HIPPO provides a physics-based model that rationalizes water's structure, dynamics, and thermodynamics by correlating them with ab initio energy decomposition.

The HIPPO water model
provides a balanced representation of a wide variety of properties of gas phase clusters, liquid water, and ice polymorphs, across a
range of temperatures and pressures. This water potential yields a rationalization of water structure, dynamics, and thermodynamics
explicitly correlated with an ab initio energy decomposition, while providing a level of accuracy comparable or superior to previous
polarizable atomic multipole force fields.

\subsection{Density calculations} 
Ordinary lammps estimate based on the varying npt volume as printed out as a lammps thermo output. 
Voronoi's cell approach, based on atomic volumes:
This analysis calculates the Voronoi tessellation of the simulation box, taking the particle positions as Voronoi cell centers. By default two quantities are output by the analysis for each particle: The volume of the particle's Voronoi cell and the number of faces the Voronoi cell has. Since the face count is equal to the number of nearest neighbors of a particle, this quantity is also called the Voronoi's coordination number \cite{ovito}.
We use oxygen-focused analysis:
It is common to focus the analysis on the oxygen atoms, as they form the main hydrogen bond network that dictates the local structure of water.
The system averaged Voronoi's density is calculated as the total mass of the simulation cell is divided by the sum of atomic ployhedral volumes.
% The modifier outputs both quantities as new particle properties, which allows them to be referenced in subsequent analysis or filtering steps \cite{ovito}.
Voronoi analysis is a suitable and effective method for separating low-density and high-density phases in water, as demonstrated by its use in characterizing structural heterogeneities, identifying low-density patches, and analyzing the density-dependent properties of liquid water and its structural changes. By defining regions of space around each water molecule, Voronoi tessellation reveals density fluctuations and local structural characteristics that are not apparent in traditional methods like radial distribution functions. 
A Voronoi diagram divides space into cells (called Voronoi polyhedra in 3D) around each water molecule, where each cell contains the points closest to its central molecule than to any other. 
The volume of these Voronoi cells directly relates to the local density. 
Low-density regions: are characterized by larger Voronoi cells and "empty voids". 
High-density regions: are associated with smaller Voronoi cells. 
Analyzing the distribution of cell volumes allows for the identification of clusters, both high-density and low-density. 
It captures higher-order structural information than average quantities or two-point correlation functions, revealing details about hydrogen bonding and the formation of local water structures like LDL and HDL \cite{JedlovszkyJCP25}.

\subsection{Lattice Mold technique \cite{Mold,LM14,LM_Espinoza16,LMVega,LM,LM23}}

In LM, no traditional order parameters are used, the reaction coordinate to drive
crystallization consists of a mold of potential wells (e.g. the Ih or cubic ice structure)
that promotes prenucleation and freezing.

The interactions between the particles of the system (water) and the “virtual ” of the mold is
modeled by a square-well (SW) potential (a continuous version of
it to obtain forces analytically) that induces the ordering of water
molecules into a solid structure (Ih ice in this case).

In our case we use a full volume mold instead of a usual smaller spherical like mold seed \cite{Mold,LM14,LM_Espinoza16,LMVega,LM,LM23}.
The advantage of this setup is that it boosts nucleation homogeneously
in the entire volume.
The mold is kept fixed during the entire simulation.
Only Oxygen atoms are considered in the Mold.
In practice, for the
TIP4P/2005 model, the imposed square-well potential is responsible for the effective interaction between oxygen
atoms and the mold wells, while hydrogen atoms have no interaction with
the wells and spontaneously arrange into the correct orientation
of the lattice.
Since we do not calculate nucelation rate, it is not necessary to adjust
mold parameters as accurately as it has been done e.g. in ref \cite{LM14}.
Our intention is not to account for experimental quantities, such as the nucleation rate,
instead we focus on the details of the mechanism of prenucleation and freezing.

LM simulations are performed in the NPT
canonical ensemble. However, we kept fixed the crystal lattice positions of the
mold along the calculations, and we let the density of the system to evolve over time at the temperature and pressure (1 bar in
this work) of interest.
%We make the simulation box large enough so
%that the emergence of the precritical nuclei and/or  barely alters the desired
%pressure.

Although it is advised in refs. \cite{LM_Espinoza16,LMVega} to chose the smallest mold that results in complete crystallization
after an induction time that is computationally affordable we 
employ full volume mold (FVM). By applying this protocol, we gurantee reversibility in prenucleation.
The prenucleation cluster induced
by the mold cannot be sustained in the absence of potential wells and
quickly dissolves once the interaction between the mold and the liquid molecules is switched off.
The main advantage of FVM is that the boosting effect of it on nucleation is homogeneous
across the full volume.
Applying smaller spherical molds affect mostly directly the immediate environment of the mold
leading to preferential (inhomogeneous) clustering in that region within the simulation cell.
Similar artifact is imposed in seeded simulations \cite{Espinoza13}.
The FVM allows the simulation of completely homogeneous nucleation with randomly generated prenucleation clusters within the full volume of the simulation cell.

However, the
minimization of the imposed constraint across the transition state is not necessarily guranteed.
Since in LM simulations the position
and orientation of the mold are imposed, this constraint must in principle 
affect the height of the barrier corresponding to the transition state
of crystallization. In this way the LM can affect that which
reaction pathway will be energetically favorable.
Keeping this in mind, however we attempt to follow phase development in SW using this constraint.
The explored prenucleation mechanism is reasonable and does not contradict prevailing theories.

The
interaction between the $j^{th}$ wells and the $i^{th}$ particles
can be approximated by the following square-well potential:

\be
V_{sw}(r_{iw_j})=-\frac{1}{2} \epsilon_D \bigg[ 1-tanh \bigg(\frac{r_{iw_j}-r_w}{\alpha} \bigg) \bigg],
\ee

where $\epsilon_D$ (in $kT$, where $k$ is the Boltzmann constant and $T$ is the temperature in K), $r_w$ (\r{A}) the depth and the radius of the wells
respectively and $\alpha$ controls the steepness of the well's walls.
These are the only adjustable parameters of the method.
This potential is continuous and differentiable and can therefore be used in MD.
We use $\epsilon_D=1.0$ ($kT$), $r_w=1.3$ \r{A} and $\alpha=0.017$ (\cite{LM14}).
Well potential is cutoffed at $6*r_w$.
These are properly optimized values at $T=150 $ K. The choice of $\epsilon_D=1.0$ $kT$ controls nucleation in such a way that we are able to follow
prenucleation for nanoseconds. Further decrease of this parameter slows down nucleation which makes freezing unfeasible. 
We follow ref. \cite{LM14} and set $r_w=1.3$ \r{A}.
According to our test results, the influence of $r_w$ on the final results is negligible within the range of $r_w=[1.0,2.0]$ \r{A}.

{\em Well occupancy test:}
Our aim is to minimize the effect of the FVM as an artificial "external potential" on prenucleation.
Therefore we select the parameter setup of $D=1.0$ ($kT$), $r_w=1.3$ \r{A}.
The choice of the width of the square well of $r_w=1.3$ $\AA$ although leads to underpopulated
well (at nearly half filling, $\sim 55$ \%), however, in this way nucleation sets in within few nanoseconds, while we can keep
the depth of the well ($D$) at the minimal level of the thermal motion ($kT$).
At around half filling of the well together with the well depth of $\sim kT$ we are able to "float" the water molecules between the well and
the inter-well regions allowing the system long enough sampling time for the prenucleation.
Using the conditions of $\epsilon_D \approx kT$ and nearly half filling of the wells allow us to avoid the undesirable premature trapping of water molecules in the wells.
On the other hand the onset time of few nanoseconds is optimal for our purpose, e.g. we can follow prenucleation
for a sufficiently long time which is still computationally affordable.

Nevertheless, the occupancy test of the well given in Eq (1) has also been done (not shown).
According to these test results, nearly $\sim 100$ \% occupancy requires the increase of $D$ and/or $r_w$,
e.g. stronger interaction of water molecules with the well, well width and/or the increase of the incubation time
beyond the affordable level.
Regardless of the well width chosen in the regime of $[1.0,2.0]$ \r{A} we are able to initialize nucleation, however,
the average onset time of nucleation could be sensitive to the choice of $\epsilon_D$ and/or $r_w$.
The more precise tuning of these parameters is beyond the scope of the present paper.

\subsection{Isothermal compressibility:}
The isothermal compressibility $\kappa_T$ is defined as:
\be
\kappa_T= \frac{1}{\rho} \left( \frac{\partial \rho}{\partial p} \right)_T
\label{kappa}
\ee
where $\rho$ is the density of the system.
It is related to volume fluctuations in the isothermal-isobaric ensemble:
\be
\kappa_T=\frac{\langle V^2 \rangle-\langle V \rangle^2}{k_B T \langle V \rangle}
\label{kappafluct}
\ee
where $\langle V \rangle$ is the average volume of the system and $k_B$ is the Boltzmann constant.
The fluctuation formula of Eq. \ref{kappafluct} has been used to estimate temperature dependence of $\kappa_T$
using the TIP4P/2005 force field.
In principle rather long simulations are needed to reduce statistical errors \cite{VegaMolPhys24,NilssonPCCP19}.
In our case we find that for a relatively large system with 49152 water molecules it is sufficient to
equilibrate until few tens of ns.
Unfortunately more accurate sampling is beyond the available computational resources.
The observed trends, especially the drop in $\kappa_T$ in the no mans's land is carefully checked in order to 
avoid possible artifacts. 

In Fig. 3 the calculated $\kappa_T$ values are shown as a function of the temperature.
The most striking feature is the sudden drop of $\kappa_T$ below $T_c \approx 228$ K. 
The crossover point takes place at around $190$ K and with abruptly increasing temperature
$\kappa_T$ peaks at $\sim 150-160$ K surprisingly. This peak has not appeared for much smaller simulation box including 2592 water molecules.
Therefore $\kappa_T$ seems to be strongly size dependent \cite{VegaMolPhys24,NilssonPCCP19}.
$\kappa_T$ drops again upon further decreasing T reaching the next crossover point at $\sim 135$ K (ultraviscous state).
The two minima in $\kappa_T$ should in principle be linked to glassy states with decreased degree of compressibility \cite{VegaMolPhys24,NilssonPCCP19}. 
However, we argue that the minimum of $\kappa_T$ at $\sim 190$ K is not necessarily associated with a glassy state.
Water can form glassy states in the "no man's land" if cooled fast enough to prevent crystallization, although the exact transition is still debated \cite{Gallo16,Amann-Winkel16,DebenedettiPT03,SciortinoPNAS17,glass,HeJCP07,Eltareb24}.

%%%%%%%%%%%%%%%%%%%%%%%%%%%%%%%%%%%%%%%%%%%%%%%%%%%%%%%%%

\end{document}